\documentclass[longauth]{aa}  
\usepackage{graphicx}
\usepackage{caption}
\usepackage{txfonts}
\usepackage{booktabs}
\usepackage{color}
\usepackage[dvipsnames]{xcolor}
\usepackage{natbib,twoopt}
\usepackage{lscape}
\usepackage{caption}
\usepackage{subcaption}
\usepackage[symbol]{footmisc} 
\usepackage{tablefootnote}
\usepackage{multirow}
\usepackage[bookmarks=true, pdfnewwindow=true, colorlinks=true, allcolors=cyan, breaklinks=true, final=true, citecolor=cyan]{hyperref}
\usepackage{orcidlink}
\bibpunct{(}{)}{;}{a}{}{,} 
\newcommand{\gpcc}{g~cm$^{-2}$}
\newcommand{\massrate}{$M_{\odot}$\,yr$^{-1}$}

\newcommand{\msun}{\mbox{$M_\odot$}}
\newcommand{\lsun}{\mbox{$L_\odot$}}

\newcommand{\kms}{\mbox{km~s$^{-1}$}}
\newcommand{\jybeam}{\mbox{Jy~beam$^{-1}$}}
\newcommand{\mjybeam}{\mbox{mJy~beam$^{-1}$}}
\newcommand{\ujybeam}{\mbox{$\mu$Jy~beam$^{-1}$}}

\newcommand{\degree}{\mbox{$^{\circ}$}}

\begin{document}

\title{Magnetic Fields in Massive Star-forming Regions (MagMaR). VII.}

\subtitle{On the dynamical importance of B-fields in massive protocluster W33~A}

%

\author{Fengwei Xu \inst{\ref{mpia},\ref{kiaa}, \thanks{fengweilookuper@gmail.com, fengwei@mpia.de (PFE fellow)}} \orcidlink{0000-0001-5950-1932} \and 
Q. Zhang\inst{\ref{cfa}} \orcidlink{0000-0003-2384-6589} \and
P. Sanhueza\inst{\ref{utda}} \orcidlink{0000-0002-7125-7685} \and
K. Wang\inst{\ref{kiaa}} \orcidlink{0000-0002-7237-3856} \and
Hauyu Baobab Liu\inst{\ref{nsysu},\ref{cag}} \orcidlink{0000-0003-2300-2626} \and
H. Beuther\inst{\ref{mpia}} \orcidlink{0000-0002-1700-090X} \and
Wenyu Jiao\inst{\ref{shao}} \orcidlink{0000-0001-9822-7817} \and
C. Wang\inst{\ref{naoc}} \and
P. C. Cort\'es\inst{\ref{jao}, \ref{nrao}} \orcidlink{0000-0002-3583-780X} \and
P. M. Koch\inst{\ref{asiaa}} \orcidlink{0000-0003-2777-5861} \and
J. M. Girart \inst{\ref{ice}, \ref{ieec}} \orcidlink{0000-0002-3829-5591} \and
M. T.\ Beltr\'an\inst{\ref{inaf}} \orcidlink{0000-0003-3315-5626} \and
J.-W. Wang\inst{\ref{eao}} \orcidlink{0000-0002-6668-974X} \and
J. Liu\inst{\ref{nju},\ref{klnju}} \orcidlink{0000-0002-4774-2998} \and
F. A. Olguin\inst{\ref{cfgp},\ref{naoj}} \orcidlink{0000-0002-8250-6827} \and
Xing Lu\inst{\ref{shao}, \ref{sklr}} \orcidlink{0000-0003-2619-9305} \and
S. Li\inst{\ref{nju},\ref{klnju}} \orcidlink{0000-0003-1275-5251} \and
Pak Shing Li\inst{\ref{shao}} \orcidlink{0000-0001-8077-7095} \and
T. Liu\inst{\ref{shao}} \orcidlink{0000-0002-5286-2564} \and
K. Morii\inst{\ref{cfa}} \orcidlink{0000-0002-6752-6061} \and
J. Hwang\inst{\ref{kuias}, \ref{kudeps}} \orcidlink{0000-0001-7866-2686} \and
H.-R. V. Chen\inst{\ref{nthu}} \orcidlink{0000-0002-9774-1846} \and
S. Jiao\inst{\ref{naoc}, \ref{mpia}} \orcidlink{0000-0002-9151-1388} \and
Y. Cheng\inst{\ref{naoj}} \orcidlink{0000-0002-8691-4588} \and
Q. Luo\inst{\ref{utia}, \ref{utda}, \ref{cfa}} \orcidlink{0000-0003-4506-3171} \and
Piyali Saha\inst{\ref{asiaa}, \ref{naoj}} \orcidlink{0000-0002-0028-1354} \and
Ji-hyun Kang\inst{\ref{kasi}} \orcidlink{0000-0001-7379-6263} \and
C. Y. Law\inst{\ref{inaf}} \orcidlink{0000-0003-1964-970X} \and
L. K. Dewangan\inst{\ref{aad}} \orcidlink{0000-0001-6725-0483} \and
O. R. Jadhav\inst{\ref{aad}, \ref{iit}} \orcidlink{0009-0001-2896-1896} \and
E. J. Chung\inst{\ref{kasi}} \orcidlink{0000-0003-0014-1527} \and
Chakali Eswaraiah\inst{\ref{iis}} \orcidlink{0000-0003-4761-6139} \and
Luis A. Zapata\inst{\ref{irya}} \orcidlink{0000-0003-2343-7937}
} 

\institute{
\label{mpia} Max Planck Institute for Astronomy, Königstuhl 17, 69117 Heidelberg, Germany \and
\label{kiaa} Kavli Institute for Astronomy and Astrophysics, Peking University, Beijing 100871, PR China \and
\label{cfa} Center for Astrophysics | Harvard \& Smithsonian, 60 Garden Street, Cambridge, MA 02138, USA \and
\label{utda} Department of Astronomy, Graduate School of Science, The University of Tokyo, 7-3-1 Hongo, Bunkyo-ku, Tokyo 113-0033, Japan. \and
\label{nsysu} Department of Physics, National Sun Yat-Sen University, No. 70, Lien-Hai Road, Kaohsiung City 80424, Taiwan, ROC \and
\label{cag} Center of Astronomy and Gravitation, National Taiwan Normal University, Taipei 116, Taiwan \and
\label{shao} Shanghai Astronomical Observatory, Chinese Academy of Sciences, 80 Nandan Road, Shanghai 200030, People's Republic of China \and
\label{naoc} National Astronomical Observatories, Chinese Academy of Sciences, Beijing 100101, PR China \and
\label{jao} Joint ALMA Observatory, Alonso de C\'ordova 3107, Vitacura, Santiago, Chile \and
\label{nrao} National Radio Astronomy Observatory, 520 Edgemont Road, Charlottesville, VA 22903, USA \and
\label{asiaa} Academia Sinica Institute of Astronomy and Astrophysics, No.1, Sec. 4., Roosevelt Road, Taipei 10617, Taiwan \and
\label{ice} Institut de Ciències de l'Espai (ICE, CSIC), Can Magrans s/n, 08193, Cerdanyola del Vallés, Catalonia, Spain \and
\label{ieec} Institut d’Estudis Espacials de Catalunya (IEEC), 08860 Castelldefels, Catalonia, Spain \and
\label{inaf} INAF–Osservatorio Astrofisico di Arcetri, Largo E.\ Fermi 5, 50125 Firenze, Italy \and
\label{eao} East Asian Observatory, 660 N. A`oh\={o}k\={u} Place, University Park, Hilo, HI 96720, USA \and
\label{nju} School of Astronomy and Space Science, Nanjing University, Nanjing, China \and
\label{klnju} Key Laboratory of Modern Astronomy and Astrophysics (Nanjing University), Ministry of Education, Nanjing, China \and 
\label{cfgp} Center for Gravitational Physics, Yukawa Institute for Theoretical Physics, Kyoto University, Kitashirakawa Oiwakecho, Sakyo-ku, Kyoto 606-8502, Japan \and
\label{naoj} National Astronomical Observatory of Japan, 2-21-1 Osawa, Mitaka, Tokyo 181-8588, Japan \and
\label{sklr} State Key Laboratory of Radio Astronomy and Technology, A20 Datun Road, Chaoyang District, Beijing, 100101, P.\ R.\ China \and
\label{kuias} Institute for Advanced Study, Kyushu University, Japan \and
\label{kudeps} Department of Earth and Planetary Sciences, Faculty of Science, Kyushu University, Nishi-ku, Fukuoka 819-0395, Japan \and
\label{nthu} Institute of Astronomy and Department of Physics, National Tsing Hua University, Hsinchu 300044, Taiwan \and
\label{utia} Institute of Astronomy, Graduate School of Science, The University of Tokyo, 2-21-1 Osawa, Mitaka, Tokyo 181-0015, Japan \and
\label{kasi} Korea Astronomy and Space Science Institute, 776 Daedeokdae-ro, Yuseong-gu, Daejeon 34055, Republic of Korea \and
\label{aad} Astronomy \& Astrophysics Division, Physical Research Laboratory, Navrangpura, Ahmedabad 380009, India \and
\label{iit} Indian Institute of Technology Gandhinagar Palaj, Gandhinagar 382355, India \and
\label{iis} Department of Physical Sciences, Indian Institute of Science Education and Research (IISER) Mohali, Knowledge City, Sector 81, SAS Nagar 140306, Punjab, India \and
\label{irya} Instituto de Radioastronomía y Astrofísica, Universidad Nacional Autónoma de México, C.P. 58089 Morelia, Michoacán, México
}

\date{Received on Nov. 29 2025; accepted on Jun. 15 2026}

\titlerunning{MagMaR. VII. W33~A}
\authorrunning{Xu et al.}

\abstract{Our understanding of magnetic fields (B-fields) in massive star formation remains incomplete. Linear polarized emission from magnetically aligned dust grains provides a good way to map the morphology of B-field on the plane of the sky. Here, we present the 1.2~mm full polarization observation of W33~A, a massive star-forming region at 2.4~kpc, obtained with the Atacama Large Millimeter/Submillimeter Array (ALMA) to achieve an angular resolution of $\sim$0\farcs3 ($\sim$730~au). W33~A is resolved into 20 dense cores and 9 filaments. It reveals various B-field structures, including two perpendicular large-scale components oriented northwest–southeast (NW-SE) and northeast-southwest (NE-SW) directions, as well as two local distinct features towards the millimeter peaks MM1 and MM2. The NW-SE component could be shaped by a molecular outflow. The NE-SW one is remarkably coherent along the main filamentary structures (F1, F-Main, Tail), all showing trans-Alfv{\'e}nic turbulence. In F-Main, the line mass exceeds the turbulent critical limits, so magnetic support is likely required to prevent radial collapse and suppress local fragmentation. In F1 and Tail, turbulence itself is sufficient to support gravity, although B-fields can potentially provide additional support. Toward MM1, the fields follow a spiral-like, infalling streamer traced by CH$_3$CN; the inferred trans-Alfv{\'e}nic turbulence in the accreting material suggest efficient magnetic damping of turbulence and a magnetically regulated, laminar accretion flow that continues to feed the core. Toward MM2, the field exhibits a hourglass geometry described by parabolic curves. Two independent methods yield a consistently strong field strength of $\sim\!8.1\pm1.9$~mG. The virial analysis shows that B-field can add 25\% (75\% from turbulence) support against gravity but is not by itself sufficient to halt collapse. Our study shows that within one protocluster, B-fields can both help stabilize gas filament against local fragmentation to facilitate mass accretion and delay gravitational collapse. The distinct evolutionary stages of MM1 and MM2 highlights the dynamic importance of B-field in high-mass star formation.}


\maketitle
%
%
\section{Magnetic fields in high-mass star formation} \label{sec:intro}

Magnetic fields (B-fields) are integral to our understanding of the formation and dynamical evolution of high-mass stars \citep[see review in][]{hull&zhang2019}. Strongly coupled with the molecular gas in the interstellar medium, B-fields restrict gas motions across field lines and thus regulate or even hinder mass accretion \citep{zhang2025}. Their presence can therefore prolong the timescale of star formation and reduce its efficiency \citep[e.g.][]{hennebelle2019}. But recent numerical simulation and observations have also suggested that B-fields can help to dissipate turbulence and channel material from envelope to disk \citep[e.g.,][]{tu2024, cortes2025}. Down to disk scales, strong B-fields can suppress the formation of Keplerian disks through magnetic braking and favor the formation of single stars rather than multiple systems \citep[e.g.,][]{seifried2011}. 

Despite their fundamental role in star formation theories \citep{shu1987}, the B-field is hard to measure directly. Zeeman splitting of molecular lines provides the most direct probe of the line-of-sight component of the field \citep[e.g.][]{crutcher1993, crutcher2019}, but in typical molecular clouds the splitting is usually smaller than the spectral linewidth and thus challenging to detect \citep[see review in][]{crutcher2012}. Alternatively, linearly polarized emission from dust grains or spectral lines produced by the Goldreich–Kylafis effect \citep{gk1981} can be used to trace the morphology of the plane-of-sky field component \citep[e.g.][]{girart1999}. On the top of those, the relative orientations (ROs) between B-fields and density structures \citep[e.g.][]{soler2013} as well as gravity \citep[e.g.][]{koch2012a, koch2012b, koch2013, koch2014} have been used to infer the relative importance of magnetic, turbulent, and gravitational forces. 

Interferometers equipped with polarimeters have greatly advanced magnetic-field mapping in high-mass star-forming regions (HMSFRs). Following the first polarized imaging of a high-mass protostar by \citet{lai2001}, full polarization observations have been extended to a large number of HMSFRs \citep[including but not limited to][]{cortes2008, tang2009, girart2009, girart2013, hull2013, qiu2014, sridharan2014, zhang2014, lihb2015, beltran2019, beuther2020, beuther2024, chen2025}. In particular, the ``Magnetic Fields in Massive Star-forming Regions'' (MagMaR) project have surveyed 30 HMSFRs, with individual case studies including G5.89$-$0.39 \citep{magmar1}, NGC 6334I/I(N) \citep{magmar2, magmar3}, IRAS~18089$-$1732 \citep{sanhueza2021}, IRAS~16547$-$4247 \citep{magmar4}, G11.92$-$0.61~MM2 \citep{magmar5}, G333.46$-$0.16 \citep{magmar6}, and G35.20$-$0.74N \citep{magmar7}, as well as statistical results \citep{liu2026}. These observations reveal a wide diversity of magnetic-field morphologies, ranging from hourglass-like to spiral and more disordered configurations. Based on the fact that only a small fraction of sources exhibit hourglass-shaped fields, \citet{hull&zhang2019} concluded that B-fields do not generally dominate the dynamics of high-mass star formation. By comparing ROs across spatial scales from 1~pc to $10^3$~au, authors like \citet{zhang2014, zhang2025} suggested that gravitational collapse at high densities can drag and reorient B-fields. The observed distribution of ROs is consistent with an evolutionary scenario in which an initially sub-Alfv{\'e}nic cloud becomes magnetically supercritical and super-Alfv{\'e}nic as it collapses to form stars.

High-mass stars are known to form predominantly in clusters or massive protoclusters \citep[e.g.][]{lada2003, zinnecker2007, motte2018, assemble2024, beuther2025}. In such environments, gravity globally dominates over magnetic forces \citep[e.g.,][]{zhang2025}; however, this dominance may not hold uniformly throughout a protocluster. The relative role of B-fields can vary across different spatial scales and density regimes. Therefore, mapping B-fields over the entire extent of massive protoclusters is crucial for understanding whether and how B-fields regulate the assembly of high-mass stellar clusters, rather than just the formation of individual high-mass stars or their immediate surroundings. These considerations motivate our study of W33~A. 

\section{W33~A: an accreting high-mass protocluster}

\begin{figure*}
\centering
\includegraphics[width=0.9\linewidth]{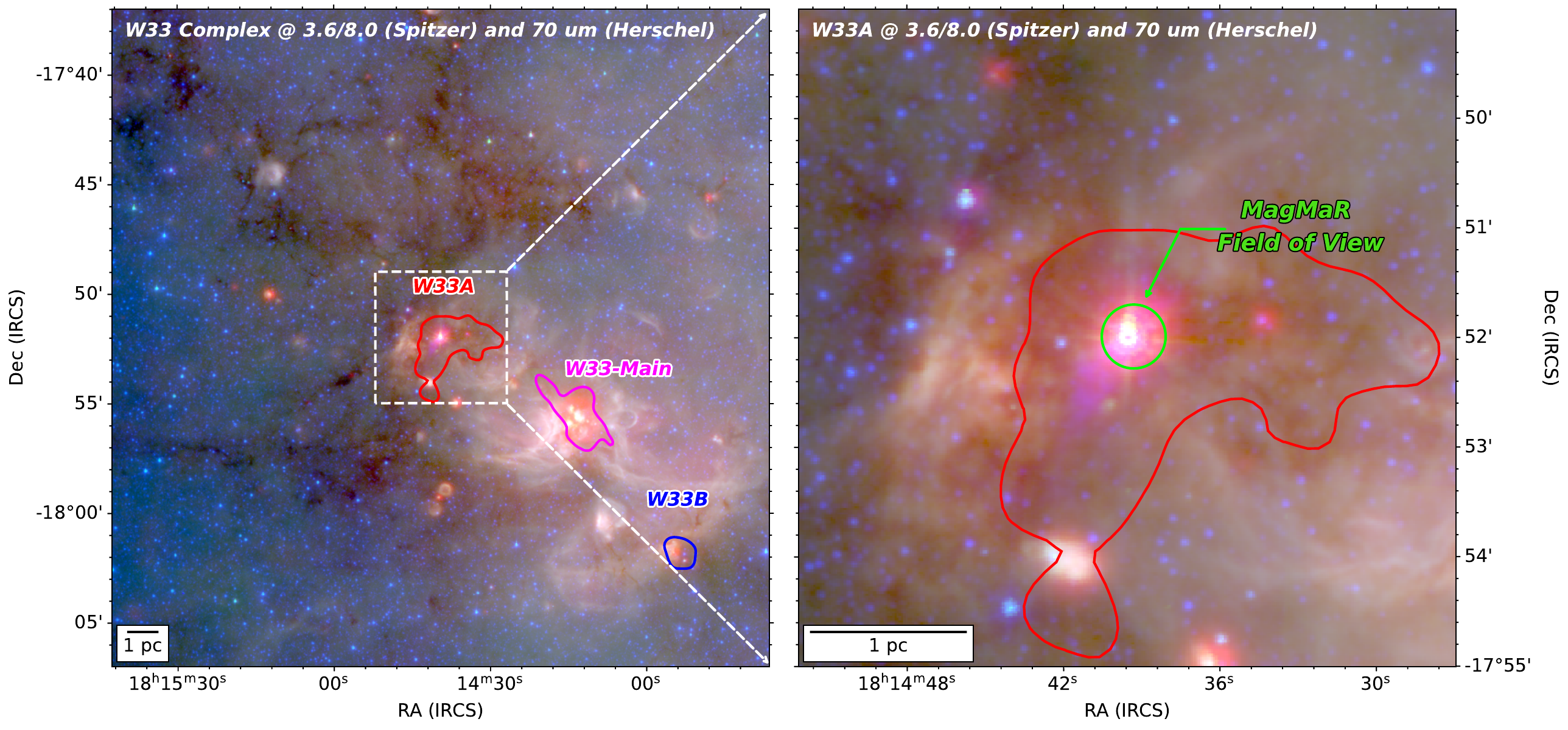}
\caption{The overview of the W33 complex. The background color map is produced with the data from the \textit{Spitzer} 3.6 and 8.0 $\mu$m and the \textit{Herschel} 70 $\mu$m. 
\textit{Left:} Three massive clumps W33~A, W33~Main, and W33~B are defined within the surface density threshold of 0.5~\gpcc\ \citep{lin2016} in red, purple and blue contours, respectively. 
\textit{Right:} The zoom-in view of W33~A which corresponds to the white dashed box on the left panel. The green circle outlines the primary beam response of 0.2 of the MagMaR observations. }
\label{fig:W33_overview}
\end{figure*}

W33~A (aka. IRAS~18117$-$1753) is located in the Scutum spiral arm with a heliocentric distance of $2.40^{+0.17}_{-0.15}$~kpc \citep{immer2013}. As shown in Fig.~\ref{fig:W33_overview}, W33~A, together with W33~Main and W33~B, is considered to be the three most prominent HMSFRs in the W33 complex. By fitting the infrared spectral energy distribution (SED), \citet{urquhart2022} estimated its molecular gas mass reservoir of $\sim\!410$~\msun\ and bolometric luminosity ($L_{\rm bol}$) of $\sim\!3.4\times10^4$~\lsun. We consider the luminosity budget with both stellar luminosity and accretion luminosity terms,
\begin{equation} \label{eq:budget}
    L_{\rm bol} \;=\; L_\star \;+\; L_{\rm acc} = L_\star + \frac{G M_\star \dot{M}}{R_\star}
\end{equation}
where $L_{\star}$ is stellar luminosity, $M_\star$ is stellar mass, $\dot{M}$ is accretion rate, and $R_\star$ is the protostellar radius. If W33~A is a zero-age main-sequence star (ZAMS) which has no accretion $\dot{M}=0$, then a B0~V star with $M_\star\sim\!15$~\msun\ is expected to account for the bolometric luminosity \citep{ekstrom2012}. However, its bremsstrahlung emission at centimeter wavelengths is too faint compared to a normal B0 ZAMS ($T_{\rm eff}\sim\!30,000$~K); the latter can produce a Lyman continuum photon rate $N_{\rm Lyc} \sim \! 10^{47}$~s$^{-1}$ \citep{sternberg2003}, which would yield optically thin radio continuum emission of 200~mJy at 5~GHz or 160~mJy at 50~GHz \citep{mezger1967}. The VLA 5~GHz observation only found a $3\sigma$ upper limit of $\sim\!5~\mathrm{mJy}$ \citep{wynn1981}. The follow-up 8.4 and 15~GHz observations detected point sources of $0.79\pm 0.05$~mJy and $1.63\pm0.05$~mJy associated with infrared sources \citep{rengarajan1996}. At 43~GHz, or 7~mm, the flux density is $4.3\pm0.5$~mJy but this measurement can also contain dust emission \citep{vdt2005}. As such, W33~A is more likely to be a massive protostar \citep{stier1984} under rapid accretion $\dot M\!\sim\!10^{-4}\!-\!10^{-3}$~\massrate\ \citep[see review in][]{beltran2016}. As predicted by evolutionary models \citep{hosokawa2009, hosokawa2010}, protostellar radii can be bloated as large as $R_\star\!\sim\!30{-}100~R_\odot$. Therefore, the effective temperature at the surface of the protostar drops below 10,000~K with many fewer Lyman continuum photons, which can explain the very faint radio emission in W33~A. Such high accretion rates also cause non-negligible accretion luminosity contribution as,
\begin{equation} \label{eq:lacc}
    L_{\rm acc} = 
    0.63\times10^4~L_{\odot}
    \left(\frac{M_\star}{10~M_{\odot}}\right) 
    \left(\frac{\dot{M}}{10^{-3}~M_{\odot} \mathrm{yr}^{-1}}\right)
    \left(\frac{R_\star}{30~R_\odot}\right)^{-1}
\end{equation}
The solution of Eqs.~(1-2) gives an instantaneous stellar mass of $M_\star\!\approx\!13.6^{+0.7}_{-0.9}$~\msun\ assuming $R_\star\!\sim\!30{-}100~R_\odot$ of bloated protostars. This range is consistent with independent measurements by radiative-transfer and kinematic inferences: $M_\star\!\sim\!8\!-\!15$~\msun\ \citep{davies2010, galvan2010, izquierdo2018, navarete2021}.

When resolved with high angular resolutions, W33~A exhibits clustered dense structures embedded in a complex gas environment. Using the {\it Very Large Array} (VLA), \citet{galvan2010} found two parsec-scale perpendicular molecular filaments in NH$_3$~(1,1) emission at $\sim5\arcsec$ resolution. Such two filaments are also seen to connect to the larger-scale gas filaments in NH$_3$~(1,1) by the {\it Green Bank Telescope} (GBT) at $\sim32\arcsec$ resolution \citep{hogge2018}. The convergence of two filaments likely promoted the formation of the two principal dense cores, MM1 and MM2, identified with the \textit{Submillimeter Array} (SMA) at $0\farcs5$ resolution \citep{galvan2010}. The brighter core, MM1, shows a rich molecular spectrum and a mean gas temperature $>300$~K determined by using CH$_3$CN K-ladder line modeling \citep{galvan2010, maud2017}. Adopting a temperature range of [100, 347]~K, the mass of MM1 is found to be [9, 32]~\msun\ \citep{galvan2010}. Using H$_2$CO line modeling at resolution of $5$\arcsec, \citet{quarks2} have calculated the gas kinetic temperature of $53\pm13$~K, indicating extensive gas heating. Towards the MM1 center, a fast ($\sim600$~\kms) bipolar jet is detected on sub-milliarcsecond ($<1$~au) scales via 3D spectro-astrometry of Br~$\gamma$ emission \citep{davies2010}, with its axis aligned with the larger-scale bipolar molecular outflow \citep{galvan2010}. Using the \textit{Australia Telescope Compact Array} (ATCA) at 5.5, 9.0, 17.0 and 22.8~GHz, the radio emission has a positive spectral index and is classified as a ionized jet candidate \citep{purser2016}. Perpendicular to the jet, the source exhibits a rotation-flattened, cool molecular envelope traced by CO absorption \citep{davies2010, navarete2021} and evidence for a Keplerian disk traced by CH$_3$CN \citep{galvan2010}. Higher-resolution CH$_3$CN observations further resolve this rotation into a spiral-in streamer feeding a disk candidate \citep{maud2017, yang2026}. A radiative-transfer model with a $\sim7\,M_\odot$ central source reproduces both dust continuum and CH$_3$CN emission \citep{izquierdo2018}. In contrast, MM2 is colder ($\sim46$~K; determined from NH$_3$ kinetic temperature) and less evolved \citep{galvan2010}. Although with a larger mass reservoir of $\sim60\,M_\odot$, it lacks hot-core lines and is detected primarily in simple species (CO, $^{13}$CO, SO). MM2 is nevertheless not a starless core: a red-shifted CO outflow likely originates from it \citep{galvan2010}. 

Despite extensive imaging and spectroscopic observations in W33~A, B-fields, as one of the key components in determining the dynamical state, have no constraints yet. The absence of such data has limited our understanding of what energetic processes control the early stages of high-mass star formation. To our best knowledge, previous polarimetric efforts have primarily focused on the NIR regime at 3, 5, and 10~$\mu$m bands \citep{hough1989, chrysostomou1996, smith2000}. More recently, \citet{kwon2025} conducted high-resolution NIR imaging polarimetry and detected centrosymmetric patterns indicative of scattered light from a bipolar cavity, as well as potential hints of a compact polarized pseudo-disk. However, to date, there are no published (sub)millimeter polarimetric observations and thus no constraints on its B-field morphology via thermal dust emission. 

Taking advantage of the superior capabilities of the {\it Atacama Large Millimeter/submillimeter Array} (ALMA), we have observed W33~A in polarized dust continuum and molecular line emission to better understand the role of the B-field in the formation of high-mass stars. This target was observed as part of the MagMaR survey. The paper is outlined as follows. In Sect.~\ref{sec:observation}, the observations, data calibration, and imaging are presented. In Sect.~\ref{sec:result}, we introduce the structural decomposition of 1.2~mm dust emission and their physical characterization. Using the full polarization data, we present B-field images and analysis methods. To understand the abundant and well-ordered field features, we discuss the field topologies and strengths in separated regions individually in Sect.~\ref{sec:discuss}. Last, we make a conclusion in Sect.~\ref{sec:conclude}. 

\section{Observations} \label{sec:observation}


The ALMA 1.2~mm full polarization observations (Project ID: 2017.1.00101.S and 2018.1.00105.S, PI: P. Sanhueza) of W33~A were taken on September 25 and 26, 2018. A total of 47 antennas of the 12-m array were used, covering baselines from 14 to 1400 meters. The data set consists of full polarization observations in Band~6 (at $\sim250.486$~GHz; 1.2~mm). The correlator setup includes three wide spectral windows of width 1875~MHz, with a spectral resolution of 1.953~MHz ($\sim$2.4~\kms), and two narrow spectral windows of width 234~MHz, with a spectral resolution of 0.488~MHz ($\sim$0.56~\kms). One wide window covers a set of closely spaced K-ladders of the CH$_3$CN~(14-13) transition (more details in Appendix~\ref{app:ch3cn}). The two narrow spectral windows are centered on HN$^{13}$C (3--2) at 261.26331~GHz and H$^{13}$CO$^{+}$ (3--2) at 260.255339~GHz, respectively. 
Data were routinely calibrated using the ALMA pipelines (\url{https://almascience.nrao.edu/processing/science-pipeline}) of Common Astronomy Software Applications \citep[CASA;][]{CASA2022} version 5.1.1. The quasar J1924$-$2914 was adopted for the calibration of flux, bandpass, and polarization. The quasar J1832$-$2039 was used for phase calibration.

Data imaging was performed using CASA 5.6.1. Line contamination was removed from the Stokes I continuum image following the procedure described in \citet{olguin2021}. Stokes I continuum was self-calibrated in phase and amplitude, while Stokes Q and U were not self-calibrated. The self-calibration solutions of Stokes I continuum were then applied to the spectral cubes. The continuum imaging was done by independently cleaning each Stokes parameter at the final self-calibrated visibility using the CASA task {\tt tclean} with Briggs weighting {\tt robust=1.0}. The resulting images have an angular resolution of 0\farcs34$\times$0\farcs27 with position angle of 81\fdg6. The sensitivities are measured from the residual maps after the final round of cleaning: $\sigma_I \approx 130$ \ujybeam\ for Stokes I and $\sigma_Q=\sigma_U \approx 30$ \ujybeam~for both Stokes Q and U. Linearly polarized dust continuum emission is detected in the inner $\sim8$\arcsec\ of the field of view, that is, the one third of the primary beam ($\sim24$\arcsec), with polarization angles having less than 1\% errors. However, we note that polarization angles beyond are still useful with only a few percent errors \citep{hull2020}. The self-calibrated line cubes were imaged using the automatic masking procedure \textit{yclean} from \citet{contreras2018}. The noise level is $\sim2~\mjybeam$ per 0.28 \kms~channel. 


W33~A is as part of ALMA Three-millimeter Observations of Massive Star-forming regions \citep[ATOMS;][]{liu2020} in a Cycle-7 ALMA project (Project ID: 2019.1.00685.S; PI: Tie Liu). The target name in their observations is I18117$-$1753. We used the 12m$+$7m combined data cube of HCO$^+$~(1-0) in this paper to trace molecular outflows at larger scales.
The combined data have synthesized beam size of $2\farcs4 \times 1\farcs9$ with position angle of $-29$\degree\ and maximum recoverable scale (MRS) of $\sim\!80\arcsec$. The sensitivity of the data cube is $11~\mjybeam$ per 0.1~\kms\ velocity channel. 

\section{Results} \label{sec:result}

\subsection{Structure decomposition and physical characterization} \label{result:structure}

We use the \textit{getsf} (version: v231026) algorithm that spatially decomposes the observed images to separate roundish sources (cores) and elongated filaments from their background emission \citep{menshchikov2021}, thanks to its good performance in dealing with image artifacts \citep[e.g.,][]{xu2023}. 

\begin{figure}[!ht]
\centering
\includegraphics[width=\linewidth]{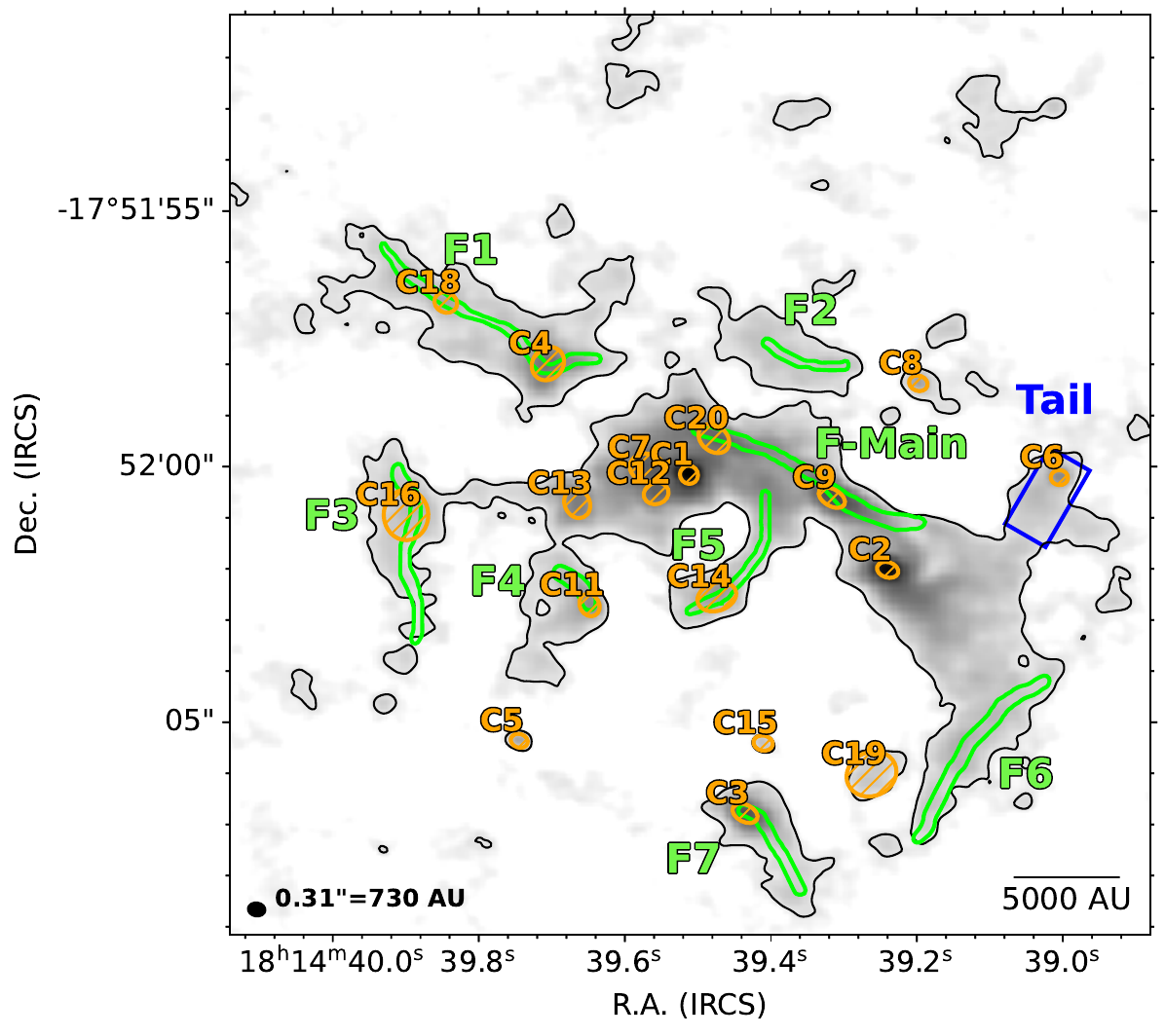}
\caption{The background gray-scale map shows the 1.2~mm continuum emission with a $4\sigma_I=0.52~\mjybeam$ contour. The \textit{getsf}-identified cores are shown in orange ellipses and filaments in green spines. A filament called ``Tail'', as outlined by the blue rectangle, is added to our filament sample. There are two cores C10 and C17 outside the current field. The synthesized beam and the scale bar are both shown in the bottom. 
\label{fig:getsf}}
\end{figure}

In W33~A, we identified 20 dense cores and nine filaments. The cores, labeled as C1--C20 in Table~\ref{tab:coreinfo}, are fitted using 2-D Gaussians with parameters including size and flux density measured. 
The filament sample consists of eight from the original \textit{getsf} catalog and an additional one recognized by eye called the \textit{Tail}, due to its interesting B-field feature. The most prominent filament, connecting MM1 and MM2, is designated as F-Main, while the others are named F1--F7 from northeast to southwest. The Tail is at the northwestern end of F-Main. The filament ID and equatorial coordinates of the geometric center of the extracted filaments (RA and Dec) are listed in Table~\ref{tab:filainfo}. In Fig.~\ref{fig:getsf}, the identified structures are labeled: dense cores are in orange elliptical footprints and filaments are in green spines (with one-beam width) or in blue boxes. 

We cross-matched the compact structures identified in our ALMA 1.2~mm map with the SMA 1.3~mm sources reported by \citet{galvan2010}. Core C1 coincides with MM1-Main, the brightest millimeter source in W33~A. The MM1-SE is resolved into two components, C7 and C12, while C20 corresponds to MM1-NW; together, cores C7, C12, and C20 are on the gas spiral reported by \citet{maud2017, izquierdo2018}. C9 coincides in position with the SMA source MM2-NE but actually lies on the F-Main filament in our data. C2 aligns with MM2-Main (the second-brightest source). A compact source near the SMA peak labeled here as C4 is visible in the SMA map but was not cataloged as an individual core by \citet{galvan2010}. Beyond the SMA detections, our ALMA image reveals additional faint cores (C3, C5, C6, C8, C10--C13, C15--C19) and elongated filaments F-Main, F1--F7, and the Tail.

\begin{table*}
\caption{Characteristics of the cores detected by ALMA 1.2~mm observations in the W33~A region.}
\label{tab:coreinfo}
\centering
\small
\begin{tabular}{cccccccccc}
\hline\hline
\\[-6pt]
Core & RA & Dec & FWHM & $R_{\rm deconv}$ & $I^{\rm peak}_{\rm \nu}$ & $S^{\rm int}_{\rm \nu}$ & $M_{\rm core}$ & $\langle{n_{\rm H_2}}\rangle$ & $T_{\rm gas}$ \\
ID & [ICRS] & [ICRS] & [$\arcsec\times\arcsec$] & [au] &[mJy\,beam$^{-1}$] & [mJy] & [M$_\odot$] & [$\times10^6$\,cm$^{-3}$] & [K] \\[3pt]
\hline
\\[-8pt]
C1 & 18:14:39.51 & -17:52:00.14 & 0.36$\times$0.35 & 660 & 48.5$\pm$1.3 & 89.4$\pm$1.6 & 8.6$\pm$2.9 & 900$\pm$300 & 270$\pm$50 \\
C2 & 18:14:39.24 & -17:52:02.00 & 0.43$\times$0.30 & 670 & 42.8$\pm$0.9 & 79.0$\pm$1.1 & 7.2$\pm$2.4 & 710$\pm$240 & - \\
C3 & 18:14:39.44 & -17:52:06.76 & 0.53$\times$0.33 & 850 & 12.6$\pm$0.3 & 32.1$\pm$0.4 & 2.3$\pm$0.8 & 110$\pm$40 & 450$\pm$90 \\
C4 & 18:14:39.71 & -17:51:57.97 & 0.69$\times$0.60 & 1450 & 4.8$\pm$0.3 & 23.0$\pm$0.5 & 1.6$\pm$0.5 & 15$\pm$5 & - \\
C5 & 18:14:39.75 & -17:52:05.34 & 0.34$\times$0.27 & 500 & 2.2$\pm$0.1 & 2.4$\pm$0.1 & 0.16$\pm$0.06 & 40$\pm$14 & - \\
C6 & 18:14:39.01 & -17:52:00.19 & 0.34$\times$0.29 & 520 & 2.9$\pm$0.3 & 3.6$\pm$0.3 & 0.24$\pm$0.08 & 51$\pm$17 & - \\
C7 & 18:14:39.57 & -17:52:00.00 & 0.68$\times$0.55 & 1380 & 5.6$\pm$1.0 & 19.8$\pm$1.1 & 1.4$\pm$0.5 & 16$\pm$5 & 130$\pm$30 \\
C8 & 18:14:39.20 & -17:51:58.35 & 0.37$\times$0.30 & 580 & 1.7$\pm$0.1 & 2.4$\pm$0.1 & 0.16$\pm$0.06 & 24$\pm$8 & - \\
C9 & 18:14:39.32 & -17:52:00.59 & 0.57$\times$0.33 & 890 & 8.4$\pm$1.0 & 17.4$\pm$1.1 & 1.2$\pm$0.4 & 51$\pm$18 & - \\
C10 & 18:14:38.46 & -17:51:57.27 & 0.51$\times$0.41 & 960 & 2.5$\pm$0.3 & 6.7$\pm$0.3 & 0.46$\pm$0.15 & 15$\pm$5 & - \\
C11 & 18:14:39.65 & -17:52:02.69 & 0.46$\times$0.38 & 850 & 2.5$\pm$0.3 & 5.2$\pm$0.4 & 0.35$\pm$0.12 & 17$\pm$6 & - \\
C12 & 18:14:39.56 & -17:52:00.51 & 0.50$\times$0.42 & 960 & 4.9$\pm$1.0 & 9.8$\pm$0.9 & 0.67$\pm$0.23 & 23$\pm$8 & 170$\pm$30 \\
C13 & 18:14:39.67 & -17:52:00.70 & 0.59$\times$0.49 & 1190 & 3.0$\pm$0.8 & 9.4$\pm$0.9 & 0.64$\pm$0.22 & 11$\pm$4 & - \\
C14 & 18:14:39.48 & -17:52:02.53 & 0.78$\times$0.54 & 1480 & 1.6$\pm$0.3 & 7.4$\pm$0.5 & 0.50$\pm$0.17 & 4.6$\pm$1.6 & - \\
C15 & 18:14:39.41 & -17:52:05.39 & 0.38$\times$0.30 & 590 & 1.1$\pm$0.1 & 1.5$\pm$0.1 & 0.10$\pm$0.03 & 15$\pm$5 & - \\
C16 & 18:14:39.90 & -17:52:00.92 & 0.99$\times$0.87 & 2190 & 1.0$\pm$0.2 & 8.8$\pm$0.5 & 0.59$\pm$0.20 & 1.7$\pm$0.6 & - \\
C17 & 18:14:38.47 & -17:51:56.54 & 1.06$\times$0.83 & 2210 & 1.3$\pm$0.2 & 13.7$\pm$0.6 & 0.9$\pm$0.3 & 2.6$\pm$0.9 & - \\
C18 & 18:14:39.85 & -17:51:56.77 & 0.46$\times$0.39 & 860 & 1.4$\pm$0.3 & 2.7$\pm$0.2 & 0.18$\pm$0.06 & 8.2$\pm$2.8 & - \\
C19 & 18:14:39.26 & -17:52:05.99 & 1.00$\times$0.86 & 2180 & 0.7$\pm$0.2 & 6.2$\pm$0.3 & 0.42$\pm$0.14 & 1.2$\pm$0.4 & - \\
C20 & 18:14:39.48 & -17:51:59.43 & 0.66$\times$0.50 & 1280 & 3.7$\pm$1.4 & 11.3$\pm$1.7 & 0.77$\pm$0.26 & 11$\pm$4 & 140$\pm$30 \\
\hline
\end{tabular}
\tablefoot{
RA and Dec are ICRS coordinates. 
FWHM gives the measured major and minor full width at half maximum. 
$R_{\rm deconv}$ is the physical size in au deconvolved from the synthesized beam. 
$I^{\rm peak}_{\rm \nu}$ and $S^{\rm int}_{\rm \nu}$ are the peak intensity and integrated flux density at 1.2~mm, respectively. 
$M_{\rm core}$ is the molecular gas mass of core assuming a dust temperature of $25\pm5~\mathrm{K}$. For a hot molecular core ($T_{\rm gas}>100~\mathrm{K}$), such mass should be a upper limit.
$\langle n_{\rm H_2}\rangle$ is the core-averaged volume density of molecular hydrogen. 
$T_{\rm gas}$ is the gas temperature by fitting CH$_3$CN K-ladders under the LTE assumption.
}
\end{table*}

\begin{table*}
\caption{Characteristics of the filaments detected by ALMA 1.2~mm observations in the W33~A region.}
\label{tab:filainfo}
\centering
\small
\begin{tabular}{ccccccccccc}
\hline
\hline
\\[-6pt]
Filament & RA & Dec & $l$ & $w$ & $\langle{I}_{\rm \nu}\rangle$ & $S^{\rm int}_{\rm \nu}$ & $\langle N_{\rm H_2}\rangle$ & $\langle n_{\rm H_2}\rangle$ & $M_{\rm fila}$ & $M/l$ \\
ID & [ICRS] & [ICRS] & [au] & [au] & [\mjybeam] & [mJy] & [$\times10^{23}$\,cm$^{-2}$] & [$\times10^{6}$\,cm$^{-3}$] & [M$_\odot$] & [M$_\odot$\,pc$^{-1}$] \\[3pt]
\hline
\\[-8pt]
F-Main & 18:14:39.32 & $-$17:52:00.61 & 5800 & 2300 & 4.0 & 83.4 & 8.8$\pm$3.0 & 25$\pm$8 & 5.6$\pm$1.9 & 200$\pm$70 \\
F1 & 18:14:39.79 & $-$17:51:57.22 & 10700 & 3700 & 1.3 & 82.6 & 2.9$\pm$1.0 & 5$\pm$1 & 5.5$\pm$1.9 & 110$\pm$40 \\
F2 & 18:14:39.37 & $-$17:51:57.81 & 5200 & 2300 & 0.9 & 12.0 & 2.0$\pm$0.7 & 5$\pm$1 & 0.8$\pm$0.3 & 30$\pm$10 \\
F3 & 18:14:39.90 & $-$17:52:01.09 & 6800 & 2500 & 0.9 & 25.2 & 2.0$\pm$0.7 & 5$\pm$1 & 1.7$\pm$0.6 & 50$\pm$20 \\
F4 & 18:14:39.67 & $-$17:52:02.37 & 4300 & 1700 & 1.4 & 16.1 & 3.1$\pm$1.0 & 12$\pm$4 & 1.1$\pm$0.4 & 50$\pm$20 \\
F5 & 18:14:39.45 & $-$17:52:02.04 & 6200 & 1700 & 1.2 & 19.5 & 2.6$\pm$0.9 & 10$\pm$3 & 1.3$\pm$0.4 & 40$\pm$10 \\
F6 & 18:14:39.12 & $-$17:52:05.30 & 8700 & 2100 & 1.1 & 32.0 & 2.4$\pm$0.8 & 7$\pm$2 & 2.1$\pm$0.7 & 50$\pm$20 \\
F7 & 18:14:39.40 & $-$17:52:07.32 & 6200 & 1900 & 2.5 & 44.8 & 5.5$\pm$1.9 & 19$\pm$6 & 3.0$\pm$1.0 & 100$\pm$30 \\
Tail & 18:14:39.02 & $-$17:52:00.63 & 4100 & 2300 & 1.1 & 16.5 & 2.4$\pm$0.8 & 7$\pm$2 & 1.1$\pm$0.4 & 60$\pm$20 \\
\hline
\end{tabular}
\tablefoot{
RA and Dec are ICRS coordinates of the filament geometric centers.
$l$ and $w$ are the filament length and width in rectangular masks as introduced in Sect.~\ref{structure:fila}. 
$\langle{I}_{\rm \nu}\rangle$ and $S^{\rm int}_{\rm \nu}$ are the mean intensity and integrated flux density within the masks, respectively. 
Assuming dust temperature of $25\pm5$~K, $\langle{N}_{\rm H_2}\rangle$ and $\langle n_{\rm H_2}\rangle$ denote the mean column and volume densities filament.
$M_{\rm fila}$ is the total filament mass, and $M/l$ represents the filament line mass. 
Uncertainties in parentheses indicate 1$\sigma$ errors.
}
\end{table*}

\subsubsection{Physical properties of dense cores} \label{structure:core}

From the measured FWHM, the intrinsic radius of dense cores was deconvolved from beam,
\begin{equation}
    \theta_{\rm deconv} = \eta\left[\left(\theta_{\rm maj}^2-\theta_{\rm bm}^2\right) \left(\theta_{\rm min}^2-\theta_{\rm bm}^2\right) \right]^{1/4}.
\end{equation}
Here $\theta_{\rm maj}$ and $\theta_{\rm min}$ are the FWHM of major and minor axis. $\theta_{\rm bm}$ is the FWHM of the synthesized beam. $\eta$ is a shape correction factor and is taken as unity \citep{rosolowsky2010}. The angular size was then converted to physical size by $R_{\rm deconv} = \theta_{\rm deconv}\,d$. 

The dust temperature of core is assumed to be the same as $25\pm5$~K, which aligns with the SED fitting result of 24~K \citep{urquhart2022} and 28~K from \citep{lin2016}, as well as $26^{+9}_{-8}$~K from NH$_3$ kinetic temperature \citep{lu2014}. For five cores C1, C3, C7, C12, and C20 with detectable CH$_3$CN lines, we also calculated their $0\farcs3$-resolution temperature by LTE line modeling of CH$_3$CN. The technical details can be found in Appendix~\ref{app:ch3cn}. As listed in Table~\ref{tab:coreinfo}, the CH$_3$CN gas temperature is typically as high as $>100$~K. Since our flux measures are from dust continuum map, the mass and density calculation below will adopt assumed dust rather than gas temperature.

We estimated the gas mass from 1.2~mm dust emission accounting for the optical-depth effects \citep{motte2018na},
\begin{equation} \label{eq:mcore_thick}
\begin{split}
M_{\rm core} &= 
- \frac{S^{\rm int}_{\rm \nu} d^2 R_{\rm gd}}{\kappa_{\rm \nu} B_{\rm \nu}(T_{\rm dust})} 
\times \frac{\Omega_{\rm beam} B_{\rm \nu} (T_{\rm dust})}{I^{\rm peak}_{\rm \nu}} \\
&\times 
\ln\left[1-\frac{I^{\rm peak}_{\rm \nu}}{\Omega_{\rm beam} B_{\rm \nu}(T_{\rm dust})}\right]
\end{split}
\end{equation}
Here $S^{\rm int}_{\rm \nu}$ and $I^{\rm peak}_{\rm \nu}$ are the integrated flux density and peak intensity at 1.2~mm for each core. $d=2.4$~kpc is the distance to W33~A. $\kappa_{\rm \nu}=1.1$ cm$^{2}$~g$^{-1}$ is the dust opacity interpolated to 1.2~mm for the case of $10^8$ cm$^{-3}$ density and initial condition of thick ice mantle \citep{ossenkopf1994}. $B_{\rm \nu} (T_{\rm dust})$ is the Planck function at 1.2~mm and at dust temperature $T_{\rm dust}$. $\Omega_{\rm beam}$ is the solid angle of the beam. $R_{\rm gd}$ is gas-to-dust ratio, which is assumed to be 100. Assuming spherical geometry of dense cores, the core-averaged volume density of molecular hydrogen can be obtained as
\begin{equation} \label{eq:volden}
    \langle n_{\rm H_2}\rangle = \frac{M_{\rm core}}{(4/3)\pi \mu m_{\rm H} R_{\rm deconv}^3}
\end{equation}
where $\mu=2.8$ is the molecular weight per H$_2$ molecule and $m_{\rm H}$ is the mass of a hydrogen atom. 

\begin{figure*}
\centering
\includegraphics[width=\linewidth]{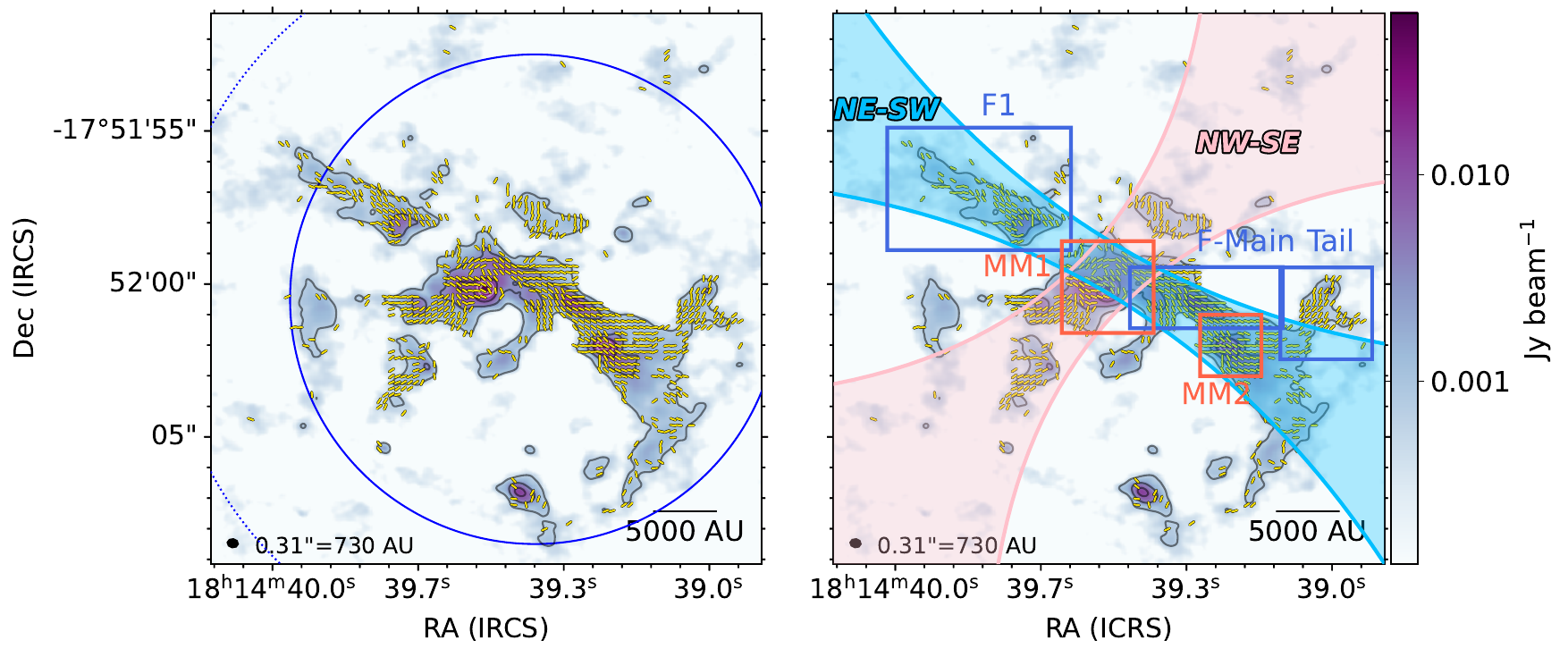}
\caption{The ALMA 1.2~mm Stokes I continuum emission (color map and contours) is overlaid with B-field directions (yellow segments) every half beam, rotated by $90^{\circ}$ from the dust polarization directions. The segments are shown only when both debiased polarization intensity $\mathrm{S/N}(P_I)>2$ and Stokes I signal-to-noise ratio $\mathrm{S/N}(I))>3$. {\it Left}: One third and FWHM of primary beam are indicated by blue solid and dashed circles. {\it Right}: Two global directions northwest-southeast (NW-SE) and northeast-southwest (NE-SW) are marked in pink and blue shades. At the NE-SW direction, F1, F-Main, and Tail subregions are outlined. Two local features are highlighted towards MM1 and MM2. The contour levels are 0.52, 2.9, 8.7, 19.4, and $36.3~\mjybeam$. The synthesized beam of 730~au ($0\farcs3$) and the 5000-au scale bar are shown on the bottom.}
\label{fig:bfield}
\end{figure*}

\subsubsection{Physical properties of filaments} \label{structure:fila}

For each filament spine, we define a rectangular mask that follows the spine orientation and tightly encloses the $\mathrm{S/N}(I)>3$ contour, by which the length ($l$) and width ($w$) of filaments are defined. Such threshold naturally encloses all associated polarization measurements. Throughout the mask, mean intensity ($\langle I_{\rm \nu}\rangle$) and integrated flux density ($S^{\rm int}_{\rm \nu}$) are obtained. The dust temperature of filaments is assumed to be $25\pm5$~K. The mean column density can be estimated as
\begin{equation} \label{eq:colden}
    \langle N_{\rm H_2}\rangle = \frac{-\ln\left[1-\langle I_{\rm \nu}\rangle/B_{\rm \nu} (T_{\rm dust})\right] R_{\rm gd}}{\kappa_{\rm \nu}}
\end{equation}
The mean volume density is calculated as $\langle n_{\rm H_2}\rangle = \langle N_{\rm H_2}\rangle / w$ assuming filament depth equal to its width. The total mass in filament $M_{\rm fila}$ is estimated using the same method as per the cores by Eq.~(\ref{eq:mcore_thick}). The filament line mass is defined as $M_{\rm fila}/l$. These filament properties are listed in Table~\ref{tab:filainfo}.

The uncertainty budgets of mass and density of cores and filaments both take into account the uncertainties of absolute flux density (10\%), dust temperature (20\%), distance (7\%; parallax), and dust-to-gas ratio \citep[24\%;][]{giannetti2017}, leading to a combined uncertainty of 34\%. Besides, missing flux can lead to an underestimation of mass and density, but more severely for filaments than cores.

\subsection{Dust polarization and B-field} \label{result:bfield}

Because the polarized intensity and polarized percentage are defined as positive values, the measured quantities of the two parameters are biased toward larger values \citep{vaillancourt2006}. So we followed the method introduced in \citet{plaszczynski2014} to estimate the debiased polarized intensity $\hat{P_I}$ as
\begin{equation} \label{eq:debiasp}
    \hat{P_I} = P_I - \sigma^2 \frac{1-\exp\left(-P_I^2/\sigma^2\right)}{2P_I},
\end{equation}
where $P_I = \sqrt{Q^2+U^2}$ is the measured polarized intensity from the Stokes Q and U maps and $\sigma=\sigma_Q=\sigma_U$ is the measured noise of Stokes Q or U map. The nominal uncertainty of $\hat{P_I}$ map $\sigma_{P_I}$ is identical to $\sigma$.

The debiased polarization fraction $\hat{f_P}$ and its uncertainty $\sigma(\hat{f_P})$ are therefore derived by,
\begin{equation} \label{eq:debiasfp}
\begin{split}
    & \hat{f_P} = \frac{\hat{P_I}}{I}, \\
    & \sigma(\hat{f_P}) = \sqrt{\left(\frac{\sigma_{P_I}^2}{I^2}+\frac{\sigma_I^2(Q^2+U^2)}{I^4}\right)}
\end{split}
\end{equation}
where $I$ and $\sigma_I$ are the Stokes I intensity map and its measured noise, respectively. The debiased polarization intensity and fraction maps are presented in Fig.~\ref{fig:pol}.

The polarization angle $\psi$ and its uncertainty $\sigma_\psi$ are estimated as \citep{nk1993},
\begin{equation} \label{eq:pa}
   \psi = \frac{1}{2} \arctan \left(\frac{U}{Q}\right) \quad (\mathrm{radians}),
\end{equation}
and as
\begin{equation} \label{eq:err_pa}
    \sigma_\psi = \frac{1}{2}\sqrt{\frac{\sigma^2}{Q^2+U^2}} \quad (\mathrm{radians}).
\end{equation}

Assuming that the shortest axis of a fraction of irregular dust grains is aligned with the B-field, the plane-of-sky (POS) B-field orientation can be traced by rotating the observed position angle (P.A.) of linearly polarized dust emission by 90$^{\circ}$ \citep{lazarian2007a, lazarian2007b}.

Fig.~\ref{fig:bfield} shows the B-field morphology on the plane of the sky (POS). The B-field morphology in W33~A is complex, likely due to the presence of multiple dominant structures within a highly clustered environment. To interpret the B-field patterns, it is essential to isolate individual components that may each trace distinct physical processes. From visual inspection of Fig.~\ref{fig:bfield}, the POS B-field basically reveals two global, coherent components that are oriented nearly perpendicular to each other. One aligns with the NE-SW gas filamentary structure traced by the Stokes I map and we refer to it as the NE-SW component. The other, almost orthogonal to the NE-SW component, is denoted as the NW-SE component. The two main components intersect at the location of the massive dense core MM1. In Sect.~\ref{discuss:outflow}, we will give a more physical explanation on the component separation. 

The NE-SW component follows F1 and F-Main filaments and dense cores C18, C4, C20, and C9 in a line. Towards the SW end, it shows an almost 90\degree\ upward but coherent warp. We identified three subregions of interest with highly organized and coherent B-field morphologies. They are outlined by blue rectangles in Fig.~\ref{fig:bfield}. The NW-SE component goes through F2, F4, C11, and C13 showing seemingly pinched feature from the MM1 center. Besides the two main perpendicular components, there are two additional localized features towards MM1 and MM2, which significantly deviate from the global patterns; these regions are outlined by red rectangles in Fig.~\ref{fig:bfield}. 

The decomposition, physical characterization, and analyses of POS B-field will be discussed in Sect.~\ref{sec:discuss}. The relation between the B-field and outflow are discussed in Sect.~\ref{discuss:outflow}. We present analyses of three subregions which show coherent B-field direction in Sect.~\ref{discuss:coherent}. We discuss two localized features towards MM1 and MM2 in Sections \ref{discuss:mm1} and \ref{discuss:mm2}, respectively.

\subsection{Measurements of B-fields} \label{result:method}

The Davis–Chandrasekhar–Fermi \citep[DCF;][]{davis1951, chandrasekhar1953} method and its modified form have been widely used to estimate the POS B-field strength. The physical scenario is that a POS B-field ($B_{\rm pos}$) is perturbed by POS turbulent velocity dispersion $\delta v^{\rm nt}_{\rm pos}$, resulting in an intrinsic P.A. dispersion of B-field ($\delta \psi_{\rm int}$). With the small angle approximation, the POS B-field strength $B_{\rm pos}$ is estimated by,
\begin{equation} \label{eq:dcf}
\begin{split}
    B_{\rm pos} &= Q_c \sqrt{4\pi\rho} \frac{\delta v^{\rm nt}_{\rm pos}}{\delta\psi_{\rm int}} \\
    & \approx\ 2.2 \left(\frac{Q_c}{0.5}\right) \left(\frac{\langle n_{\rm H_2}\rangle}{10^6\,\mathrm{cm}^{-3}}\right)^{0.5} \left(\frac{\delta v^{\rm nt}_{\rm los}}{1\,\mathrm{km/s}}\right) \left(\frac{\delta\psi_{\rm int}}{10^{\circ}}\right)^{-1}\quad \mathrm{mG}.
\end{split}
\end{equation}
Here $Q_c$ is a correction factor due to the line of sight and beam integration effects from MHD simulation. We adopted the value of 0.28 measured at physical scales of $0.2{-}1~\mathrm{pc}$ \citep{liu2021}. The molecular hydrogen volume density ($\langle n_{\rm H_2}\rangle$) and turbulent velocity dispersion ($\delta v_{\rm pos}$) are both averaged quantities from the region of interest. Here we assumed POS velocity dispersion the same as LOS one ($\delta v^{\rm nt}_{\rm pos}\sim\delta v^{\rm nt}_{\rm los}$). The LOS turbulent velocity dispersion is estimated from H$^{13}$CO$^+$~(3-2) line fitting with thermal component subtracted,
\begin{equation}
    \delta v^{\rm nt}_{\rm los} = \sqrt{\delta v_{\rm los}^2 - \frac{k_B T}{m_{\rm mol}}}
\end{equation}
where $m_{\rm mol}=30 m_{\rm p}$ is the mass of the H$^{13}$CO$^+$ molecule.

The total B-field is estimated by correcting projection effect $B_{\rm tot} = B_{\rm pos} /\cos{i}$. The Alfv{\'e}nic velocity is the characteristic propagation speed of transverse MHD disturbances - i.e., Alfv{\'e}n waves - along B-field lines in a magnetized fluid. It is given by
\begin{equation} \label{eq:va}
    v_A = \frac{B_{\rm tot}}{\sqrt{4\pi\,\rho}} \approx\ 1.3
    \left(\frac{B_{\rm tot}}{1~\mathrm{mG}}\right)
    \left(\frac{\langle n_{\rm H_2} \rangle}{10^{6}~\mathrm{cm^{-3}}}\right)^{-0.5} \quad \mathrm{km\,s^{-1}}
\end{equation}
When assuming DCF, the Alfv{\'e}nic Mach number $\mathcal{M}_A$ can be directly estimated from the field angle dispersion as
\begin{equation} \label{eq:ma}
    \mathcal{M}_A = \frac{\delta v^{\rm nt}_{\rm los}}{v_A} = \frac{\delta v^{\rm nt}_{\rm los}}{B_{\rm tot}/\sqrt{4\pi\rho}} \approx\ 0.8 \left(\frac{\delta\psi_{\rm int}}{10^\circ}\right) \left(\frac{Q_c}{0.5}\right)^{-1}\cos{i}
\end{equation}
where the Alfv{\'e}nic Mach number can be used to gauge the relative energy balance between turbulence and the B-field. For $\mathcal{M}_A < 1$, the gas is sub-Alfv{\'e}nic which means B-field is stronger than turbulent energy; while for $\mathcal{M}_A > 1$, the gas is super-Alfv{\'e}nic, meaning the B-field energy density is weaker than the turbulent energy.

The intrinsic B-field angle dispersion ($\delta \psi_{\rm int}$) in the DCF method (Eq.~\ref{eq:dcf}) is usually convolved with pixelwise measurement uncertainties. Generally, we have an independent measurement set $\{\psi_i\}$ of B-field P.A. Rather than estimating the dispersion directly from a P.A. histogram of $\{\psi_i\}$, we adopt a forward statistical model to infer intrinsic angle dispersion. The likelihood of an individual measurement $\psi_i$ can be written as,
\begin{equation} \label{eq:angle_model}
p(\psi_i) =
(1 - f_{\rm bg}) \,
\mathcal{N}
\left(
0,\,
\delta \psi_{\rm int}^2 + \sigma_{\psi,i}^2
\right)
+
f_{\rm bg} \,
\mathcal{U}(-90^\circ,\,90^\circ).
\end{equation}
where $\mathcal{N}$ is the Gaussian distribution with $\delta \psi_{\rm int}$ and measure uncertainties, while $\mathcal{U}$ is a uniform distribution over the interval $[-90^\circ,90^\circ]$. The background fraction is $f_{\rm bg}$. By maximizing the total likelihood over all selected pixels, the intrinsic dispersion $\delta \psi_{\rm int}$ can be derived. The uncertainty of $\delta \psi_{\rm int}$ is estimated using a profile-likelihood approach: for each fixed value of $\delta \psi_{\rm int}$, the background fraction is re-optimized, and the $1\sigma$ confidence interval is defined by $\Delta(-\ln \mathcal{L})=0.5$. This method both incorporates the heterogeneous angle uncertainties of individual polarization measurements and mitigates the influence of unrelated or strongly misaligned segments. Besides, it can avoid the influence of histogram bin choice. We adopt this method for all the estimates of the intrinsic B-field dispersion in the following DCF analysis.

\section{Discussion} \label{sec:discuss}

\subsection{Magnetic fields at outflow direction} \label{discuss:outflow}

\subsubsection{Association with the MM1 outflow} \label{outflow:associate}

The alignment between bipolar outflow and B-field is often used as a diagnostic of how dynamically important B-fields are in star-forming regions. Most surveys conclude that aside from a few cases protostellar outflows and fields measured on $\sim\!1000$~au scales are consistent with random alignment \citep{hull&zhang2019, huang2024}. Recent high-resolution ALMA observations have shown that outflow/jet feedback can reshape the local field morphology (e.g., Serpens SMM1, \citealt{hull2017a}; B335, \citealt{maury2018}; Serpens Emb 8(N), \citealt{lego2019};  L1448 IRS 2, \citealt{kwon2019}; HH211, \citealt{choi2025}). Although B-fields can appear relatively well ordered and roughly parallel to the outflow cavity walls, such configurations are generally interpreted as the result of outflow-driven shaping rather than evidence of a dynamically dominant field.

\begin{figure*}
\centering
\includegraphics[width=\linewidth]{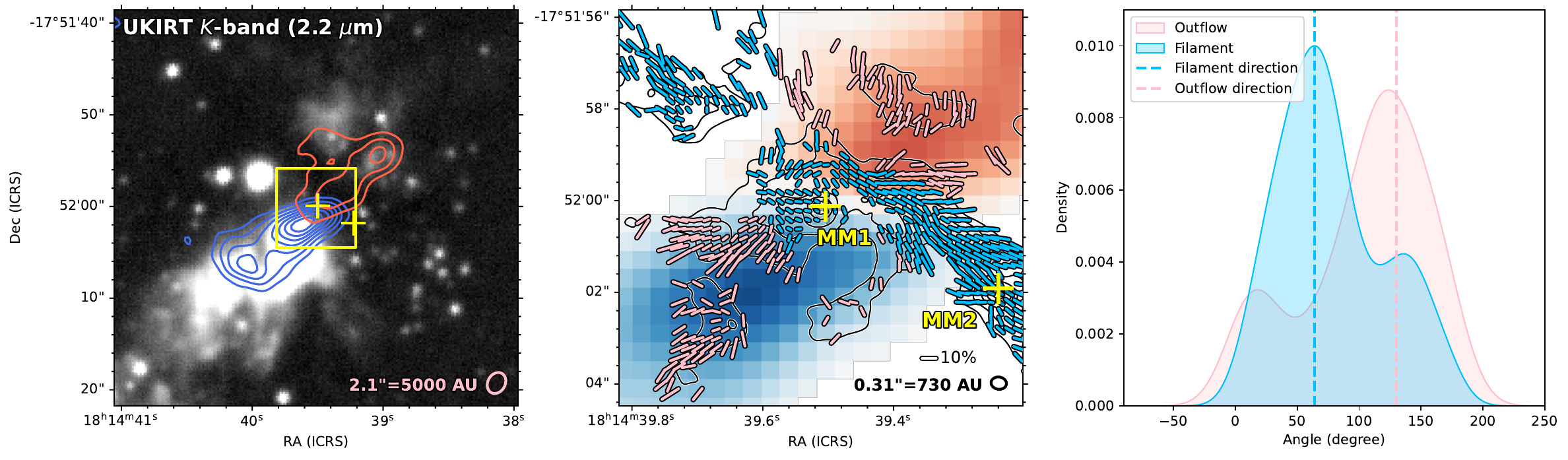}
\caption{\textit{Left}: The UKIRT $K$-band (2.2~$\mu$m) image is overlaid with the moment-0 contours of the high-velocity wings of HCO$^+$~(1-0). The integration is from -2 to -18~\kms\ for blue and from 58 to 78~\kms\ for red wings, respectively. The contour levels are 0.5, 0.7, 0.9, 1.1, 1.3, and 1.5 \jybeam~$\cdot$~\kms\ for blue and 0.3, 0.5, and 0.7 \jybeam~$\cdot$~\kms\ for red wings, respectively. The yellow box indicates the zoomed-in field of the middle panel while the yellow crosses mark the positions of MM1 and MM2.
\textit{Middle}: The background shows the HCO$^+$~(1-0) high-velocity wings but are color-coded for integrated intensity. The pink and blue segments are the 1.2~mm B-field segment groups that are associated with the outflow and the main gas filament, respectively. The polarization fraction $\hat{f}_P$ scaled by square-root is indicated by the length of segments, with a 10\% scale bar shown on the bottom right. Black contours of Stokes I map are the same as Fig.~\ref{fig:bfield}.
\textit{Right}: The kernel density estimation (KDE) of B-field angle distribution for the outflow (pink) and the filament (blue) groups. The dashed vertical lines indicate the directions of outflow and filament.
}
\label{fig:outflow}
\end{figure*}

In the case of W33~A, the NIR polarimetry indicates centrosymmetric patterns of polarization segments surrounding the W33~A infrared sources, mostly at northwest-southeast direction \citep[see Fig.~2 in][]{kwon2025}. Rotated by $90^\circ$, those segments can trace back to the infrared source A \citep[see Fig.~4 and 5 in][]{kwon2025}. It is most likely that the jet/outflow at this direction evacuates the bipolar cavity walls across a spatial extent of $\sim$1~pc and their surface or volume dust scatters the star light. Therefore, these segments only trace the illumination geometry rather than the B-field direction. However, such geometry inspires the isolation of NW-SE B-field direction as found in Sect.~\ref{result:bfield}. We retrieve the UKIRT $K$-band (2.2~$\mu$m) photometric data \citep{lucas2008} in Fig.~\ref{fig:outflow} Left, and overlap it with ALMA HCO$^+$~(1-0) red and blue wings from the ATOMS survey, to demonstrate their good spatial correspondence. The P.A. of the HCO$^+$ outflow is $49^\circ$, similar to the P.A. of $56^\circ$ of the CO outflow reported by \citet{galvan2010}. To isolate the B-field segments mainly influenced by the outflow, we generated a mask from the moment-0 map of HCO$^+$~(1-0) high-velocity wings (-2 to -18~\kms\ for blue and from 58 to 78~\kms\ for red wings), but excluded those regions dominated by other local processes (i.e., MM1 in-spiral and the F-Main filament). As a result in the middle panel of Fig.~\ref{fig:outflow}, the B-field segments are separated into two groups: those spatially related with the outflow are highlighted in pink, and the others are in blue. The pink and blue groups are essentially NW-SE and NE-SW components identified by visual inspection in Sect.~\ref{result:bfield}, but it is now better-defined here. In the right panel, two groups are also shown to demonstrate their distinct distributions and their good alignment with corresponding physical structures, i.e., outflow and filament.

\begin{figure*}
\centering
\includegraphics[height=0.4\linewidth]{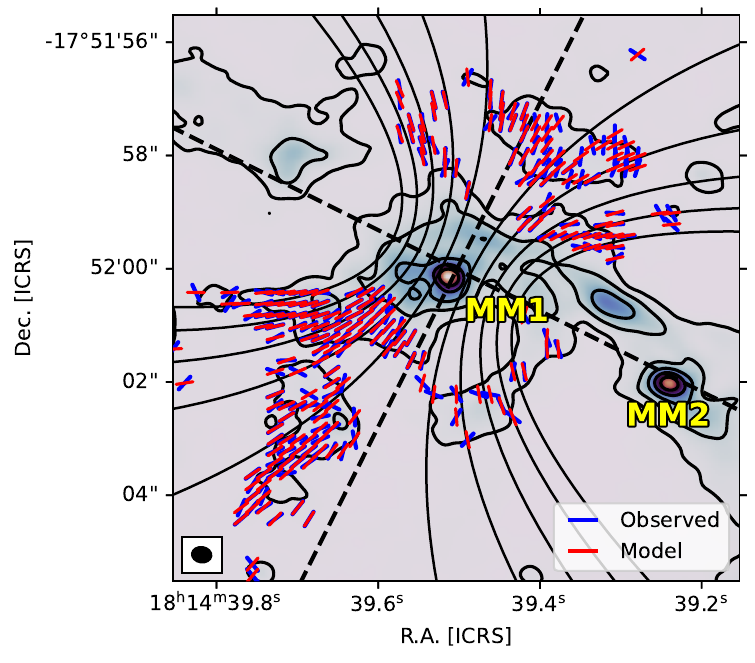}
\includegraphics[height=0.4\linewidth]{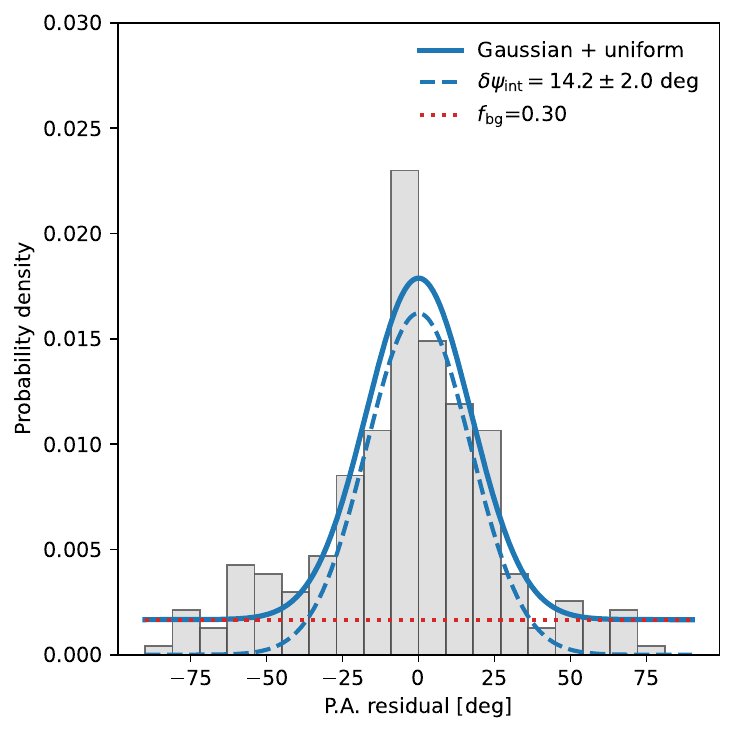}
\caption{\textit{Left}: B-fields associated with outflow are fitted by parabolic curves. The optimized parabolic models are show in blue and red segments. Black dashed lines mark the coordinate system while the solid lines indicate the parabolic curves every step of $0\farcs2$. Background color map and contours show 1.2~mm Stokes I continuum emission, with same levels as Fig.~\ref{fig:bfield}. \textit{Right}: Residual angle distribution and fitted model. Nyquist-sampled data points are used. The intrinsic angle dispersion, inferred from the forward modeling, is labeled as $\delta \psi_{\rm int}$. The uniform background contribution $f_{\rm bg}$ is indicated as red dashed line.}
\label{fig:outflow_modelfit}
\end{figure*}

\subsubsection{Estimates of B-field strength and energy balance} \label{outflow:estimate}

The B-fields associated with outflow (pink in Fig.~\ref{fig:outflow}) can be described with a series of parabolic functions $y = g (Cx^2 + 1)$ in a Cartesian coordinate system rotated by a specific angle. $x$ and $y$ are the POS distance in arcsecond offset from the center of the coordinate system \citep{girart2006, rao2009, qiu2014}. The fitting result gives the P.A. of the coordinate system $\theta_{\rm PA} = 63.6(^{+4.6}_{-7.1})^\circ$, the center of symmetry of the B-field (RA, Dec)$_{\rm ICRS}$ = ($18^\mathrm{h}14^\mathrm{m}39^\mathrm{s}\!.50 \pm 00^\mathrm{s}\!.02, -17^\mathrm{d}52^\mathrm{m}00^\mathrm{s}\!.04 \pm 00^\mathrm{s}\!.33$), and curvature $C=7.4^{+3.8}_{-1.4}\times10^{-4}$~pix$^{-1}$. More details of the parabolic fitting methodology are described in Appendix~\ref{app:parabola-fit}. As shown in Fig.~\ref{fig:outflow_modelfit}, the model agrees mostly well with the observed fields except for the western half of F2 and C14. The residual distribution of position angle is shown on the right panel. By forward model (Eq.~\ref{eq:angle_model}), the intrinsic dispersion gives $\delta\psi_{\rm obs}=14.2 \pm 2.0~\mathrm{deg}$. 

The mean volume density of the outflow shells are estimated as the average value of F2 and C13 regions, because they are clean from field distortions by local physics like MM1 and C11, and they are representative for both sides of the outflow. The mean volume density gives $(6\pm 2)\times10^6$~cm$^{-3}$. Substituting these parameters into Eq.~(\ref{eq:dcf}), the POS B-field strength is estimated to be $2.5~\mathrm{mG}$. Following the modeling work of \citet{dewit2010}, we adopt an inclination $i=60^\circ$ and the total B-field strength along the outflow path is $B = B_{\rm pos} / \cos(i) = 5.1~\mathrm{mG}$. These measurements and calculations are summarized in Table~\ref{tab:dcf}. 

\begin{table*}
\caption{Summary of the B-field strength and related parameters derived from the DCF method.}
\renewcommand{\arraystretch}{1.2}
\label{tab:dcf}
\centering
\small
\begin{tabular}{c|cccccccc}
\hline\hline
\\[-6pt]
Region & $\delta\psi_{\rm int}$ & $\delta v^{\rm nt}_{\rm los}$ & $\langle n_{\rm H_2}\rangle$ & $i$ & $B_{\rm pos}$ & $B_{\rm tot}$ & $v_A$ & $\mathcal{M}_A$ \\[2pt]
& [$^\circ$] & [km\,s$^{-1}$] & [$\times10^6$\,cm$^{-3}$] & [$^\circ$] & [mG] & [mG] & [km\,s$^{-1}$] & \\[2pt]
\hline
\\[-8pt]
NW-SE & $14.2\pm2.0$ & $1.20\pm0.10$ & $6\pm2$ & 60$^{\rm (d)}$ & $2.5^{+0.6}_{-0.6}$ & $5.1^{+1.3}_{-1.1}$ & $2.7^{+0.9}_{-0.7}$ & $0.6^{+0.1}_{-0.1}$ \\
F1 & $14.4\pm3.0$ & $0.46\pm0.02$ & $5\pm2$ & 30$^{\rm (e)}$ & $0.9^{+0.3}_{-0.2}$ & $1.0^{+0.3}_{-0.3}$ & $0.6^{+0.3}_{-0.2}$ & $1.0^{+0.2}_{-0.2}$ \\
F-Main & $15.7\pm1.7$ & $0.44\pm0.01$ & $26\pm9$ & 30$^{\rm (e)}$ & $1.8^{+0.4}_{-0.4}$ & $2.0^{+0.4}_{-0.4}$ & $0.5^{+0.2}_{-0.1}$ & $1.1^{+0.1}_{-0.1}$ \\
Tail & $12.0\pm1.8$ & $0.31\pm0.03$ & $7\pm2$ & 30$^{\rm (e)}$ & $0.8^{+0.2}_{-0.2}$ & $1.0^{+0.2}_{-0.2}$ & $0.5^{+0.1}_{-0.1}$ & $0.8^{+0.1}_{-0.1}$ \\
\multirow{2}{*}{MM1-spiral} & $12.5\pm1.6^{\rm (a)}$ & \multirow{2}{*}{$1.24\pm0.03$} & \multirow{2}{*}{$10\pm4$} & \multirow{2}{*}{30$^{\rm (e)}$} & $3.8^{+0.9}_{-0.9}$ & $4.4^{+1.1}_{-1.1}$ & $1.8^{+0.7}_{-0.5}$ & $0.9^{+0.1}_{-0.1}$ \\
 & $15.7\pm2.6^{\rm (b)}$ &  &  &  & $3.1^{+0.9}_{-0.8}$ & $3.5^{+1.0}_{-0.9}$ & $1.5^{+0.6}_{-0.4}$ & $1.1^{+0.2}_{-0.2}$ \\
MM2 & $13.8\pm1.8$ & $1.02\pm0.04$ & $48\pm16^{\rm (c)}$ & 30$^{\rm (e)}$ & $6.3^{+1.4}_{-1.3}$ & $7.2^{+1.7}_{-1.5}$ & $1.4^{+0.4}_{-0.3}$ & $1.0^{+0.1}_{-0.1}$ \\
\hline
\end{tabular}
\tablefoot{
$\delta\psi_{\rm int}$: intrinsic dispersion of polarization position angle with local angle uncertainty removed. For the MM1-spiral, there are two methods: (a) After removing a smoothed B-field with a kernel size of three beams. (b) Original dispersion without removing any smoothed field (upper limit). See Sect.~\ref{discuss:mm1} for more details. 
$\delta v^{\rm nt}_{\rm los}$: turbulent velocity dispersion measured from H$^{13}$CO$^+$~(3-2) line, averaged over the region of interest.
$n$(H$_2$): volume density of molecular hydrogen measured from the 1.2~mm dust continuum. (c) this value is recalculated as the mean density over the MM2 region rather than the denest part as calculated in Table~\ref{tab:coreinfo}. 
$i$: inclination angle; (d) retrieved from \citet{dewit2010}, (e) is assumed due to the moderate POS inclination angle for W33~A.
$B_{\rm pos}$: B-field strength in the plane of the sky. 
$B_{\rm tot}$: total B-field strength corrected by inclination angle. 
$v_A$: Alfv{\'e}nic speed. 
$\mathcal{M}_A$: Alfv{\'e}nic Mach number.
}
\end{table*}

We compare the pressure, i.e., the energy density of outflow and B-field. The magnetic pressure writes \citep{MHDequations},
\begin{equation} \label{eq:pmag}
    P_B = \frac{B^2}{8\pi} \approx 4.0 \times10^{-8}\,\left(\frac{B}{1~\mathrm{mG}}\right)^2  \left[\mathrm{erg}\cdot\mathrm{cm}^{-3}\right].
\end{equation}
For a conical cavity of half–opening $\theta$ with bulk axial speed $v$ and local density $\rho$, outflow's expansion or ram pressure on the cavity wall can be estimated as \citep[e.g.,][]{lauvzikas2024},
\begin{equation} \label{eq:pram}
\begin{split}
    P_{\rm exp} &= \rho v^2 \sin^2\theta \\
    & \approx 4.7 \times 10^{-6} \left(\frac{n}{10^6~\mathrm{cm}^{-3}}\right) \left(\frac{v}{10~\mathrm{km/s}}\right)^2 \sin^2\theta \left[\mathrm{erg}\cdot\mathrm{cm}^{-3}\right].
\end{split}
\end{equation}
In the case of the NW-SE outflow, the intensity-weighted velocity shift of the outflow wings relative to the systematic velocity is $52$~\kms\ after inclination angle correction. The gas density is assumed to be the same as that of outflow cavity: $(6\pm 2)\times10^6$~cm$^{-3}$. We used a half opening angle $\theta=10^\circ$ measured by \citet{dewit2010}. Substituting these values, the B-field pressure to confine the outflow wall is $P_B \approx 1.0\times10^{-6}\,\mathrm{erg} \cdot \mathrm{cm}^{-3}$, while the outflow expansion pressure is $P_{\rm exp} \approx 2.3\times10^{-5}\,\mathrm{erg} \cdot \mathrm{cm}^{-3}$. Therefore, the B-field can only provide $5\%$ of the pressure to confine the outflow. Magnetic pressure is proportional to the density by $P_B \sim B^2 \sim \rho$ if the DCF assumption holds, while expansion pressure is also proportional to density $P_{\rm exp} \sim \rho$. That means, even if the density is uncertain, the ratio $P_{B}/P_{\rm exp}$ remains small. So the B-field morphology is more likely to be shaped by the outflow. Such a configuration is also observed in B335 \citep{maury2018}, where the expansion of a magnetized outflow cavity is responsible for the field orientation switching from perpendicular to almost parallel to the cavity axis \citep{galli2006, masson2016}. 

\subsection{Coherent B-fields along the filaments} \label{discuss:coherent}

\begin{figure*}
\centering
\includegraphics[width=0.8\linewidth]{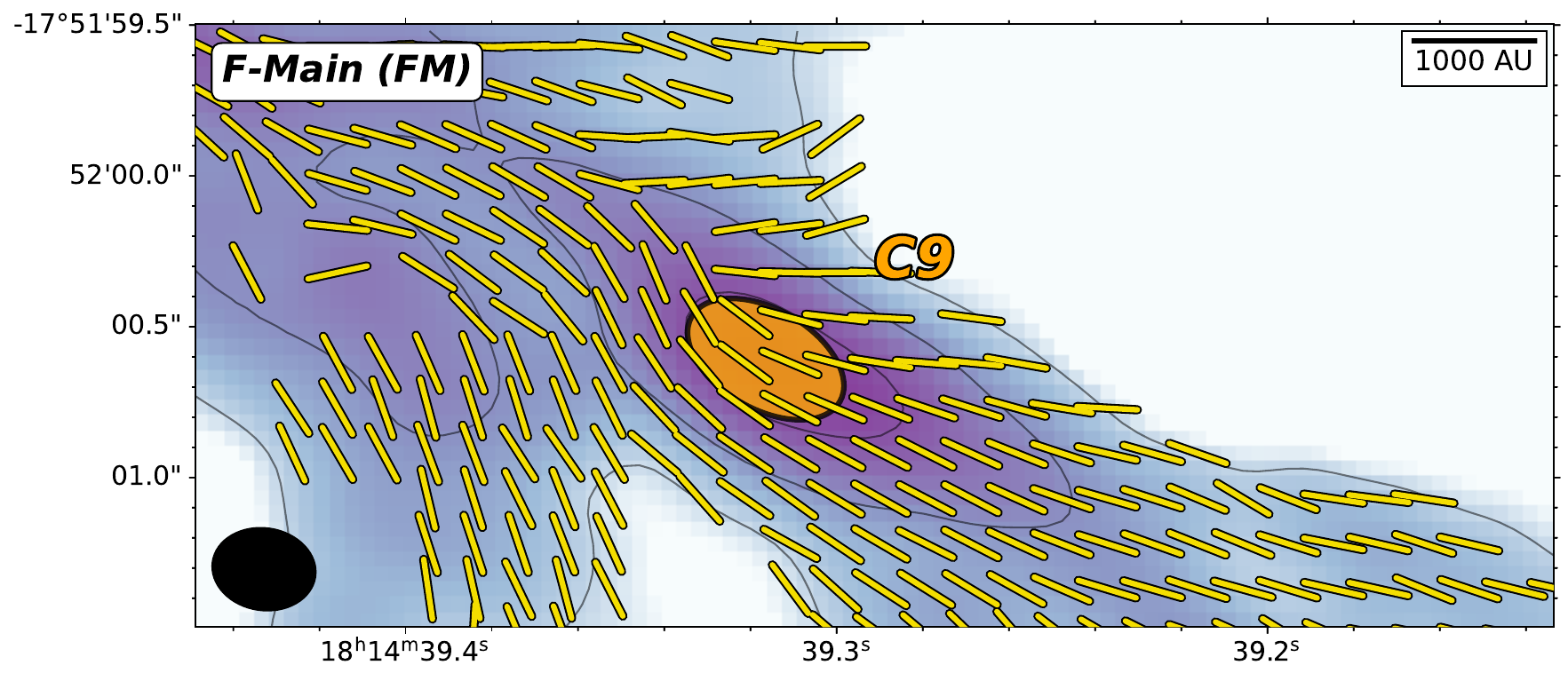}
\includegraphics[height=0.3\linewidth]{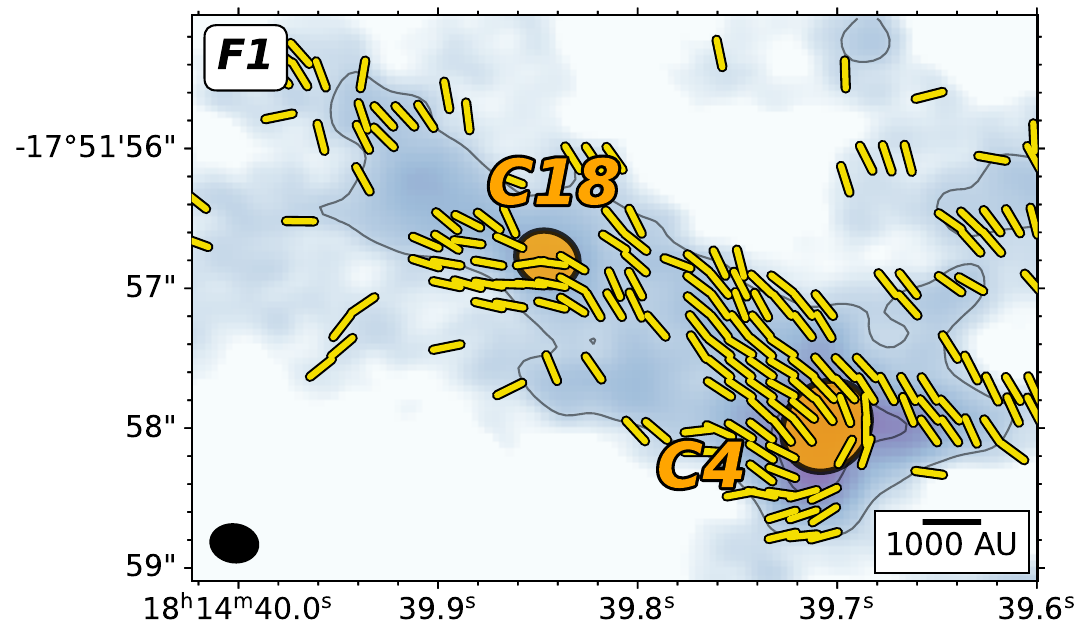}
\includegraphics[height=0.3\linewidth]{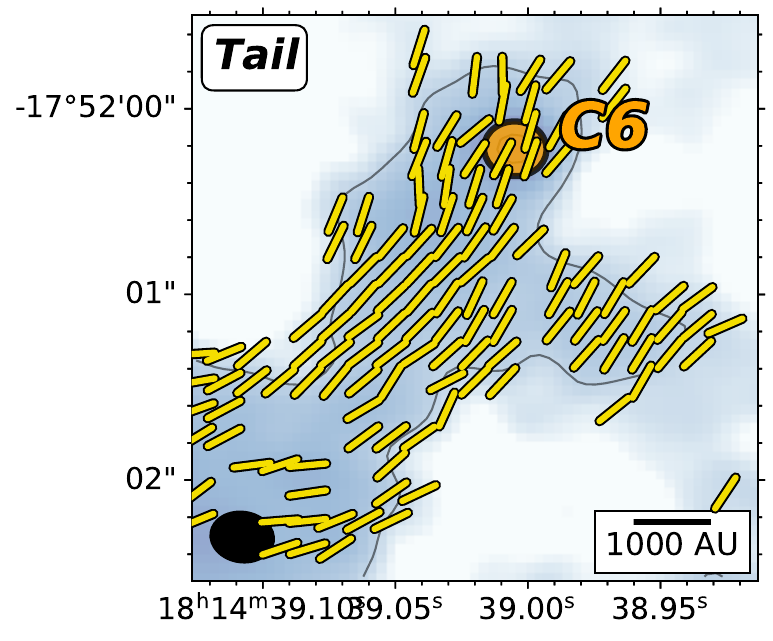}
\includegraphics[width=\linewidth]{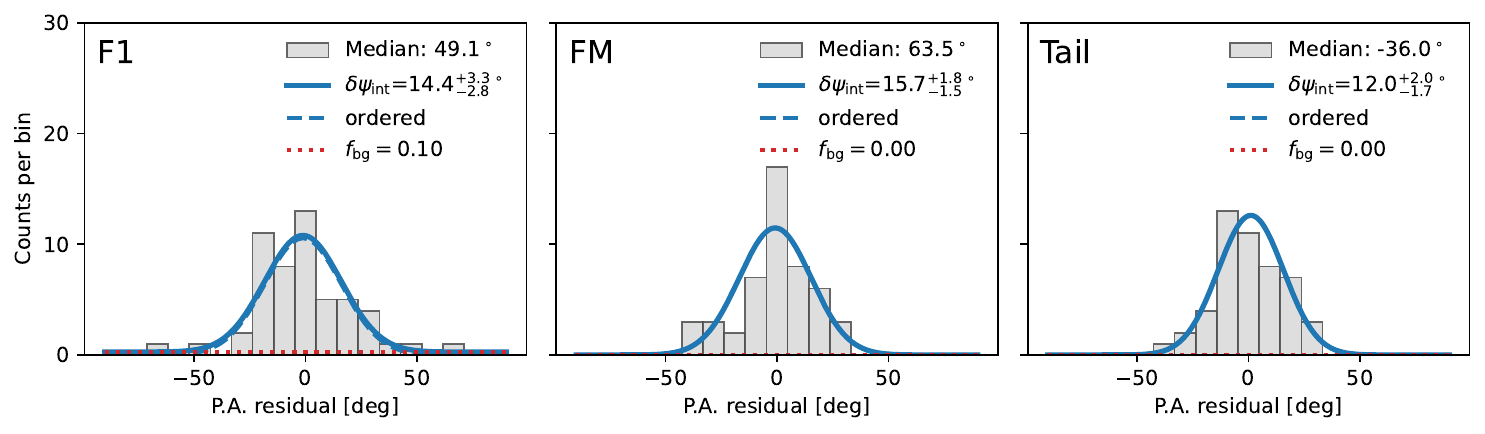}
\caption{{\it Top \& Middle}: Three filaments F-Main, F1, and Tail are zoomed-in to show the coherent B-fields along the filament spines. The dense cores associated with the filaments are shown in yellow shaded ellipses. {\it Bottom}: The distributions of P.A. that is shifted by a median value within [$-90^\circ$, $+90^\circ$] range. Nyquist-sampled measurement (every half beam) is presented. The intrinsic angle dispersion, inferred from the forward modeling, is labeled as $\delta \psi_{\rm int}$. 
}
\label{fig:coherent_fit}
\end{figure*}

The other global feature of B-fields in W33~A is a coherence along the elongation of the main filament. Along its path, there are three subregions of interest, namely F-Main (FM), F1 and Tail. A closer view is shown in Fig.~\ref{fig:coherent_fit}.

We used the DCF method to estimate the B-field strengths. For each region, P.A. is shifted by a median value into the range of [$-90^\circ$, $+90^\circ$]. The distribution is fitted using forward modeling (Eq.~\ref{eq:angle_model}) to obtain intrinsic dispersion $\delta \psi_{\rm int}$. The small $f_{\rm bg}$ further confirms the dominance of ordered field in the three regions. The velocity dispersion and volume density of the three subregions were previously derived in Sect.~\ref{result:structure} and summarized in Table~\ref{tab:dcf}. Substituting these quantities into Eq.~\ref{eq:dcf}, $B_{\rm pos}$ is obtain. The inclination angle of the main filament is unknown, but we assumed it to be more likely oriented close to the plane of the sky because of its elongated morphology; otherwise, the filament will be much shorter due to projection. So the inclination angle of $30^{\circ}$ is adopted. With this assumption, we further derived the total B-field strength, $B_{\rm tot}$, as well as the corresponding Alfv{\'e}nic velocity and Alfv{\'e}nic Mach number, as listed in Table~\ref{tab:dcf}. Among these filaments, the turbulent motion is all trans-Alfv{\'e}nic, consistent with the coherent B-fields.

To quantify the radial stability of the filaments, we idealize them as cylinders and compare their observed line masses with an effective critical line mass \citep{ostriker1964, wang2014, hacar2023},
\begin{equation}\label{eq:mline_crit}
    m_{\rm crit}(\sigma_{\rm tot}) =
    \frac{2\sigma^2_{\rm tot}}{G}
    \approx\
    465
    \left(\frac{\sigma_{\rm tot}}{1\,\mathrm{km~s^{-1}}}\right)^2
    \,M_\odot\,{\rm pc}^{-1},
\end{equation}
where $\sigma_{\rm tot}$ represents the effective one-dimensional support against radial collapse. It consists of three contributions,
\begin{equation}
    \sigma_{\rm tot}^2 =
    c_s^2 + (\delta v^{\rm nt}_{\rm los})^2 + v_A^2,
\end{equation}
where $c_s$, $\delta v^{\rm nt}_{\rm los}$, and $v_A$ describe the thermal, non-thermal turbulent, and magnetic contributions, respectively. Since the observed plane-of-sky B-fields are largely aligned with the filament axes, the measured field is expected to contribute to support against radial contraction. We caution, however, that the three-dimensional field geometry is unknown, and therefore the magnetic contribution should be regarded as an approximate estimate. The sound speed is
\begin{equation} \label{eq:cs}
    c_s =
    \sqrt{\frac{k_B T}{\mu m_{\rm H}}}
    \approx\
    0.19
    \left(\frac{T}{10\,\mathrm{K}}\right)^{0.5}
    \quad \mathrm{km\,s^{-1}},
\end{equation}
where $k_B$ is the Boltzmann constant and $\mu=2.33$ is the mean molecular weight per particle.

\begin{table}
\caption{Observed and critical line masses of the three filaments.}
\label{tab:mlin_crit}
\centering
\small
\begin{tabular}{c|cccc}
\hline\hline
\\[-6pt]
Region & $m_{\rm obs}$ & $m_{\rm crit}^{\rm therm}$ & $m_{\rm crit}^{\rm turb}$ & $m_{\rm crit}^{\rm mag}$ \\
 & [M$_\odot$\,pc$^{-1}$] & [M$_\odot$\,pc$^{-1}$] & [M$_\odot$\,pc$^{-1}$] & [M$_\odot$\,pc$^{-1}$] \\[3pt]
\hline
\\[-8pt]
F1      & $110\pm40$ & $41\pm4$ & $140\pm30$ & $290\pm80$ \\
F-Main  & $200\pm70$ & $41\pm4$ & $130\pm30$ & $250\pm80$ \\
Tail    & $60\pm20$ & $41\pm4$ & $83\pm20$ & $200\pm60$ \\
\hline
\end{tabular}
\tablefoot{
$m_{\rm obs}$: observed line mass.
$m_{\rm crit}^{\rm therm}$: thermal critical line mass.
$m_{\rm crit}^{\rm turb}$: turbulent critical line mass.
$m_{\rm crit}^{\rm mag}$: magnetic critical line mass.
Uncertainties in parentheses denote 1$\sigma$ errors.
}
\end{table}

Substituting the physical parameters of each subregion, we derived three types of critical line mass that account for thermal, turbulent, and magnetic supports against self-gravity (Table~\ref{tab:mlin_crit}). By comparing them with the observed line masses, we find that for F1 and Tail, the turbulent support is sufficient to maintain stability against gravitational collapse, while B-fields may provide an additional stabilizing contribution. In contrast, the FM filament line mass $200\pm70~M_\odot\,{\rm pc}^{-1}$ is larger than or approximately equal to the critical line mass due to turbulence, indicating that the B-field may play an important role in stabilizing the filament. The magnetic support is also consistent with the observed scarcity of dense fragments in all the three filaments (see Fig.~\ref{fig:coherent_fit}). Otherwise, FM should contain more dense structures than F1 and Tail. But further high-resolution observations are indeed required to confirm whether the elongated core C9 (the only core in FM) has fragmentation inside or simply shows a smooth density profile. We also noticed that FM shows some large-scale bending. So, the angle dispersion caused by turbulent motion could be even smaller if the bending is removed. In this case, the B-fields may be underestimated in FM and magnetic support could be even larger, maintaining our conclusions.

The magnetic configuration of these regions resembles the poloidal field morphology observed in B335 \citep{maury2018, yen2020}, where the equatorial B-fields are pinched inward by gravitational inflow. As shown in Fig.~\ref{fig:velocity}, we find no significant LOS velocity gradient along F-Main. This absence does not necessarily rule out longitudinal gas motions, because projection effects can potentially dilute the observed velocity gradient by a factor of $\sin i$, where $i$ is the inclination angle. If longitudinal inflow is present, the inflowing material could drag and amplify the B-field, producing the observed field morphology approximately parallel to the filament axes \citep{gomez2018}. Such dynamically important B-fields can increase the maximum stable line mass of a filament \citep{fiege2000, tomisaka2014, kirk2015, pillsworth2024}. Observationally, such magnetically regulated, high line-mass filaments with parallel fields have been identified in several star-forming regions, including Musca \citep{cox2016}, DR21 \citep{ching2022}, and Mon R2 \citep{hwang2022}. As such, the poloidal-dominated B-fields in W33~A can stabilize the massive inflowing gas against local fragmentation, thereby enabling more efficient mass accretion onto dense cores, for example MM1. Such magnetically moderated accretion may represent a crucial mechanism preceding the formation of dense stellar clusters in high-mass star-forming environments. 

\subsection{Spiral-in B-fields in MM1} \label{discuss:mm1}

\begin{figure*}
    \centering
    \includegraphics[height=0.4\linewidth]{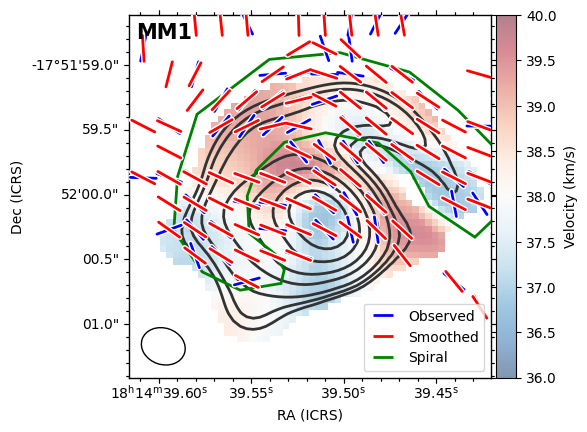}
    \includegraphics[height=0.4\linewidth]{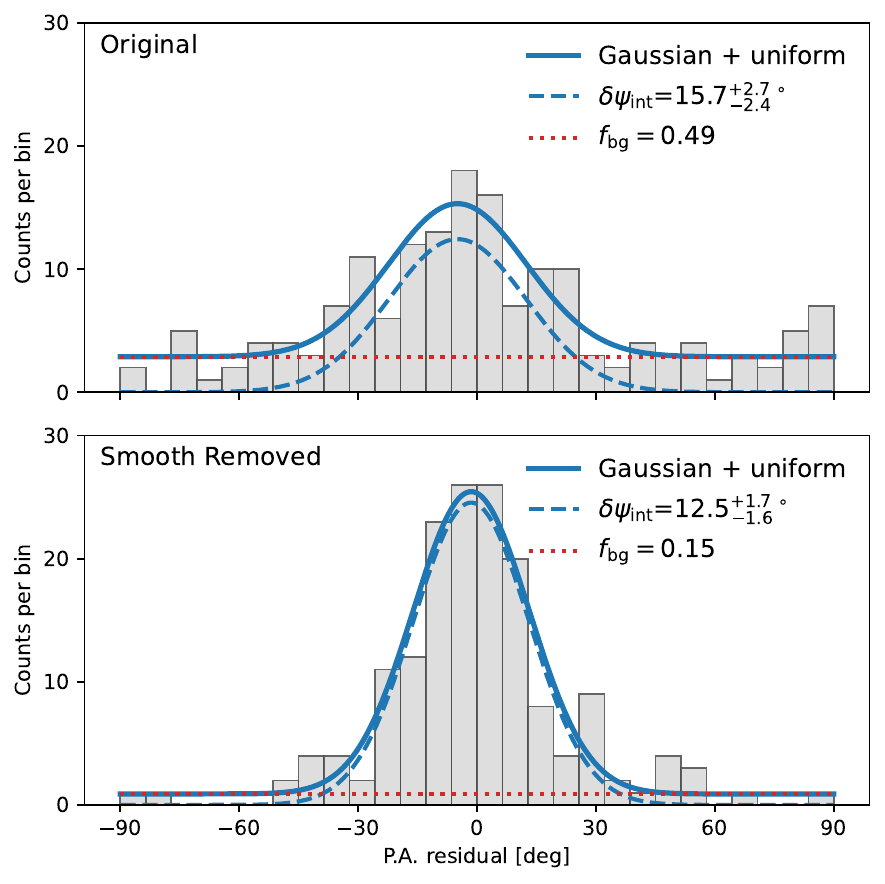}
    \caption{
    \textit{Left}: The background color map and black contours show the moment-1 and moment-0 of the CH$_3$CN $K=3$ line towards MM1. The contour levels are 0.8, 1.0, 1.3, 2.0, 2.8, 4.0, 5.4, and 8.0~\jybeam~\kms. The spiral-in path is highlighted by the green polygon. Observed and smoothed B-field segments are overlaid in blue and red segments, respectively. 
    \textit{Right}: The angle distributions are shown in gray histograms. The intrinsic angle dispersion, inferred from the forward modeling, is labeled as $\delta \psi_{\rm int}$. The upper and lower ones show the cases before and after removal of smoothed field, respectively.}
    \label{fig:mm1_spiral}
\end{figure*}

Toward the MM1 region, CH$_3$CN lines trace a clear spiral-in structure, delineated by the hand-drawn green polygon in the left panel of Fig.~\ref{fig:mm1_spiral}. This spiral feature was first reported by \citet{maud2017} and later reproduced by a three-dimensional physical model \citep{izquierdo2018}. In our new polarization data, the observed B-field segments (shown in blue) follow the same spiral-in pattern. Along the spiral path, the field orientation rotates by nearly 180\degree, indicating that the subtraction of a uniform field component is not sufficient. So, we used two methods to constrain intrinsic angle dispersion. First, we only shift the original angle distribution in the range of [$-90^\circ$, $+90^\circ$] without any uniform field removal. In this case, since we didn't remove any larger-scale field, the forward modeling gives an intrinsic dispersion $15.7\pm2.6^\circ$, which should be set as the upper limit of the angle dispersion.

The second method is to remove a smoothed field which describes the spiral geometry. We smoothed the Stokes $Q$ and $U$ maps with a Gaussian kernel FWHM of three synthesized beams and then removed it from the original Q and U maps. The smoothed field is shown in red segments in the top panel. The technical details are described in Appendix~\ref{app:remove_smooth}. The choice of three beams is a balance between avoiding too small a kernel - which would reproduce most of the original field morphology and artificially suppress the residual angle dispersion - and avoiding too large a kernel, which would over-smooth the underlying spiral geometry and fail to capture the large–scale curvature of the field. After the removal of the smoothed field, the residual angle distribution is shown on the bottom right. The intrinsic angle dispersion is $12.5\pm1.6^{\circ}$. Because the smoothing procedure absorbs any angular variations on scales comparable to or larger than the 3-beam kernel, subtracting the smoothed field removes part of the true turbulent component. As a result, the residual angles provide a lower limit to the intrinsic dispersion used in the DCF analysis. 

We used the two angle dispersions obtained above to constrain B-field strengths. Because CH$_3$CN gas shows extended and spatially consistent emission with 1.2~mm dust continuum emission, hot gas and dust are thought to be well mixed here. So, the dust temperature in the MM1 spiral is assumed to be equal to the gas temperature measured from CH$_3$CN, which is $130\pm30~\mathrm{K}$ \citep{maud2017}. The spiral has a mean intensity of $6.1~\mjybeam$ at 1.2~mm, and the mean column density is calculated to be $2.1(\pm1.0)\times10^{23}$~cm$^{-2}$ using Eq.~(\ref{eq:colden}). The width of MM1 spiral is measured as the width of the green polygon in Fig.~\ref{fig:mm1_spiral}, which yields 0.007~pc. If the spiral is cylinder-like, its depth should be the same as its width and the volume density is calculated as $10(\pm4)\times10^{6}$~cm$^{-3}$. Adopting the DCF method, we obtain the POS field strength of $3.1{-}3.8$~mG and total strength $3.5{-}4.4$~mG. These calculations and results are listed in Table~\ref{tab:dcf}. 

The MM1 spiral is approximately trans-Alfv{\'e}nic with $0.9 \lesssim \mathcal{M}_A < 1.1$. The MM1 spiral exhibits an evident velocity gradient, which has been interpreted as an infalling streamer \citep{maud2017, izquierdo2018}. Such a trans-Alfv{\'e}nic streamer represents a high-mass analog of the recently detected magnetically regulated infalling streamer in SVS13A \citep{cortes2025} and HOPS-182 \citep{Huang2026}, where magnetic field is strong enough to help confine and guide the infalling gas and efficiently remove the angular momentum.

The observed alignment between the gas kinematics and the B-field morphology indicates a strong dynamical coupling between the two. One possible interpretation is that the flow is partially guided along B-field lines, producing the observed spiral morphology. Alternatively, the B-field may be dynamically shaped by the infalling gas, such that the field is stretched and bent to follow the gas motion. In this case, the observed alignment reflects the response of the B-field to a flow with comparable kinetic and magnetic energy densities. Recent numerical simulations by \citet{tu2024} reproduce such laminar, magnetically coupled inflows, demonstrating that magnetic diffusion together with gravitational bending of field lines can naturally produce similar accretion channels.

\subsection{Hourglass-shaped B-fields in MM2} \label{discuss:mm2}

\begin{figure*}
\centering
\includegraphics[height=0.4\linewidth]{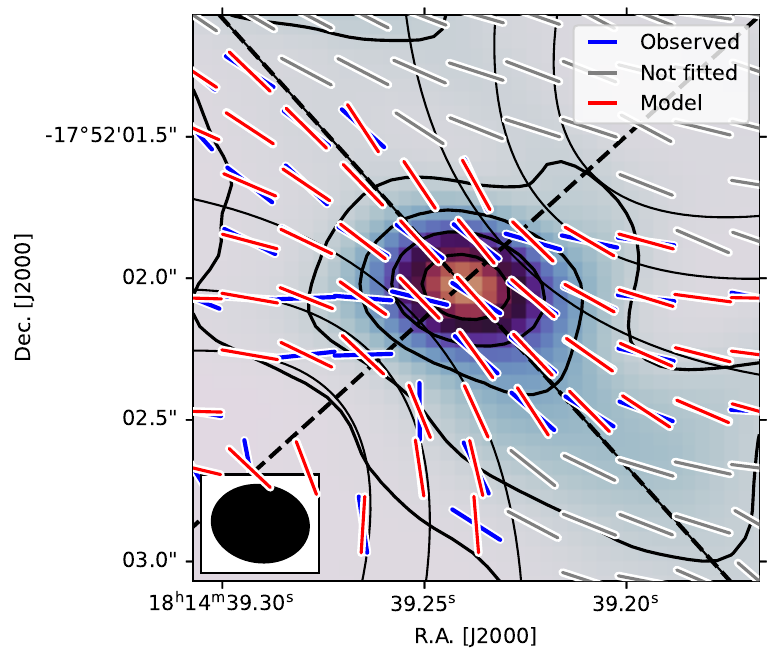}
\includegraphics[height=0.4\linewidth]{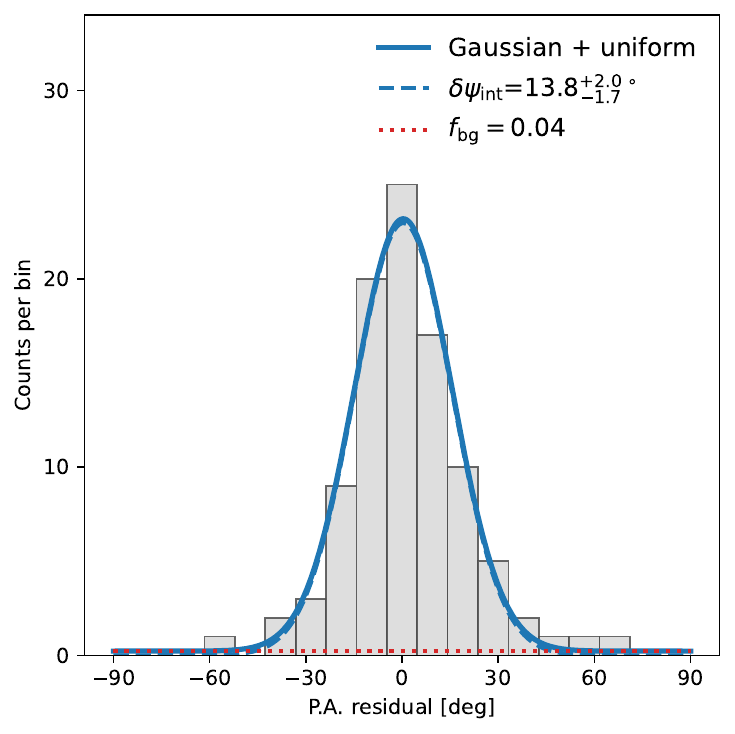}
\caption{\textit{Left}: The MM2 B-fields and fitted parabolic models are shown in blue and red segments. Black dashed lines mark the coordinate system while the solid lines indicate the parabolic curves every step of $0\farcs2$. The background color map and contours are the same as Fig.~\ref{fig:bfield}. \textit{Right}: The distribution of residual angles from parabolic model-fitting is shown in the gray histogram. The intrinsic angle dispersion, inferred from the forward modeling, is labeled as $\delta \psi_{\rm int}$. }
\label{fig:mm2_modelfit}
\end{figure*}

Fig.~\ref{fig:mm2_modelfit} shows the B-fields at the position of MM2. At the northwestern and southwestern corners, a set of gray segments coherently follow the NE-SW global direction (i.e., blue shade in Fig.~\ref{fig:bfield}), while the rest of segments in blue significantly deviate from the direction, showing a pinched or an hourglass shape. This morphology can be again fitted by parabolic functions as described in Appendix~\ref{app:parabola-fit}. The best fitting model is presented in red segments, providing the center of symmetry of (RA, Dec)$_{\rm ICRS}$ = ($\rm 18^h14^m39^s\!.24 \pm 00^s\!.02, -17^d52^m01^s\!.77 \pm 00^s\!.25$), position angle of $47(\pm7)^\circ$ and curvature $C=5.3^{+1.8}_{-1.8}\times10^{-3}$~pix$^{-1}$. We have modeled the residual position angle distribution with intrinsic dispersion of $13\farcs8$. The mass and density of MM2 are recalculated using a 0.6\arcsec-radius photometric circle because the B-field measurements are more extended than the \textit{getsf} dense core, but better defined by the farthest parabolic curve with $g=0\farcs6$ or equivalent to 1400~au. The 1.2~mm integrated flux density and peak intensity are 107.2~mJy and $53.0~\mjybeam$, which gives a mass of $10.8 \pm 3.7~\msun$ using Eq.~(\ref{eq:mcore_thick}) under the same assumed temperature of 25~K as Sect.~\ref{structure:core}. The volume density is also recalculated using Eq.~(\ref{eq:volden}) to be $4.0(\pm2.0)\times10^7$~cm$^{-3}$. The DCF method gives POS B-field of $6.3\pm1.4~\mathrm{mG}$ and total B-field of $7.2\pm1.6~\mathrm{mG}$.

If B-field is pulled by gravitational collapse to form a hourglass shape, its strength could be estimated from the curvature of field lines considering the force balance between gravity and B-field \citep[e.g.,][]{schleuning1998, tang2009, qiu2014}
\begin{equation} \label{eq:curve}
    B \approx\ 1.4~\mathrm{mG}~\left(\frac{R_c}{0.1~\mathrm{pc}}\right)^{0.5} \left(\frac{D_c}{0.01~\mathrm{pc}}\right)^{-1} \left(\frac{M}{1~M_\odot}\right)^{0.5} \left(\frac{n}{10^6~\mathrm{cm}^{-3}}\right)^{0.5}
\end{equation}
where $R_c$ is the curvature radius, $D_c$ is the distance from the field line to the symmetry center, $M$ is the object mass, and $n$ is the average volume density. For the farthest parabolic curve in MM2, the curvature is $R_c \sim 940~\mathrm{au}$ and $D_c \sim 1400~\mathrm{au}$, we obtain a field strength of $8.9\pm3.5$~mG. Note that if the collapse has started or gravity has already dominated the system, B-fields should be smaller than this value because the force balance does not hold any more. Above all, the values obtained from the two methods are remarkably consistent with each other, yielding a mean value of $8.1\pm1.9$~mG in MM2.

The normalized mass-to-flux ratio $\lambda \equiv (M/\Phi)/(M/\Phi)_{\rm crit}$, with magnetic flux $\Phi=\pi R^2 B$ and $(M/\Phi)_{\rm crit}=1/(2\pi\sqrt{G})$ writes,
\begin{equation}
    \lambda \;=\; 2\pi \sqrt{G}\,\frac{M}{\pi R^2 B}
    \approx\ 4.6\,
    \left( \frac{M}{1~M_\odot} \right)
    \left( \frac{B}{1~\mathrm{mG}} \right)^{-1}
    \left( \frac{R}{1000~\mathrm{au}} \right)^{-2}.
\end{equation}
The uncertainty of $\lambda$ is propagated from mass and B-field strength uncertainties. We found that MM2 is supercritical, $\lambda = 3.2\pm1.4 > 1$, indicating that the B-field may add support against gravity but is not by itself sufficient to halt collapse. 

To evaluate the dynamical state of MM2, we computed its virial parameter by including both kinetic and magnetic support \citep{bertoldi1992},
\begin{equation} \label{eq:virial}
    \alpha_{\rm tot} = \alpha_{\rm kin} + \alpha_{B} = \frac{5R}{GM}\,\left((\delta v^{\rm nt}_{\rm int})^{2} + \frac{v_{\rm A}^{2}}{6}\right).
\end{equation}
For MM2, we obtain $\alpha_{\rm kin}\sim0.75\pm0.26$ and $\alpha_{B}\sim0.25\pm0.11$, giving a total virial parameter of $\alpha_{\rm tot}\simeq1.0\pm0.28$. Uncertainties are propagated from $M$, $\delta v^{\rm nt}_{\rm int}$, and $v_A$. These values indicate that turbulent motions alone cannot support the core, while the B-field provides a dynamically important—yet still insufficient—level of support. The magnetic contribution implied by the virial analysis is consistent with the observed hourglass morphology, which naturally arises when gravity acts on an initially ordered field. Because $\alpha_{\rm tot}<2$, MM2 remains gravitationally bound and is likely contracting. However, with substantial magnetic moderation, the collapse is expected to proceed more slowly than in a purely hydrodynamic case. 

To put into the context of W33~A core evolution, MM1 is considerably more evolved than MM2, exhibiting an energetic jet/outflow and a rich ``hot molecular line forest'' at millimeter wavelengths (features also observed in our ALMA data; see the top right panel in Fig.~\ref{fig:ch3cn}). In contrast, although MM2 appears to host a larger molecular gas reservoir, it is deficient in hot molecular lines; only a weak CH$_3$CN signal is recovered in our stacked line cube (see the top left panel in Fig.~\ref{fig:ch3cn}). Molecular outflows are likewise absent toward MM2. We have checked HCO$^+$ (1-0) and further CS~(5-4) at 244.9356~GHz and found no extended emission at high-velocity line wings towards MM2. Although \citet{galvan2010} mentioned that a red-shifted CO outflow in the north could originate from MM2, it is not clearly detected in our high-resolution data. Based on these contrasts, if MM1 and MM2 in the same protocluster formed contemporaneously, MM1 must have evolved more rapidly than MM2 \citep{galvan2010}.

Our new magnetic-field data could provide a physical pathway for
this divergence. In our proposed model both the spiral and F-Main filament is stabilized by B-fields from local fragmentation, which provides an efficient gas accretion flow towards MM1, accelerating evolutionary progression. By contrast, in MM2 the stronger, more ordered B-field likely raises the effective support to inhibit further gravitational collapse and delay the onset of vigorous star formation. If such magnetically moderated picture and causality holds, a tale of two dense cores in W33~A provides a good example to argue that B-fields are dynamically important to high-mass star formation. 

\subsection{Caveats}

For the DCF method (Eq.~\ref{eq:dcf}), we have implicitly assumed an approximate equipartition between turbulent magnetic energy and turbulent kinetic energy. However, this assumption is strictly valid only under sub- to trans-Alfv{\'e}nic conditions. In particular, numerical studies have shown that even in super-Alfv{\'e}nic regimes, the observed polarization angle dispersion can remain comparable to that expected in trans- or sub-Alfv{\'e}nic cases, especially when projection effects and limited angular resolution are taken into account. Therefore, an inferred sub-Alfv{\'e}nic or trans-Alfv{\'e}nic Mach number does not uniquely imply that the B-field dynamically dominates over turbulence (e.g. Sect.~\ref{discuss:mm1}). Instead, it should be interpreted as an indication that the B-field is at least dynamically important, but not necessarily dominant. In this context, the DCF-derived B-field strength and Alfv{\'e}nic Mach number should be regarded as order-of-magnitude estimates rather than precise measurements.

\section{Conclusions} \label{sec:conclude}

Magnetic fields (B-fields) are integral to our understanding of the formation and dynamical evolution of high-mass stars. But our knowledge of B-fields in massive protocluster formation remains incomplete. Linear polarized emission from magnetically aligned dust grains provides a good way to map the morphology B-fields on the plane of sky. In this paper, we present 1.2~mm full polarization observation of W33~A, a massive star-forming region at 2.4~kpc, by the \textit{Atacama Large Millimeter/submillimeter Array} (ALMA). Our findings are summarized below.

\begin{itemize}
\item With an angular resolution of $\sim$0\farcs3 ($\sim$730~au), W33~A is resolved into 20 dense cores and 9 filaments.
\item From linear polarized dust emission at 1.2~mm, the B-fields are mapped across the W33~A massive protocluster. The field shows various features: two large-scale perpendicular components at northwest–southeast (NW-SE) and northeast–southwest (NE-SW) directions and two distinct local features towards millimeter peaks MM1 and MM2. 
\item The NW-SE component is 2D-spatially associated with a bipolar jet and a molecular outflow direction. The inferred B-field in this region is well fitted by parabolic functions and the residual angle dispersion was used to estimate a B-field strength of $5.1^{+1.3}_{-1.1}$ mG using the Davis Chandrasekhar Fermi (DCF) technique. Comparison between outflow expansion pressure and magnetic pressure suggest that NW-SE B-fields are likely to be shaped by the outflow. 
\item The NE-SW component are remarkably coherent along the filaments F1, F-Main, and Tail. In these filaments, the turbulent motion is trans-Alfv{\'e}nic. In F-Main, the observed line masses exceed the turbulent critical limits and magnetic support is likely required to prevent radial collapse and suppress local fragmentation. While in F1 and Tail, turbulence itself can already support against gravitational collapse although B-field can potentially provide additional support.
\item Toward MM1, the B-fields follow a spiral-like, infalling gas streamer traced by CH$_3$CN. After removing the large-scale fields by a smoothing kernel, the residual angle dispersion inferred a trans-Alfv{\'e}nic turbulence in the accreting material. This is consistent with predictions from recent numerical simulations where with an efficient magnetic damping of turbulence, magnetically regulated, laminar accretion flows can continue to feed cores.
\item Toward MM2, the field exhibits an hourglass geometry which seems to be decoupled from the large-scale fields. Described by parabolic curves, two independent methods yield a consistently strong field strength estimate of $8.1\pm1.9$~mG. The normalized mass-to-flux ratio $\lambda \sim 3.2 \pm 1.4$ indicates that the B-field can add support to self-gravity but is insufficient to support the core against gravity on its own. By virial parameter analyses, it is found that both turbulence and B-fields may provide support to delay collapse and potentially star formation. 
\item Given that the clearly different evolutionary stages of MM1 and MM2, the tale of two demonstrates that B-fields may both facilitate mass delivery in MM1 through magnetically stabilized gas filaments (like F-Main) and inhibit collapse in MM2 within the same massive protocluster. Our work highlights the dynamic importance of B-fields in high-mass star formation. 
\end{itemize}

%
\begin{acknowledgements}
We thank the anonymous referee for improving and polishing the paper. F.W.X. thanks Daniel Seifried, Gary Fuller, and Han-Tsung Lee for helpful discussion on B-fields, and thanks Yisheng Qiu, Tianwei Zhang, Thomas M\"uller, Peter Schilke, Roya Hamedani Golshan, and Siqi Zheng for their great help on spectral line fitting. 
PS was partially supported by a Grant-in-Aid for Scientific Research (KAKENHI Number JP23H01221 and JP26H02066) of JSPS.
H.B.L. is supported by the National Science and Technology Council (NSTC) of Taiwan (Grant Nos. 113-2112-M-110-022-MY3).
J.M.G. acknowledges support by grant PID2023-146675NB-I00 (MCI-AEI-FEDER, UE). This work was also partly supported by the Spanish program Unidad de Excelencia María de Maeztu CEX2020-001058-M, financed by MCIN/AEI/10.13039/501100011033, and by the MaX-CSIC Excellence Award MaX4-SOMMA-ICE. 
J.L. was partially supported by Grant-in-Aid for Scientific Research (KAKENHI Number JP25K17445) of the Japan Society for the Promotion of Science (JSPS). 
M.T.B. and C.Y.L acknowledges financial support through the INAF Large Grant The role of MAGnetic fields in MAssive star formation (MAGMA). 
P.S.L. acknowledges the supports by the NSFC through grant No. 1241101426 and the National Key R\&D Program of China (No. 2022YFA1603101).
X.L. acknowledges support from the Strategic Priority Research Program of the Chinese Academy of Sciences (CAS) Grant No.\ XDB0800300, the National Key R\&D Program of China (No.\ 2022YFA1603101), State Key Laboratory of Radio Astronomy and Technology (CAS), the National Natural Science Foundation of China (NSFC) through grant Nos.\ 12273090 and 12322305, the Natural Science Foundation of Shanghai (No.\ 23ZR1482100), and the CAS ``Light of West China'' Program No.\ xbzg-zdsys-202212. 
Y.C. was partially supported by Grant-in-Aid for Scientific Research (KAKENHI  number JP24K17103 and 26K00748) of the JSPS.
Q.Y.L. acknowledges the support by JSPS KAKENHI Grant Number JP23K20035. 
This paper makes use of the following ALMA data: ADS/JAO.ALMA\#2017.1.00101.S ALMA is a partnership of ESO (representing its member states), NSF (USA) and NINS (Japan), together with NRC (Canada), NSTC and ASIAA (Taiwan), and KASI (Republic of Korea), in cooperation with the Republic of Chile. The Joint ALMA Observatory is operated by ESO, auI/NRAO and NAOJ. 
PSL acknowledges the support by the National Key R\&D Program of China (No. 2022YFA1603100) and National Natural Science Foundation of China (NSFC) (No. 1241101426).
LAZ acknowledges financial support from CONACyT-280775, UNAM-PAPIIT IN110618, and IN112323 grants, Mexico.
The paper has used the following software packages: \textit{astropy} \citep{Astropy2022}, \textit{scipy} \citep{Scipy2020}, \textit{numpy} \citep{Numpy2020}, and \textit{spectuner} \citep{qiu2025a, qiu2025b}. 
\end{acknowledgements}

%
%

\bibliographystyle{aa} 
\bibliography{w33a} 

@ARTICLE{Astropy2022,
       author = {{Astropy Collaboration} and {Price-Whelan}, Adrian M. and {Lim}, Pey Lian and {Earl}, Nicholas and {Starkman}, Nathaniel and {Bradley}, Larry and {Shupe}, David L. and {Patil}, Aarya A. and {Corrales}, Lia and {Brasseur}, C.~E. and {N{\"o}the}, Maximilian and {Donath}, Axel and {Tollerud}, Erik and {Morris}, Brett M. and {Ginsburg}, Adam and {Vaher}, Eero and {Weaver}, Benjamin A. and {Tocknell}, James and {Jamieson}, William and {van Kerkwijk}, Marten H. and {Robitaille}, Thomas P. and {Merry}, Bruce and {Bachetti}, Matteo and {G{\"u}nther}, H. Moritz and {Aldcroft}, Thomas L. and {Alvarado-Montes}, Jaime A. and {Archibald}, Anne M. and {B{\'o}di}, Attila and {Bapat}, Shreyas and {Barentsen}, Geert and {Baz{\'a}n}, Juanjo and {Biswas}, Manish and {Boquien}, M{\'e}d{\'e}ric and {Burke}, D.~J. and {Cara}, Daria and {Cara}, Mihai and {Conroy}, Kyle E. and {Conseil}, Simon and {Craig}, Matthew W. and {Cross}, Robert M. and {Cruz}, Kelle L. and {D'Eugenio}, Francesco and {Dencheva}, Nadia and {Devillepoix}, Hadrien A.~R. and {Dietrich}, J{\"o}rg P. and {Eigenbrot}, Arthur Davis and {Erben}, Thomas and {Ferreira}, Leonardo and {Foreman-Mackey}, Daniel and {Fox}, Ryan and {Freij}, Nabil and {Garg}, Suyog and {Geda}, Robel and {Glattly}, Lauren and {Gondhalekar}, Yash and {Gordon}, Karl D. and {Grant}, David and {Greenfield}, Perry and {Groener}, Austen M. and {Guest}, Steve and {Gurovich}, Sebastian and {Handberg}, Rasmus and {Hart}, Akeem and {Hatfield-Dodds}, Zac and {Homeier}, Derek and {Hosseinzadeh}, Griffin and {Jenness}, Tim and {Jones}, Craig K. and {Joseph}, Prajwel and {Kalmbach}, J. Bryce and {Karamehmetoglu}, Emir and {Ka{\l}uszy{\'n}ski}, Miko{\l}aj and {Kelley}, Michael S.~P. and {Kern}, Nicholas and {Kerzendorf}, Wolfgang E. and {Koch}, Eric W. and {Kulumani}, Shankar and {Lee}, Antony and {Ly}, Chun and {Ma}, Zhiyuan and {MacBride}, Conor and {Maljaars}, Jakob M. and {Muna}, Demitri and {Murphy}, N.~A. and {Norman}, Henrik and {O'Steen}, Richard and {Oman}, Kyle A. and {Pacifici}, Camilla and {Pascual}, Sergio and {Pascual-Granado}, J. and {Patil}, Rohit R. and {Perren}, Gabriel I. and {Pickering}, Timothy E. and {Rastogi}, Tanuj and {Roulston}, Benjamin R. and {Ryan}, Daniel F. and {Rykoff}, Eli S. and {Sabater}, Jose and {Sakurikar}, Parikshit and {Salgado}, Jes{\'u}s and {Sanghi}, Aniket and {Saunders}, Nicholas and {Savchenko}, Volodymyr and {Schwardt}, Ludwig and {Seifert-Eckert}, Michael and {Shih}, Albert Y. and {Jain}, Anany Shrey and {Shukla}, Gyanendra and {Sick}, Jonathan and {Simpson}, Chris and {Singanamalla}, Sudheesh and {Singer}, Leo P. and {Singhal}, Jaladh and {Sinha}, Manodeep and {Sip{\H{o}}cz}, Brigitta M. and {Spitler}, Lee R. and {Stansby}, David and {Streicher}, Ole and {{\v{S}}umak}, Jani and {Swinbank}, John D. and {Taranu}, Dan S. and {Tewary}, Nikita and {Tremblay}, Grant R. and {de Val-Borro}, Miguel and {Van Kooten}, Samuel J. and {Vasovi{\'c}}, Zlatan and {Verma}, Shresth and {de Miranda Cardoso}, Jos{\'e} Vin{\'\i}cius and {Williams}, Peter K.~G. and {Wilson}, Tom J. and {Winkel}, Benjamin and {Wood-Vasey}, W.~M. and {Xue}, Rui and {Yoachim}, Peter and {Zhang}, Chen and {Zonca}, Andrea and {Astropy Project Contributors}},
        title = "{The Astropy Project: Sustaining and Growing a Community-oriented Open-source Project and the Latest Major Release (v5.0) of the Core Package}",
      journal = {\apj},
         year = 2022,
        month = aug,
       volume = {935},
       number = {2},
          eid = {167},
        pages = {167},
          doi = {10.3847/1538-4357/ac7c74},
archivePrefix = {arXiv},
       eprint = {2206.14220},
 primaryClass = {astro-ph.IM},
       adsurl = {https://ui.adsabs.harvard.edu/abs/2022ApJ...935..167A}
}

@ARTICLE{CASA2022,
       author = {{CASA Team} and {Bean}, Ben and {Bhatnagar}, Sanjay and {Castro}, Sandra and {Donovan Meyer}, Jennifer and {Emonts}, Bjorn and {Garcia}, Enrique and {Garwood}, Robert and {Golap}, Kumar and {Gonzalez Villalba}, Justo and {Harris}, Pamela and {Hayashi}, Yohei and {Hoskins}, Josh and {Hsieh}, Mingyu and {Jagannathan}, Preshanth and {Kawasaki}, Wataru and {Keimpema}, Aard and {Kettenis}, Mark and {Lopez}, Jorge and {Marvil}, Joshua and {Masters}, Joseph and {McNichols}, Andrew and {Mehringer}, David and {Miel}, Renaud and {Moellenbrock}, George and {Montesino}, Federico and {Nakazato}, Takeshi and {Ott}, Juergen and {Petry}, Dirk and {Pokorny}, Martin and {Raba}, Ryan and {Rau}, Urvashi and {Schiebel}, Darrell and {Schweighart}, Neal and {Sekhar}, Srikrishna and {Shimada}, Kazuhiko and {Small}, Des and {Steeb}, Jan-Willem and {Sugimoto}, Kanako and {Suoranta}, Ville and {Tsutsumi}, Takahiro and {van Bemmel}, Ilse M. and {Verkouter}, Marjolein and {Wells}, Akeem and {Xiong}, Wei and {Szomoru}, Arpad and {Griffith}, Morgan and {Glendenning}, Brian and {Kern}, Jeff},
        title = "{CASA, the Common Astronomy Software Applications for Radio Astronomy}",
      journal = {\pasp},
         year = 2022,
        month = nov,
       volume = {134},
       number = {1041},
          eid = {114501},
        pages = {114501},
          doi = {10.1088/1538-3873/ac9642},
archivePrefix = {arXiv},
       eprint = {2210.02276},
 primaryClass = {astro-ph.IM},
       adsurl = {https://ui.adsabs.harvard.edu/abs/2022PASP..134k4501C}
}

@ARTICLE{Scipy2020,
       author = {{Virtanen}, Pauli and {Gommers}, Ralf and {Oliphant}, Travis E. and {Haberland}, Matt and {Reddy}, Tyler and {Cournapeau}, David and {Burovski}, Evgeni and {Peterson}, Pearu and {Weckesser}, Warren and {Bright}, Jonathan and {van der Walt}, St{\'e}fan J. and {Brett}, Matthew and {Wilson}, Joshua and {Millman}, K. Jarrod and {Mayorov}, Nikolay and {Nelson}, Andrew R.~J. and {Jones}, Eric and {Kern}, Robert and {Larson}, Eric and {Carey}, C.~J. and {Polat}, {\.I}lhan and {Feng}, Yu and {Moore}, Eric W. and {VanderPlas}, Jake and {Laxalde}, Denis and {Perktold}, Josef and {Cimrman}, Robert and {Henriksen}, Ian and {Quintero}, E.~A. and {Harris}, Charles R. and {Archibald}, Anne M. and {Ribeiro}, Ant{\^o}nio H. and {Pedregosa}, Fabian and {van Mulbregt}, Paul and {SciPy 1. 0 Contributors}},
        title = "{SciPy 1.0: fundamental algorithms for scientific computing in Python}",
      journal = {Nature Methods},
         year = 2020,
        month = feb,
       volume = {17},
        pages = {261-272},
          doi = {10.1038/s41592-019-0686-2},
archivePrefix = {arXiv},
       eprint = {1907.10121},
 primaryClass = {cs.MS},
       adsurl = {https://ui.adsabs.harvard.edu/abs/2020NatMe..17..261V}
}

@Article{Numpy2020,
 title         = {Array programming with {NumPy}},
 author        = {Charles R. Harris and K. Jarrod Millman and St{\'{e}}fan J.
                 van der Walt and Ralf Gommers and Pauli Virtanen and David
                 Cournapeau and Eric Wieser and Julian Taylor and Sebastian
                 Berg and Nathaniel J. Smith and Robert Kern and Matti Picus
                 and Stephan Hoyer and Marten H. van Kerkwijk and Matthew
                 Brett and Allan Haldane and Jaime Fern{\'{a}}ndez del
                 R{\'{i}}o and Mark Wiebe and Pearu Peterson and Pierre
                 G{\'{e}}rard-Marchant and Kevin Sheppard and Tyler Reddy and
                 Warren Weckesser and Hameer Abbasi and Christoph Gohlke and
                 Travis E. Oliphant},
 year          = {2020},
 month         = sep,
 journal       = {Nature},
 volume        = {585},
 number        = {7825},
 pages         = {357--362},
 doi           = {10.1038/s41586-020-2649-2},
 publisher     = {Springer Science and Business Media {LLC}},
 url           = {https://doi.org/10.1038/s41586-020-2649-2}
}

@ARTICLE{magmar1,
       author = {{Fern{\'a}ndez-L{\'o}pez}, M. and {Sanhueza}, P. and {Zapata}, L.~A. and {Stephens}, I. and {Hull}, C. and {Zhang}, Q. and {Girart}, J.~M. and {Koch}, P.~M. and {Cort{\'e}s}, P. and {Silva}, A. and {Tatematsu}, K. and {Nakamura}, F. and {Guzm{\'a}n}, A.~E. and {Nguyen Luong}, Q. and {Guzm{\'a}n Ccolque}, E. and {Tang}, Y. -W. and {Chen}, H. -R.~V.},
        title = "{Magnetic Fields in Massive Star-forming Regions (MagMaR). I. Linear Polarized Imaging of the Ultracompact H II Region G5.89-0.39}",
      journal = {\apj},
         year = 2021,
        month = may,
       volume = {913},
       number = {1},
          eid = {29},
        pages = {29},
          doi = {10.3847/1538-4357/abf2b6},
archivePrefix = {arXiv},
       eprint = {2104.03331},
 primaryClass = {astro-ph.GA},
       adsurl = {https://ui.adsabs.harvard.edu/abs/2021ApJ...913...29F}
}

@ARTICLE{magmar2,
       author = {{Cort{\'e}s}, Paulo C. and {Sanhueza}, Patricio and {Houde}, Martin and {Mart{\'\i}n}, Sergio and {Hull}, Charles L.~H. and {Girart}, Josep M. and {Zhang}, Qizhou and {Fernandez-Lopez}, Manuel and {Zapata}, Luis A. and {Stephens}, Ian W. and {Li}, Hua-bai and {Wu}, Benjamin and {Olguin}, Fernando and {Lu}, Xing and {Guzm{\'a}n}, Andres E. and {Nakamura}, Fumitaka},
        title = "{Magnetic Fields in Massive Star-forming Regions (MagMaR). II. Tomography through Dust and Molecular Line Polarization in NGC 6334I(N)}",
      journal = {\apj},
         year = 2021,
        month = dec,
       volume = {923},
       number = {2},
          eid = {204},
        pages = {204},
          doi = {10.3847/1538-4357/ac28a1},
archivePrefix = {arXiv},
       eprint = {2109.09270},
 primaryClass = {astro-ph.GA},
       adsurl = {https://ui.adsabs.harvard.edu/abs/2021ApJ...923..204C}
}

@ARTICLE{magmar3,
       author = {{Cort{\'e}s}, Paulo C. and {Girart}, Josep M. and {Sanhueza}, Patricio and {Liu}, Junhao and {Mart{\'\i}n}, Sergio and {Stephens}, Ian W. and {Beuther}, Henrik and {Koch}, Patrick M. and {Fern{\'a}ndez-L{\'o}pez}, M. and {S{\'a}nchez-Monge}, {\'A}lvaro and {Wang}, Jia-Wei and {Morii}, Kaho and {Li}, Shanghuo and {Saha}, Piyali and {Zhang}, Qizhou and {Rebolledo}, David and {Zapata}, Luis A. and {Kang}, Ji-hyun and {Jiao}, Wenyu and {Kim}, Jongsoo and {Cheng}, Yu and {Hwang}, Jihye and {Chung}, Eun Jung and {Choudhury}, Spandan and {Lyo}, A. -Ran and {Olguin}, Fernando},
        title = "{MagMaR III. Resisting the Pressure, Is the Magnetic Field Overwhelmed in NGC6334I?}",
      journal = {\apj},
         year = 2024,
        month = sep,
       volume = {972},
       number = {1},
          eid = {115},
        pages = {115},
          doi = {10.3847/1538-4357/ad59a7},
archivePrefix = {arXiv},
       eprint = {2406.14663},
 primaryClass = {astro-ph.GA},
       adsurl = {https://ui.adsabs.harvard.edu/abs/2024ApJ...972..115C}
}

@ARTICLE{magmar4,
       author = {{Zapata}, Luis A. and {Fern{\'a}ndez-L{\'o}pez}, Manuel and {Sanhueza}, Patricio and {Girart}, Josep M. and {Rodr{\'\i}guez}, Luis F. and {Cort{\'e}s}, Paulo and {Koch}, Patrick and {Beltr{\'a}n}, Maria T. and {Pattle}, Kate and {Beuther}, Henrik and {Saha}, Piyali and {Jiao}, Wenyu and {Xu}, Fengwei and {Lu}, Xing Walker and {Olguin}, Fernando and {Li}, Shanghuo and {Stephens}, Ian W. and {Kang}, Ji-hyun and {Cheng}, Yu and {Choudhury}, Spandan and {Morii}, Kaho and {Chung}, Eun Jung and {Wang}, Jia-Wei and {Hwang}, Jihye and {Lyo}, A. -Ran and {Zhang}, Q. and {Chen}, Huei-Ru Vivien},
        title = "{Magnetic Fields in Massive Star-forming Regions (MagMaR). IV. Tracing the Magnetic Fields in the O-type Protostellar System IRAS 16547{\textendash}4247}",
      journal = {\apj},
         year = 2024,
        month = oct,
       volume = {974},
       number = {2},
          eid = {257},
        pages = {257},
          doi = {10.3847/1538-4357/ad701d},
archivePrefix = {arXiv},
       eprint = {2408.10199},
 primaryClass = {astro-ph.SR},
       adsurl = {https://ui.adsabs.harvard.edu/abs/2024ApJ...974..257Z}
}

@ARTICLE{magmar5,
       author = {{Sanhueza}, Patricio and {Liu}, Junhao and {Morii}, Kaho and {Girart}, Josep Miquel and {Zhang}, Qizhou and {Stephens}, Ian W. and {Jackson}, James M. and {Cort{\'e}s}, Paulo C. and {Koch}, Patrick M. and {Cyganowski}, Claudia J. and {Saha}, Piyali and {Beuther}, Henrik and {Zhang}, Suinan and {Beltr{\'a}n}, Maria T. and {Cheng}, Yu and {Olguin}, Fernando A. and {Lu}, Xing and {Choudhury}, Spandan and {Pattle}, Kate and {Fern{\'a}ndez-L{\'o}pez}, Manuel and {Hwang}, Jihye and {Kang}, Ji-hyun and {Karoly}, Janik and {Ginsburg}, Adam and {Lyo}, A. -Ran and {Taniguchi}, Kotomi and {Jiao}, Wenyu and {Eswaraiah}, Chakali and {Luo}, Qiu-yi and {Wang}, Jia-Wei and {Commer{\c{c}}on}, Beno{\^\i}t and {Li}, Shanghuo and {Xu}, Fengwei and {Chen}, Huei-Ru Vivien and {Zapata}, Luis A. and {Chung}, Eun Jung and {Nakamura}, Fumitaka and {Panigrahy}, Sandhyarani and {Sakai}, Takeshi},
        title = "{Magnetic Fields in Massive Star-forming Regions (MagMaR). V. The Magnetic Field at the Onset of High-mass Star Formation}",
      journal = {\apj},
         year = 2025,
        month = feb,
       volume = {980},
       number = {1},
          eid = {87},
        pages = {87},
          doi = {10.3847/1538-4357/ad9d40},
archivePrefix = {arXiv},
       eprint = {2412.08790},
 primaryClass = {astro-ph.GA},
       adsurl = {https://ui.adsabs.harvard.edu/abs/2025ApJ...980...87S}
}

@ARTICLE{magmar6,
       author = {{Saha}, Piyali and {Sanhueza}, Patricio and {Padovani}, Marco and {Girart}, Josep M. and {Cort{\'e}s}, Paulo C. and {Morii}, Kaho and {Liu}, Junhao and {S{\'a}nchez-Monge}, {\'A}. and {Galli}, Daniele and {Basu}, Shantanu and {Koch}, Patrick M. and {Beltr{\'a}n}, Maria T. and {Li}, Shanghuo and {Beuther}, Henrik and {Stephens}, Ian W. and {Nakamura}, Fumitaka and {Zhang}, Qizhou and {Jiao}, Wenyu and {Fern{\'a}ndez-L{\'o}pez}, M. and {Hwang}, Jihye and {Chung}, Eun Jung and {Pattle}, Kate and {Zapata}, Luis A. and {Xu}, Fengwei and {Olguin}, Fernando A. and {Kang}, Ji-hyun and {Karoly}, Janik and {Law}, Chi-Yan and {Wang}, Jia-Wei and {Csengeri}, Timea and {Lu}, Xing and {Cheng}, Yu and {Kim}, Jongsoo and {Choudhury}, Spandan and {Chen}, Huei-Ru Vivien and {Hull}, Charles L.~H.},
        title = "{Magnetic Fields in Massive Star-forming Regions (MagMaR): Unveiling an Hourglass Magnetic Field in G333.46{\textendash}0.16 Using ALMA}",
      journal = {\apjl},
         year = 2024,
        month = sep,
       volume = {972},
       number = {1},
          eid = {L6},
        pages = {L6},
          doi = {10.3847/2041-8213/ad660c},
archivePrefix = {arXiv},
       eprint = {2407.16654},
 primaryClass = {astro-ph.GA},
       adsurl = {https://ui.adsabs.harvard.edu/abs/2024ApJ...972L...6S}
}

@ARTICLE{sanhueza2021,
       author = {{Sanhueza}, Patricio and {Girart}, Josep Miquel and {Padovani}, Marco and {Galli}, Daniele and {Hull}, Charles L.~H. and {Zhang}, Qizhou and {Cortes}, Paulo and {Stephens}, Ian W. and {Fern{\'a}ndez-L{\'o}pez}, Manuel and {Jackson}, James M. and {Frau}, Pau and {Kock}, Patrick M. and {Wu}, Benjamin and {Zapata}, Luis A. and {Olguin}, Fernando and {Lu}, Xing and {Silva}, Andrea and {Tang}, Ya-Wen and {Sakai}, Takeshi and {Guzm{\'a}n}, Andr{\'e}s E. and {Tatematsu}, Ken'ichi and {Nakamura}, Fumitaka and {Chen}, Huei-Ru Vivien},
        title = "{Gravity-driven Magnetic Field at  1000 au Scales in High-mass Star Formation}",
      journal = {\apjl},
         year = 2021,
        month = jul,
       volume = {915},
       number = {1},
          eid = {L10},
        pages = {L10},
          doi = {10.3847/2041-8213/ac081c},
archivePrefix = {arXiv},
       eprint = {2106.03866},
 primaryClass = {astro-ph.GA},
       adsurl = {https://ui.adsabs.harvard.edu/abs/2021ApJ...915L..10S}
}

@ARTICLE{olguin2021,
       author = {{Olguin}, Fernando A. and {Sanhueza}, Patricio and {Guzm{\'a}n}, Andr{\'e}s E. and {Lu}, Xing and {Saigo}, Kazuya and {Zhang}, Qizhou and {Silva}, Andrea and {Chen}, Huei-Ru Vivien and {Li}, Shanghuo and {Ohashi}, Satoshi and {Nakamura}, Fumitaka and {Sakai}, Takeshi and {Wu}, Benjamin},
        title = "{Digging into the Interior of Hot Cores with ALMA (DIHCA). I. Dissecting the High-mass Star-forming Core G335.579-0.292 MM1}",
      journal = {\apj},
         year = 2021,
        month = mar,
       volume = {909},
       number = {2},
          eid = {199},
        pages = {199},
          doi = {10.3847/1538-4357/abde3f},
archivePrefix = {arXiv},
       eprint = {2101.08284},
 primaryClass = {astro-ph.GA},
       adsurl = {https://ui.adsabs.harvard.edu/abs/2021ApJ...909..199O}
}

@ARTICLE{contreras2018,
       author = {{Contreras}, Yanett and {Sanhueza}, Patricio and {Jackson}, James M. and {Guzm{\'a}n}, Andr{\'e}s E. and {Longmore}, Steven and {Garay}, Guido and {Zhang}, Qizhou and {Nguyễn-Lu'o'ng}, Quang and {Tatematsu}, Ken'ichi and {Nakamura}, Fumitaka and {Sakai}, Takeshi and {Ohashi}, Satoshi and {Liu}, Tie and {Saito}, Masao and {Gomez}, Laura and {Rathborne}, Jill and {Whitaker}, Scott},
        title = "{Infall Signatures in a Prestellar Core Embedded in the High-mass 70 {\ensuremath{\mu}}m Dark IRDC G331.372-00.116}",
      journal = {\apj},
         year = 2018,
        month = jul,
       volume = {861},
       number = {1},
          eid = {14},
        pages = {14},
          doi = {10.3847/1538-4357/aac2ec},
archivePrefix = {arXiv},
       eprint = {1805.01802},
 primaryClass = {astro-ph.GA},
       adsurl = {https://ui.adsabs.harvard.edu/abs/2018ApJ...861...14C}
}

@ARTICLE{immer2013,
       author = {{Immer}, K. and {Reid}, M.~J. and {Menten}, K.~M. and {Brunthaler}, A. and {Dame}, T.~M.},
        title = "{Trigonometric parallaxes of massive star forming regions: G012.88+0.48 and W33}",
      journal = {\aap},
         year = 2013,
        month = may,
       volume = {553},
          eid = {A117},
        pages = {A117},
          doi = {10.1051/0004-6361/201220793},
archivePrefix = {arXiv},
       eprint = {1304.2041},
 primaryClass = {astro-ph.GA},
       adsurl = {https://ui.adsabs.harvard.edu/abs/2013A&A...553A.117I}
}

@ARTICLE{wynn1981,
       author = {{Wynn-Williams}, C.~G. and {Beichman}, C.~A. and {Downes}, D.},
        title = "{VLA observations of W 33, W 44, and GL 2591}",
      journal = {\aj},
         year = 1981,
        month = apr,
       volume = {86},
        pages = {565-568},
          doi = {10.1086/112916},
       adsurl = {https://ui.adsabs.harvard.edu/abs/1981AJ.....86..565W}
}

@ARTICLE{stier1984,
       author = {{Stier}, M.~T. and {Jaffe}, D.~T. and {Rengarajan}, T.~N. and {Fazio}, G.~G. and {Maxson}, C.~W. and {McBreen}, B. and {Loughran}, L. and {Serio}, S. and {Sciortino}, S.},
        title = "{Far-infrared and CO observations of the W 33 complex.}",
      journal = {\apj},
         year = 1984,
        month = aug,
       volume = {283},
        pages = {573-579},
          doi = {10.1086/162342},
       adsurl = {https://ui.adsabs.harvard.edu/abs/1984ApJ...283..573S}
}

@ARTICLE{hosokawa2009,
       author = {{Hosokawa}, Takashi and {Omukai}, Kazuyuki},
        title = "{Evolution of Massive Protostars with High Accretion Rates}",
      journal = {\apj},
         year = 2009,
        month = jan,
       volume = {691},
       number = {1},
        pages = {823-846},
          doi = {10.1088/0004-637X/691/1/823},
archivePrefix = {arXiv},
       eprint = {0806.4122},
 primaryClass = {astro-ph},
       adsurl = {https://ui.adsabs.harvard.edu/abs/2009ApJ...691..823H}
}

@ARTICLE{hosokawa2010,
       author = {{Hosokawa}, Takashi and {Yorke}, Harold W. and {Omukai}, Kazuyuki},
        title = "{Evolution of Massive Protostars Via Disk Accretion}",
      journal = {\apj},
         year = 2010,
        month = sep,
       volume = {721},
       number = {1},
        pages = {478-492},
          doi = {10.1088/0004-637X/721/1/478},
archivePrefix = {arXiv},
       eprint = {1005.2827},
 primaryClass = {astro-ph.SR},
       adsurl = {https://ui.adsabs.harvard.edu/abs/2010ApJ...721..478H}
}

@ARTICLE{izquierdo2018,
       author = {{Izquierdo}, Andr{\'e}s F. and {Galv{\'a}n-Madrid}, Roberto and {Maud}, Luke T. and {Hoare}, Melvin G. and {Johnston}, Katharine G. and {Keto}, Eric R. and {Zhang}, Qizhou and {de Wit}, Willem-Jan},
        title = "{Radiative transfer modelling of W33A MM1: 3D structure and dynamics of a complex massive star-forming region}",
      journal = {\mnras},
         year = 2018,
        month = aug,
       volume = {478},
       number = {2},
        pages = {2505-2525},
          doi = {10.1093/mnras/sty1096},
archivePrefix = {arXiv},
       eprint = {1804.09204},
 primaryClass = {astro-ph.GA},
       adsurl = {https://ui.adsabs.harvard.edu/abs/2018MNRAS.478.2505I}
}

@ARTICLE{davies2010,
       author = {{Davies}, Ben and {Lumsden}, Stuart L. and {Hoare}, Melvin G. and {Oudmaijer}, Ren{\'e} D. and {de Wit}, Willem-Jan},
        title = "{The circumstellar disc, envelope and bipolar outflow of the massive young stellar object W33A}",
      journal = {\mnras},
         year = 2010,
        month = mar,
       volume = {402},
       number = {3},
        pages = {1504-1515},
          doi = {10.1111/j.1365-2966.2009.16077.x},
archivePrefix = {arXiv},
       eprint = {0911.4592},
 primaryClass = {astro-ph.SR},
       adsurl = {https://ui.adsabs.harvard.edu/abs/2010MNRAS.402.1504D}
}

@ARTICLE{galvan2010,
       author = {{Galv{\'a}n-Madrid}, Roberto and {Zhang}, Qizhou and {Keto}, Eric and {Ho}, Paul T.~P. and {Zapata}, Luis A. and {Rodr{\'\i}guez}, Luis F. and {Pineda}, Jaime E. and {V{\'a}zquez-Semadeni}, Enrique},
        title = "{From the Convergence of Filaments to Disk-outflow Accretion: Massive Star Formation in W33A}",
      journal = {\apj},
         year = 2010,
        month = dec,
       volume = {725},
       number = {1},
        pages = {17-28},
          doi = {10.1088/0004-637X/725/1/17},
archivePrefix = {arXiv},
       eprint = {1004.2466},
 primaryClass = {astro-ph.GA},
       adsurl = {https://ui.adsabs.harvard.edu/abs/2010ApJ...725...17G}
}

@ARTICLE{ekstrom2012,
       author = {{Ekstr{\"o}m}, S. and {Georgy}, C. and {Eggenberger}, P. and {Meynet}, G. and {Mowlavi}, N. and {Wyttenbach}, A. and {Granada}, A. and {Decressin}, T. and {Hirschi}, R. and {Frischknecht}, U. and {Charbonnel}, C. and {Maeder}, A.},
        title = "{Grids of stellar models with rotation. I. Models from 0.8 to 120 M$_{{\ensuremath{\odot}}}$ at solar metallicity (Z = 0.014)}",
      journal = {\aap},
         year = 2012,
        month = jan,
       volume = {537},
          eid = {A146},
        pages = {A146},
          doi = {10.1051/0004-6361/201117751},
archivePrefix = {arXiv},
       eprint = {1110.5049},
 primaryClass = {astro-ph.SR},
       adsurl = {https://ui.adsabs.harvard.edu/abs/2012A&A...537A.146E}
}

@ARTICLE{dewit2010,
       author = {{de Wit}, W.~J. and {Hoare}, M.~G. and {Oudmaijer}, R.~D. and {Lumsden}, S.~L.},
        title = "{The origin of mid-infrared emission in massive young stellar objects: multi-baseline VLTI observations of W33A}",
      journal = {\aap},
         year = 2010,
        month = jun,
       volume = {515},
          eid = {A45},
        pages = {A45},
          doi = {10.1051/0004-6361/200913209},
archivePrefix = {arXiv},
       eprint = {0912.2869},
 primaryClass = {astro-ph.SR},
       adsurl = {https://ui.adsabs.harvard.edu/abs/2010A&A...515A..45D}
}

@ARTICLE{maud2017,
       author = {{Maud}, L.~T. and {Hoare}, M.~G. and {Galv{\'a}n-Madrid}, R. and {Zhang}, Q. and {de Wit}, W.~J. and {Keto}, E. and {Johnston}, K.~G. and {Pineda}, J.~E.},
        title = "{The ALMA view of W33A: a spiral filament feeding the candidate disc in MM1-Main}",
      journal = {\mnras},
         year = 2017,
        month = may,
       volume = {467},
       number = {1},
        pages = {L120-L124},
          doi = {10.1093/mnrasl/slx010},
archivePrefix = {arXiv},
       eprint = {1701.06958},
 primaryClass = {astro-ph.SR},
       adsurl = {https://ui.adsabs.harvard.edu/abs/2017MNRAS.467L.120M}
}

@ARTICLE{menshchikov2021,
       author = {{Men'shchikov}, A.},
        title = "{Multiscale, multiwavelength extraction of sources and filaments using separation of the structural components: getsf}",
      journal = {\aap},
         year = 2021,
        month = may,
       volume = {649},
          eid = {A89},
        pages = {A89},
          doi = {10.1051/0004-6361/202039913},
archivePrefix = {arXiv},
       eprint = {2102.11565},
 primaryClass = {astro-ph.IM},
       adsurl = {https://ui.adsabs.harvard.edu/abs/2021A&A...649A..89M}
}

@ARTICLE{motte2018na,
       author = {{Motte}, F. and {Nony}, T. and {Louvet}, F. and {Marsh}, K.~A. and {Bontemps}, S. and {Whitworth}, A.~P. and {Men'shchikov}, A. and {Nguyen Luong}, Q. and {Csengeri}, T. and {Maury}, A.~J. and {Gusdorf}, A. and {Chapillon}, E. and {K{\"o}nyves}, V. and {Schilke}, P. and {Duarte-Cabral}, A. and {Didelon}, P. and {Gaudel}, M.},
        title = "{The unexpectedly large proportion of high-mass star-forming cores in a Galactic mini-starburst}",
      journal = {Nature Astronomy},
         year = 2018,
        month = apr,
       volume = {2},
        pages = {478-482},
          doi = {10.1038/s41550-018-0452-x},
archivePrefix = {arXiv},
       eprint = {1804.02392},
 primaryClass = {astro-ph.GA},
       adsurl = {https://ui.adsabs.harvard.edu/abs/2018NatAs...2..478M}
}

@ARTICLE{assemble2024,
       author = {{Xu}, Fengwei and {Wang}, Ke and {Liu}, Tie and {Tang}, Mengyao and {Evans}, II, Neal J. and {Palau}, Aina and {Morii}, Kaho and {He}, Jinhua and {Sanhueza}, Patricio and {Liu}, Hong-Li and {Stutz}, Amelia and {Zhang}, Qizhou and {Chen}, Xi and {Li}, Pak Shing and {G{\'o}mez}, Gilberto C. and {V{\'a}zquez-Semadeni}, Enrique and {Li}, Shanghuo and {Mai}, Xiaofeng and {Lu}, Xing and {Liu}, Meizhu and {Chen}, Li and {Li}, Chuanshou and {Shi}, Hongqiong and {Ren}, Zhiyuan and {Li}, Di and {Garay}, Guido and {Bronfman}, Leonardo and {Dewangan}, Lokesh and {Juvela}, Mika and {Lee}, Chang Won and {Zhang}, S. and {Yue}, Nannan and {Wang}, Chao and {Ge}, Yifei and {Jiao}, Wenyu and {Luo}, Qiuyi and {Zhou}, J. -W. and {Tatematsu}, Ken'ichi and {Chibueze}, James O. and {Su}, Keyun and {Sun}, Shenglan and {Ristorcelli}, I. and {Toth}, L. Viktor},
        title = "{The ALMA Survey of Star Formation and Evolution in Massive Protoclusters with Blue Profiles (ASSEMBLE): Core Growth, Cluster Contraction, and Primordial Mass Segregation}",
      journal = {\apjs},
         year = 2024,
        month = jan,
       volume = {270},
       number = {1},
          eid = {9},
        pages = {9},
          doi = {10.3847/1538-4365/acfee5},
archivePrefix = {arXiv},
       eprint = {2309.14684},
 primaryClass = {astro-ph.GA},
       adsurl = {https://ui.adsabs.harvard.edu/abs/2024ApJS..270....9X}
}

@ARTICLE{ossenkopf1994,
       author = {{Ossenkopf}, V. and {Henning}, Th.},
        title = "{Dust opacities for protostellar cores.}",
      journal = {\aap},
         year = 1994,
        month = nov,
       volume = {291},
        pages = {943-959},
       adsurl = {https://ui.adsabs.harvard.edu/abs/1994A&A...291..943O}
}

@ARTICLE{tielens1997,
       author = {{Tielens}, A.~G.~G.~M. and {Charnley}, S.~B.},
        title = "{Circumstellar and Interstellar Synthesis of Organic Molecules}",
      journal = {Origins of Life and Evolution of the Biosphere},
         year = 1997,
        month = jan,
       volume = {27},
        pages = {23-51},
          doi = {10.1023/A:1006513928588},
       adsurl = {https://ui.adsabs.harvard.edu/abs/1997OLEB...27...23T}
}

@ARTICLE{bisschop2007,
       author = {{Bisschop}, S.~E. and {J{\o}rgensen}, J.~K. and {van Dishoeck}, E.~F. and {de Wachter}, E.~B.~M.},
        title = "{Testing grain-surface chemistry in massive hot-core regions}",
      journal = {\aap},
         year = 2007,
        month = apr,
       volume = {465},
       number = {3},
        pages = {913-929},
          doi = {10.1051/0004-6361:20065963},
archivePrefix = {arXiv},
       eprint = {astro-ph/0702066},
 primaryClass = {astro-ph},
       adsurl = {https://ui.adsabs.harvard.edu/abs/2007A&A...465..913B}
}

@ARTICLE{millar1997,
       author = {{Millar}, T.~J.},
        title = "{Models of hot molecular cores.}",
      journal = {IAU Symposium},
         year = 1997,
        month = jan,
       volume = {178},
        pages = {75-88},
       adsurl = {https://ui.adsabs.harvard.edu/abs/1997IAUS..178...75M}
}

@ARTICLE{charnley1992,
       author = {{Charnley}, S.~B. and {Tielens}, A.~G.~G.~M. and {Millar}, T.~J.},
        title = "{On the Molecular Complexity of the Hot Cores in Orion A: Grain Surface Chemistry as ``The Last Refuge of the Scoundrel''}",
      journal = {\apjl},
         year = 1992,
        month = nov,
       volume = {399},
        pages = {L71},
          doi = {10.1086/186609},
       adsurl = {https://ui.adsabs.harvard.edu/abs/1992ApJ...399L..71C}
}

@ARTICLE{charnley1995,
       author = {{Charnley}, S.~B.},
        title = "{Hot Core Chemistry}",
      journal = {\apss},
         year = 1995,
        month = feb,
       volume = {224},
       number = {1-2},
        pages = {251-254},
          doi = {10.1007/BF00667853},
       adsurl = {https://ui.adsabs.harvard.edu/abs/1995Ap&SS.224..251C}
}

@ARTICLE{moller2017,
       author = {{M{\"o}ller}, T. and {Endres}, C. and {Schilke}, P.},
        title = "{eXtended CASA Line Analysis Software Suite (XCLASS)}",
      journal = {\aap},
         year = 2017,
        month = feb,
       volume = {598},
          eid = {A7},
        pages = {A7},
          doi = {10.1051/0004-6361/201527203},
archivePrefix = {arXiv},
       eprint = {1508.04114},
 primaryClass = {astro-ph.IM},
       adsurl = {https://ui.adsabs.harvard.edu/abs/2017A&A...598A...7M}
}

@ARTICLE{moller2023,
       author = {{M{\"o}ller}, T. and {Schilke}, P. and {S{\'a}nchez-Monge}, {\'A}. and {Schmiedeke}, A. and {Meng}, F.},
        title = "{The physical and chemical structure of Sagittarius B2. VII. Dust and ionized gas contributions to the full molecular line survey of 47 hot cores}",
      journal = {\aap},
         year = 2023,
        month = aug,
       volume = {676},
          eid = {A121},
        pages = {A121},
          doi = {10.1051/0004-6361/202346903},
archivePrefix = {arXiv},
       eprint = {2307.06222},
 primaryClass = {astro-ph.GA},
       adsurl = {https://ui.adsabs.harvard.edu/abs/2023A&A...676A.121M}
}

@ARTICLE{kalenskii2000,
       author = {{Kalenski{\u{i}}}, S.~V. and {Promislov}, V.~G. and {Alakoz}, A.~V. and {Winnberg}, A. and {Johansson}, L.~E.~B.},
        title = "{Determination of Molecular Gas Properties Using Methyl Cyanide Lines}",
      journal = {Astronomy Reports},
         year = 2000,
        month = nov,
       volume = {44},
       number = {11},
        pages = {725-737},
          doi = {10.1134/1.1320498},
archivePrefix = {arXiv},
       eprint = {astro-ph/9910386},
 primaryClass = {astro-ph},
       adsurl = {https://ui.adsabs.harvard.edu/abs/2000ARep...44..725K}
}

@ARTICLE{qiu2025b,
       author = {{Qiu}, Yisheng and {Zhang}, Tianwei and {M{\"o}ller}, Thomas and {Jiang}, Xue-Jian and {Song}, Zihao and {Chen}, Huaxi and {Quan}, Donghui},
        title = "{Spectuner: A Framework for Automated Line Identification of Interstellar Molecules}",
      journal = {\apjs},
         year = 2025,
        month = mar,
       volume = {277},
       number = {1},
          eid = {21},
        pages = {21},
          doi = {10.3847/1538-4365/adaeba},
archivePrefix = {arXiv},
       eprint = {2408.06004},
 primaryClass = {astro-ph.GA},
       adsurl = {https://ui.adsabs.harvard.edu/abs/2025ApJS..277...21Q}
}

@ARTICLE{qiu2025a,
       author = {{Qiu}, Yisheng and {Zhang}, Tianwei and {Liu}, Tie and {Zhu}, Fengyao and {Meng}, Dezhao and {Chen}, Huaxi and {M{\"o}ller}, Thomas and {Schilke}, Peter and {Quan}, Donghui},
        title = "{Spectuner-D1: Spectral Line Fitting of Interstellar Molecules Using Deep Reinforcement Learning}",
      journal = {arXiv e-prints},
         year = 2025,
        month = nov,
          eid = {arXiv:2511.21027},
        pages = {arXiv:2511.21027},
archivePrefix = {arXiv},
       eprint = {2511.21027},
 primaryClass = {astro-ph.GA},
       adsurl = {https://ui.adsabs.harvard.edu/abs/2025arXiv251121027Q}
}

@ARTICLE{vaillancourt2006,
       author = {{Vaillancourt}, John E.},
        title = "{Placing Confidence Limits on Polarization Measurements}",
      journal = {\pasp},
         year = 2006,
        month = sep,
       volume = {118},
       number = {847},
        pages = {1340-1343},
          doi = {10.1086/507472},
archivePrefix = {arXiv},
       eprint = {astro-ph/0603110},
 primaryClass = {astro-ph},
       adsurl = {https://ui.adsabs.harvard.edu/abs/2006PASP..118.1340V}
}

@ARTICLE{plaszczynski2014,
       author = {{Plaszczynski}, S. and {Montier}, L. and {Levrier}, F. and {Tristram}, M.},
        title = "{A novel estimator of the polarization amplitude from normally distributed Stokes parameters}",
      journal = {\mnras},
         year = 2014,
        month = apr,
       volume = {439},
       number = {4},
        pages = {4048-4056},
          doi = {10.1093/mnras/stu270},
archivePrefix = {arXiv},
       eprint = {1312.0437},
 primaryClass = {astro-ph.CO},
       adsurl = {https://ui.adsabs.harvard.edu/abs/2014MNRAS.439.4048P}
}

@ARTICLE{nk1993,
       author = {{Naghizadeh-Khouei}, J. and {Clarke}, D.},
        title = "{On the statistical behaviour of the position angle of linear polarization}",
      journal = {\aap},
         year = 1993,
        month = jul,
       volume = {274},
        pages = {968},
       adsurl = {https://ui.adsabs.harvard.edu/abs/1993A&A...274..968N}
}

@ARTICLE{davis1951,
       author = {{Davis}, Jr., Leverett and {Greenstein}, Jesse L.},
        title = "{The Polarization of Starlight by Aligned Dust Grains.}",
      journal = {\apj},
         year = 1951,
        month = sep,
       volume = {114},
        pages = {206},
          doi = {10.1086/145464},
       adsurl = {https://ui.adsabs.harvard.edu/abs/1951ApJ...114..206D}
}

@ARTICLE{chandrasekhar1953,
       author = {{Chandrasekhar}, S. and {Fermi}, E.},
        title = "{Magnetic Fields in Spiral Arms.}",
      journal = {\apj},
         year = 1953,
        month = jul,
       volume = {118},
        pages = {113},
          doi = {10.1086/145731},
       adsurl = {https://ui.adsabs.harvard.edu/abs/1953ApJ...118..113C}
}

@ARTICLE{lazarian2007a,
       author = {{Lazarian}, A.},
        title = "{Tracing magnetic fields with aligned grains}",
      journal = {\jqsrt},
         year = 2007,
        month = jul,
       volume = {106},
        pages = {225-256},
          doi = {10.1016/j.jqsrt.2007.01.038},
archivePrefix = {arXiv},
       eprint = {0707.0858},
 primaryClass = {astro-ph},
       adsurl = {https://ui.adsabs.harvard.edu/abs/2007JQSRT.106..225L}
}

@ARTICLE{lazarian2007b,
       author = {{Lazarian}, A. and {Hoang}, Thiem},
        title = "{Radiative torques: analytical model and basic properties}",
      journal = {\mnras},
         year = 2007,
        month = jul,
       volume = {378},
       number = {3},
        pages = {910-946},
          doi = {10.1111/j.1365-2966.2007.11817.x},
archivePrefix = {arXiv},
       eprint = {0707.0886},
 primaryClass = {astro-ph},
       adsurl = {https://ui.adsabs.harvard.edu/abs/2007MNRAS.378..910L}
}

@ARTICLE{gomez2018,
       author = {{G{\'o}mez}, Gilberto C. and {V{\'a}zquez-Semadeni}, Enrique and {Zamora-Avil{\'e}s}, Manuel},
        title = "{The magnetic field structure in molecular cloud filaments}",
      journal = {\mnras},
         year = 2018,
        month = nov,
       volume = {480},
       number = {3},
        pages = {2939-2944},
          doi = {10.1093/mnras/sty2018},
archivePrefix = {arXiv},
       eprint = {1801.03169},
 primaryClass = {astro-ph.GA},
       adsurl = {https://ui.adsabs.harvard.edu/abs/2018MNRAS.480.2939G}
}

@ARTICLE{muller2001,
       author = {{M{\"u}ller}, H.~S.~P. and {Thorwirth}, S. and {Roth}, D.~A. and {Winnewisser}, G.},
        title = "{The Cologne Database for Molecular Spectroscopy, CDMS}",
      journal = {\aap},
         year = 2001,
        month = apr,
       volume = {370},
        pages = {L49-L52},
          doi = {10.1051/0004-6361:20010367},
       adsurl = {https://ui.adsabs.harvard.edu/abs/2001A&A...370L..49M}
}

@ARTICLE{rosolowsky2010,
       author = {{Rosolowsky}, Erik and {Dunham}, Miranda K. and {Ginsburg}, Adam and {Bradley}, E. Todd and {Aguirre}, James and {Bally}, John and {Battersby}, Cara and {Cyganowski}, Claudia and {Dowell}, Darren and {Drosback}, Meredith and {Evans}, II, Neal J. and {Glenn}, Jason and {Harvey}, Paul and {Stringfellow}, Guy S. and {Walawender}, Josh and {Williams}, Jonathan P.},
        title = "{The Bolocam Galactic Plane Survey. II. Catalog of the Image Data}",
      journal = {\apjs},
         year = 2010,
        month = may,
       volume = {188},
       number = {1},
        pages = {123-138},
          doi = {10.1088/0067-0049/188/1/123},
archivePrefix = {arXiv},
       eprint = {0909.2871},
 primaryClass = {astro-ph.GA},
       adsurl = {https://ui.adsabs.harvard.edu/abs/2010ApJS..188..123R}
}

@ARTICLE{lu2014,
       author = {{Lu}, Xing and {Zhang}, Qizhou and {Liu}, Hauyu Baobab and {Wang}, Junzhi and {Gu}, Qiusheng},
        title = "{Very Large Array Observations of Ammonia in High-mass Star Formation Regions}",
      journal = {\apj},
         year = 2014,
        month = aug,
       volume = {790},
       number = {2},
          eid = {84},
        pages = {84},
          doi = {10.1088/0004-637X/790/2/84},
archivePrefix = {arXiv},
       eprint = {1405.7933},
 primaryClass = {astro-ph.GA},
       adsurl = {https://ui.adsabs.harvard.edu/abs/2014ApJ...790...84L}
}

@ARTICLE{hull2020,
       author = {{Hull}, Charles L.~H. and {Le Gouellec}, Valentin J.~M. and {Girart}, Josep M. and {Tobin}, John J. and {Bourke}, Tyler L.},
        title = "{Understanding the Origin of the Magnetic Field Morphology in the Wide-binary Protostellar System BHR 71}",
      journal = {\apj},
         year = 2020,
        month = apr,
       volume = {892},
       number = {2},
          eid = {152},
        pages = {152},
          doi = {10.3847/1538-4357/ab5809},
archivePrefix = {arXiv},
       eprint = {1910.07290},
 primaryClass = {astro-ph.SR},
       adsurl = {https://ui.adsabs.harvard.edu/abs/2020ApJ...892..152H}
}

@ARTICLE{lego2019,
       author = {{Le Gouellec}, Valentin J.~M. and {Hull}, Charles L.~H. and {Maury}, Ana{\"e}lle J. and {Girart}, Josep M. and {Tychoniec}, {\L}ukasz and {Kristensen}, Lars E. and {Li}, Zhi-Yun and {Louvet}, Fabien and {Cortes}, Paulo C. and {Rao}, Ramprasad},
        title = "{Characterizing Magnetic Field Morphologies in Three Serpens Protostellar Cores with ALMA}",
      journal = {\apj},
         year = 2019,
        month = nov,
       volume = {885},
       number = {2},
          eid = {106},
        pages = {106},
          doi = {10.3847/1538-4357/ab43c2},
archivePrefix = {arXiv},
       eprint = {1909.00046},
 primaryClass = {astro-ph.GA},
       adsurl = {https://ui.adsabs.harvard.edu/abs/2019ApJ...885..106L}
}

@ARTICLE{hull&zhang2019,
       author = {{Hull}, Charles L.~H. and {Zhang}, Qizhou},
        title = "{Interferometric observations of magnetic fields in forming stars}",
      journal = {Frontiers in Astronomy and Space Sciences},
         year = 2019,
        month = mar,
       volume = {6},
          eid = {3},
        pages = {3},
          doi = {10.3389/fspas.2019.00003},
archivePrefix = {arXiv},
       eprint = {1903.03177},
 primaryClass = {astro-ph.SR},
       adsurl = {https://ui.adsabs.harvard.edu/abs/2019FrASS...6....3H}
}

@ARTICLE{lucas2008,
       author = {{Lucas}, P.~W. and {Hoare}, M.~G. and {Longmore}, A. and {Schr{\"o}der}, A.~C. and {Davis}, C.~J. and {Adamson}, A. and {Bandyopadhyay}, R.~M. and {de Grijs}, R. and {Smith}, M. and {Gosling}, A. and {Mitchison}, S. and {G{\'a}sp{\'a}r}, A. and {Coe}, M. and {Tamura}, M. and {Parker}, Q. and {Irwin}, M. and {Hambly}, N. and {Bryant}, J. and {Collins}, R.~S. and {Cross}, N. and {Evans}, D.~W. and {Gonzalez-Solares}, E. and {Hodgkin}, S. and {Lewis}, J. and {Read}, M. and {Riello}, M. and {Sutorius}, E.~T.~W. and {Lawrence}, A. and {Drew}, J.~E. and {Dye}, S. and {Thompson}, M.~A.},
        title = "{The UKIDSS Galactic Plane Survey}",
      journal = {\mnras},
         year = 2008,
        month = nov,
       volume = {391},
       number = {1},
        pages = {136-163},
          doi = {10.1111/j.1365-2966.2008.13924.x},
archivePrefix = {arXiv},
       eprint = {0712.0100},
 primaryClass = {astro-ph},
       adsurl = {https://ui.adsabs.harvard.edu/abs/2008MNRAS.391..136L}
}

@ARTICLE{hull2017a,
       author = {{Hull}, Charles L.~H. and {Girart}, Josep M. and {Tychoniec}, {\L}ukasz and {Rao}, Ramprasad and {Cort{\'e}s}, Paulo C. and {Pokhrel}, Riwaj and {Zhang}, Qizhou and {Houde}, Martin and {Dunham}, Michael M. and {Kristensen}, Lars E. and {Lai}, Shih-Ping and {Li}, Zhi-Yun and {Plambeck}, Richard L.},
        title = "{ALMA Observations of Dust Polarization and Molecular Line Emission from the Class 0 Protostellar Source Serpens SMM1}",
      journal = {\apj},
         year = 2017,
        month = oct,
       volume = {847},
       number = {2},
          eid = {92},
        pages = {92},
          doi = {10.3847/1538-4357/aa7fe9},
archivePrefix = {arXiv},
       eprint = {1707.03827},
 primaryClass = {astro-ph.GA},
       adsurl = {https://ui.adsabs.harvard.edu/abs/2017ApJ...847...92H}
}

@ARTICLE{maury2018,
       author = {{Maury}, A.~J. and {Girart}, J.~M. and {Zhang}, Q. and {Hennebelle}, P. and {Keto}, E. and {Rao}, R. and {Lai}, S. -P. and {Ohashi}, N. and {Galametz}, M.},
        title = "{Magnetically regulated collapse in the B335 protostar? I. ALMA observations of the polarized dust emission}",
      journal = {\mnras},
         year = 2018,
        month = jun,
       volume = {477},
       number = {2},
        pages = {2760-2765},
          doi = {10.1093/mnras/sty574},
archivePrefix = {arXiv},
       eprint = {1803.00028},
 primaryClass = {astro-ph.GA},
       adsurl = {https://ui.adsabs.harvard.edu/abs/2018MNRAS.477.2760M}
}

@ARTICLE{qiu2014,
       author = {{Qiu}, Keping and {Zhang}, Qizhou and {Menten}, Karl M. and {Liu}, Hauyu B. and {Tang}, Ya-Wen and {Girart}, Josep M.},
        title = "{Submillimeter Array Observations of Magnetic Fields in G240.31+0.07: An Hourglass in a Massive Cluster-forming Core}",
      journal = {\apjl},
         year = 2014,
        month = oct,
       volume = {794},
       number = {1},
          eid = {L18},
        pages = {L18},
          doi = {10.1088/2041-8205/794/1/L18},
archivePrefix = {arXiv},
       eprint = {1409.5608},
 primaryClass = {astro-ph.GA},
       adsurl = {https://ui.adsabs.harvard.edu/abs/2014ApJ...794L..18Q}
}

@ARTICLE{girart2006,
       author = {{Girart}, Josep M. and {Rao}, Ramprasad and {Marrone}, Daniel P.},
        title = "{Magnetic Fields in the Formation of Sun-Like Stars}",
      journal = {Science},
         year = 2006,
        month = aug,
       volume = {313},
       number = {5788},
        pages = {812-814},
          doi = {10.1126/science.1129093},
archivePrefix = {arXiv},
       eprint = {astro-ph/0609177},
 primaryClass = {astro-ph},
       adsurl = {https://ui.adsabs.harvard.edu/abs/2006Sci...313..812G}
}

@ARTICLE{liu2020,
       author = {{Liu}, Tie and {Evans}, Neal J. and {Kim}, Kee-Tae and {Goldsmith}, Paul F. and {Liu}, Sheng-Yuan and {Zhang}, Qizhou and {Tatematsu}, Ken'ichi and {Wang}, Ke and {Juvela}, Mika and {Bronfman}, Leonardo and {Cunningham}, Maria R. and {Garay}, Guido and {Hirota}, Tomoya and {Lee}, Jeong-Eun and {Kang}, Sung-Ju and {Li}, Di and {Li}, Pak-Shing and {Mardones}, Diego and {Qin}, Sheng-Li and {Ristorcelli}, Isabelle and {Tej}, Anandmayee and {Toth}, L. Viktor and {Wu}, Jing-Wen and {Wu}, Yue-Fang and {Yi}, Hee-weon and {Yun}, Hyeong-Sik and {Liu}, Hong-Li and {Peng}, Ya-Ping and {Li}, Juan and {Li}, Shang-Huo and {Lee}, Chang Won and {Shen}, Zhi-Qiang and {Baug}, Tapas and {Wang}, Jun-Zhi and {Zhang}, Yong and {Issac}, Namitha and {Zhu}, Feng-Yao and {Luo}, Qiu-Yi and {Soam}, Archana and {Liu}, Xun-Chuan and {Xu}, Feng-Wei and {Wang}, Yu and {Zhang}, Chao and {Ren}, Zhiyuan and {Zhang}, Chao},
        title = "{ATOMS: ALMA Three-millimeter Observations of Massive Star-forming regions - I. Survey description and a first look at G9.62+0.19}",
      journal = {\mnras},
         year = 2020,
        month = aug,
       volume = {496},
       number = {3},
        pages = {2790-2820},
          doi = {10.1093/mnras/staa1577},
archivePrefix = {arXiv},
       eprint = {2006.01549},
 primaryClass = {astro-ph.GA},
       adsurl = {https://ui.adsabs.harvard.edu/abs/2020MNRAS.496.2790L}
}

@ARTICLE{kwon2025,
       author = {{Kwon}, Jungmi and {Tamura}, Motohide and {Kudo}, Tomoyuki},
        title = "{Near-infrared Polarimetry of the Massive Young Star W33A in the Godzilla Nebula: Detection of Linear and Circular Polarization}",
      journal = {\aj},
         year = 2025,
        month = aug,
       volume = {170},
       number = {2},
          eid = {107},
        pages = {107},
          doi = {10.3847/1538-3881/ade712},
       adsurl = {https://ui.adsabs.harvard.edu/abs/2025AJ....170..107K}
}

@ARTICLE{hough1989,
       author = {{Hough}, J.~H. and {Whittet}, D.~C.~B. and {Sato}, S. and {Yamashita}, T. and {Tamura}, M. and {Nagata}, T. and {Aitken}, D.~K. and {Roche}, P.~F.},
        title = "{Spectropolarimetry of the 3-{\ensuremath{\mu}}m ice feature in molecular clouds. II. GL 2591, GL 2136, W33A and Elias 29 ({\ensuremath{\rho}} Ophiuchi dark cloud).}",
      journal = {\mnras},
         year = 1989,
        month = nov,
       volume = {241},
        pages = {71-88},
          doi = {10.1093/mnras/241.1.71},
       adsurl = {https://ui.adsabs.harvard.edu/abs/1989MNRAS.241...71H}
}

@ARTICLE{chrysostomou1996,
       author = {{Chrysostomou}, A. and {Hough}, J.~H. and {Whittet}, D.~C.~B. and {Aitken}, D.~K. and {Roche}, P.~F. and {Lazarian}, A.},
        title = "{Interstellar Polarization from CO and XCN Mantled Grains: A Severe Test for Grain Alignment Mechanisms}",
      journal = {\apjl},
         year = 1996,
        month = jul,
       volume = {465},
        pages = {L61},
          doi = {10.1086/310125},
       adsurl = {https://ui.adsabs.harvard.edu/abs/1996ApJ...465L..61C}
}

@ARTICLE{smith2000,
       author = {{Smith}, Craig H. and {Wright}, Christopher M. and {Aitken}, David K. and {Roche}, Patrick F. and {Hough}, James H.},
        title = "{Studies in mid-infrared spectropolarimetry - II. An atlas of spectra}",
      journal = {\mnras},
         year = 2000,
        month = feb,
       volume = {312},
       number = {2},
        pages = {327-361},
          doi = {10.1046/j.1365-8711.2000.03158.x},
       adsurl = {https://ui.adsabs.harvard.edu/abs/2000MNRAS.312..327S}
}

@ARTICLE{rao2009,
       author = {{Rao}, Ramprasad and {Girart}, Josep M. and {Marrone}, Daniel P. and {Lai}, Shih-Ping and {Schnee}, Scott},
        title = "{IRAS 16293: A ``Magnetic'' Tale of Two Cores}",
      journal = {\apj},
         year = 2009,
        month = dec,
       volume = {707},
       number = {2},
        pages = {921-935},
          doi = {10.1088/0004-637X/707/2/921},
archivePrefix = {arXiv},
       eprint = {0910.5269},
 primaryClass = {astro-ph.GA},
       adsurl = {https://ui.adsabs.harvard.edu/abs/2009ApJ...707..921R}
}

@ARTICLE{masson2016,
       author = {{Masson}, J. and {Chabrier}, G. and {Hennebelle}, P. and {Vaytet}, N. and {Commer{\c{c}}on}, B.},
        title = "{Ambipolar diffusion in low-mass star formation. I. General comparison with the ideal magnetohydrodynamic case}",
      journal = {\aap},
         year = 2016,
        month = mar,
       volume = {587},
          eid = {A32},
        pages = {A32},
          doi = {10.1051/0004-6361/201526371},
archivePrefix = {arXiv},
       eprint = {1509.05630},
 primaryClass = {astro-ph.SR},
       adsurl = {https://ui.adsabs.harvard.edu/abs/2016A&A...587A..32M}
}

@ARTICLE{galli2006,
       author = {{Galli}, Daniele and {Lizano}, Susana and {Shu}, Frank H. and {Allen}, Anthony},
        title = "{Gravitational Collapse of Magnetized Clouds. I. Ideal Magnetohydrodynamic Accretion Flow}",
      journal = {\apj},
         year = 2006,
        month = aug,
       volume = {647},
       number = {1},
        pages = {374-381},
          doi = {10.1086/505257},
archivePrefix = {arXiv},
       eprint = {astro-ph/0604573},
 primaryClass = {astro-ph},
       adsurl = {https://ui.adsabs.harvard.edu/abs/2006ApJ...647..374G}
}

@ARTICLE{liu2021,
       author = {{Liu}, Junhao and {Zhang}, Qizhou and {Commer{\c{c}}on}, Beno{\^\i}t and {Valdivia}, Valeska and {Maury}, Ana{\"e}lle and {Qiu}, Keping},
        title = "{Calibrating the Davis-Chandrasekhar-Fermi Method with Numerical Simulations: Uncertainties in Estimating the Magnetic Field Strength from Statistics of Field Orientations}",
      journal = {\apj},
         year = 2021,
        month = oct,
       volume = {919},
       number = {2},
          eid = {79},
        pages = {79},
          doi = {10.3847/1538-4357/ac0cec},
archivePrefix = {arXiv},
       eprint = {2106.09934},
 primaryClass = {astro-ph.GA},
       adsurl = {https://ui.adsabs.harvard.edu/abs/2021ApJ...919...79L}
}

@ARTICLE{choi2025,
       author = {{Choi}, Youngwoo and {Kwon}, Woojin and {Looney}, Leslie W. and {Stephens}, Ian W. and {Li}, Zhi-Yun and {van der Tak}, Floris F.~S. and {Tobin}, John J.},
        title = "{The Hourglass-shaped Magnetic Fields and Dust Filaments in the HH 211 Protostellar Envelope}",
      journal = {arXiv e-prints},
         year = 2025,
        month = sep,
          eid = {arXiv:2509.25764},
        pages = {arXiv:2509.25764},
archivePrefix = {arXiv},
       eprint = {2509.25764},
 primaryClass = {astro-ph.SR},
       adsurl = {https://ui.adsabs.harvard.edu/abs/2025arXiv250925764C}
}

@ARTICLE{kwon2019,
       author = {{Kwon}, Woojin and {Stephens}, Ian W. and {Tobin}, John J. and {Looney}, Leslie W. and {Li}, Zhi-Yun and {van der Tak}, Floris F.~S. and {Crutcher}, Richard M.},
        title = "{Highly Ordered and Pinched Magnetic Fields in the Class 0 Protobinary System L1448 IRS 2}",
      journal = {\apj},
         year = 2019,
        month = jul,
       volume = {879},
       number = {1},
          eid = {25},
        pages = {25},
          doi = {10.3847/1538-4357/ab24c8},
archivePrefix = {arXiv},
       eprint = {1805.07348},
 primaryClass = {astro-ph.SR},
       adsurl = {https://ui.adsabs.harvard.edu/abs/2019ApJ...879...25K}
}

@ARTICLE{huang2024,
       author = {{Huang}, Bo and {Girart}, Josep M. and {Stephens}, Ian W. and {Fern{\'a}ndez L{\'o}pez}, Manuel and {Arce}, Hector G. and {Carpenter}, John M. and {Cortes}, Paulo and {Cox}, Erin G. and {Friesen}, Rachel and {Le Gouellec}, Valentin J.~M. and {Hull}, Charles L.~H. and {Karnath}, Nicole and {Kwon}, Woojin and {Li}, Zhi-Yun and {Looney}, Leslie W. and {Megeath}, S. Thomas and {Myers}, Philip C. and {Murillo}, Nadia M. and {Pineda}, Jaime E. and {Sadavoy}, Sarah and {S{\'a}nchez-Monge}, {\'A}lvaro and {Sanhueza}, Patricio and {Tobin}, John J. and {Zhang}, Qizhou and {Jackson}, James M. and {Segura-Cox}, Dominique},
        title = "{On the Magnetic Field Properties of Protostellar Envelopes in Orion}",
      journal = {\apjl},
         year = 2024,
        month = mar,
       volume = {963},
       number = {1},
          eid = {L31},
        pages = {L31},
          doi = {10.3847/2041-8213/ad27d4},
archivePrefix = {arXiv},
       eprint = {2402.07267},
 primaryClass = {astro-ph.GA},
       adsurl = {https://ui.adsabs.harvard.edu/abs/2024ApJ...963L..31H}
}

@ARTICLE{cortes2025,
       author = {{Cortes}, P.~C. and {Pineda}, J.~E. and {Hsieh}, T. -H. and {Tobin}, J.~J. and {Saha}, P. and {Girart}, J.~M. and {Le Gouellec}, V.~J.~M. and {Stephens}, I.~W. and {Looney}, L.~W. and {Koumpia}, E. and {Valdivia-Mena}, M.~T. and {Cacciapuoti}, L. and {Gieser}, C. and {Offner}, S.~S.~R. and {Caselli}, P. and {Sanhueza}, P. and {Segura-Cox}, D. and {Fernandez-Lopez}, M. and {Morii}, K. and {Huang}, B. and {Alves}, F.~O. and {Zhang}, Q. and {Kwon}, W. and {Hull}, C.~L.~H. and {Li}, Z.~Y.},
        title = "{First results from ALPPS: a sub-Alfv{\'e}nic streamer in SVS13A}",
      journal = {arXiv e-prints},
         year = 2025,
        month = sep,
          eid = {arXiv:2509.21701},
        pages = {arXiv:2509.21701},
          doi = {10.48550/arXiv.2509.21701},
archivePrefix = {arXiv},
       eprint = {2509.21701},
 primaryClass = {astro-ph.GA},
       adsurl = {https://ui.adsabs.harvard.edu/abs/2025arXiv250921701C}
}

@ARTICLE{purser2016,
       author = {{Purser}, S.~J.~D. and {Lumsden}, S.~L. and {Hoare}, M.~G. and {Urquhart}, J.~S. and {Cunningham}, N. and {Purcell}, C.~R. and {Brooks}, K.~J. and {Garay}, G. and {G{\'u}zman}, A.~E. and {Voronkov}, M.~A.},
        title = "{A search for ionized jets towards massive young stellar objects}",
      journal = {\mnras},
         year = 2016,
        month = jul,
       volume = {460},
       number = {1},
        pages = {1039-1053},
          doi = {10.1093/mnras/stw1027},
archivePrefix = {arXiv},
       eprint = {1605.01200},
 primaryClass = {astro-ph.GA},
       adsurl = {https://ui.adsabs.harvard.edu/abs/2016MNRAS.460.1039P}
}

@ARTICLE{ostriker1964,
       author = {{Ostriker}, J.},
        title = "{The Equilibrium of Polytropic and Isothermal Cylinders.}",
      journal = {\apj},
         year = 1964,
        month = oct,
       volume = {140},
        pages = {1056},
          doi = {10.1086/148005},
       adsurl = {https://ui.adsabs.harvard.edu/abs/1964ApJ...140.1056O}
}

@INPROCEEDINGS{hacar2023,
       author = {{Hacar}, A. and {Clark}, S.~E. and {Heitsch}, F. and {Kainulainen}, J. and {Panopoulou}, G.~V. and {Seifried}, D. and {Smith}, R.},
        title = "{Initial Conditions for Star Formation: a Physical Description of the Filamentary ISM}",
    booktitle = {Protostars and Planets VII},
         year = 2023,
       editor = {{Inutsuka}, S. and {Aikawa}, Y. and {Muto}, T. and {Tomida}, K. and {Tamura}, M.},
       series = {Astronomical Society of the Pacific Conference Series},
       volume = {534},
        month = jul,
        pages = {153},
          doi = {10.48550/arXiv.2203.09562},
archivePrefix = {arXiv},
       eprint = {2203.09562},
 primaryClass = {astro-ph.GA},
       adsurl = {https://ui.adsabs.harvard.edu/abs/2023ASPC..534..153H}
}

@ARTICLE{wang2014,
       author = {{Wang}, Ke and {Zhang}, Qizhou and {Testi}, Leonardo and {van der Tak}, Floris and {Wu}, Yuefang and {Zhang}, Huawei and {Pillai}, Thushara and {Wyrowski}, Friedrich and {Carey}, Sean and {Ragan}, Sarah E. and et al.},
        title = "{Hierarchical fragmentation and differential star formation in the Galactic `Snake': infrared dark cloud G11.11-0.12}",
      journal = {\mnras},
         year = 2014,
        month = apr,
       volume = {439},
       number = {4},
        pages = {3275-3293},
          doi = {10.1093/mnras/stu127},
archivePrefix = {arXiv},
       eprint = {1401.4157},
 primaryClass = {astro-ph.GA},
       adsurl = {https://ui.adsabs.harvard.edu/abs/2014MNRAS.439.3275W}
}

@ARTICLE{tomisaka2014,
       author = {{Tomisaka}, Kohji},
        title = "{Magnetohydrostatic Equilibrium Structure and Mass of Filamentary Isothermal Cloud Threaded by Lateral Magnetic Field}",
      journal = {\apj},
         year = 2014,
        month = apr,
       volume = {785},
       number = {1},
          eid = {24},
        pages = {24},
          doi = {10.1088/0004-637X/785/1/24},
archivePrefix = {arXiv},
       eprint = {1402.3033},
 primaryClass = {astro-ph.GA},
       adsurl = {https://ui.adsabs.harvard.edu/abs/2014ApJ...785...24T}
}

@ARTICLE{pillsworth2024,
       author = {{Pillsworth}, Rachel and {Pudritz}, Ralph E.},
        title = "{Necessary conditions for the formation of filaments and star clusters in the cold neutral medium}",
      journal = {\mnras},
         year = 2024,
        month = feb,
       volume = {528},
       number = {1},
        pages = {209-233},
          doi = {10.1093/mnras/stae002},
archivePrefix = {arXiv},
       eprint = {2401.01737},
 primaryClass = {astro-ph.GA},
       adsurl = {https://ui.adsabs.harvard.edu/abs/2024MNRAS.528..209P}
}

@ARTICLE{fiege2000,
       author = {{Fiege}, Jason D. and {Pudritz}, Ralph E.},
        title = "{Helical fields and filamentary molecular clouds - I}",
      journal = {\mnras},
         year = 2000,
        month = jan,
       volume = {311},
       number = {1},
        pages = {85-104},
          doi = {10.1046/j.1365-8711.2000.03066.x},
archivePrefix = {arXiv},
       eprint = {astro-ph/9901096},
 primaryClass = {astro-ph},
       adsurl = {https://ui.adsabs.harvard.edu/abs/2000MNRAS.311...85F}
}

@ARTICLE{cox2016,
       author = {{Cox}, N.~L.~J. and {Arzoumanian}, D. and {Andr{\'e}}, Ph. and {Rygl}, K.~L.~J. and {Prusti}, T. and {Men'shchikov}, A. and {Royer}, P. and {K{\'o}sp{\'a}l}, {\'A}. and {Palmeirim}, P. and {Ribas}, A. and et al.},
        title = "{Filamentary structure and magnetic field orientation in Musca}",
      journal = {\aap},
         year = 2016,
        month = may,
       volume = {590},
          eid = {A110},
        pages = {A110},
          doi = {10.1051/0004-6361/201527068},
       adsurl = {https://ui.adsabs.harvard.edu/abs/2016A&A...590A.110C}
}

@ARTICLE{ching2022,
       author = {{Ching}, Tao-Chung and {Qiu}, Keping and {Li}, Di and {Ren}, Zhiyuan and {Lai}, Shih-Ping and {Berry}, David and {Pattle}, Kate and {Furuya}, Ray and {Ward-Thompson}, Derek and {Johnstone}, Doug and {Koch}, Patrick M. and {Lee}, Chang Won and {Hoang}, Thiem and {Hasegawa}, Tetsuo and {Kwon}, Woojin and {Bastien}, Pierre and {Eswaraiah}, Chakali and {Wang}, Jia-Wei and {Kim}, Kyoung Hee and {Hwang}, Jihye and {Soam}, Archana and {Lyo}, A. -Ran and {Liu}, Junhao and {Le Gouellec}, Valentin J.~M. and {Arzoumanian}, Doris and {Whitworth}, Anthony and {Di Francesco}, James and {Poidevin}, Fr{\'e}d{\'e}rick and {Liu}, Tie and {Coud{\'e}}, Simon and {Tahani}, Mehrnoosh and {Liu}, Hong-Li and {Onaka}, Takashi and {Li}, Dalei and {Tamura}, Motohide and {Chen}, Zhiwei and {Tang}, Xindi and {Kirchschlager}, Florian and {Bourke}, Tyler L. and {Byun}, Do-Young and {Chen}, Mike and {Chen}, Huei-Ru Vivien and {Chen}, Wen Ping and {Cho}, Jungyeon and {Choi}, Yunhee and {Choi}, Youngwoo and {Choi}, Minho and {Chrysostomou}, Antonio and {Chung}, Eun Jung and {Dai}, Y. Sophia and {Diep}, Pham Ngoc and {Doi}, Yasuo and {Duan}, Yan and {Duan}, Hao-Yuan and {Eden}, David and {Fanciullo}, Lapo and {Fiege}, Jason and {Fissel}, Laura M. and {Franzmann}, Erica and {Friberg}, Per and {Friesen}, Rachel and {Fuller}, Gary and {Gledhill}, Tim and {Graves}, Sarah and {Greaves}, Jane and {Griffin}, Matt and {Gu}, Qilao and {Han}, Ilseung and {Hayashi}, Saeko and {Houde}, Martin and {Hull}, Charles L.~H. and {Inoue}, Tsuyoshi and {Inutsuka}, Shu-ichiro and {Iwasaki}, Kazunari and {Jeong}, Il-Gyo and {K{\"o}nyves}, Vera and {Kang}, Ji-hyun and {Kang}, Miju and {Karoly}, Janik and {Kataoka}, Akimasa and {Kawabata}, Koji and {Kemper}, Francisca and {Kim}, Jongsoo and {Kim}, Mi-Ryang and {Kim}, Shinyoung and {Kim}, Hyosung and {Kim}, Kee-Tae and {Kim}, Gwanjeong and {Kirk}, Jason and {Kobayashi}, Masato I.~N. and {Kusune}, Takayoshi and {Kwon}, Jungmi and {Lacaille}, Kevin and {Law}, Chi-Yan and {Lee}, Sang-Sung and {Lee}, Hyeseung and {Lee}, Jeong-Eun and {Lee}, Chin-Fei and {Lee}, Yong-Hee and {Li}, Guangxing and {Li}, Hua-bai and {Lin}, Sheng-Jun and {Liu}, Sheng-Yuan and {Lu}, Xing and {Mairs}, Steve and {Matsumura}, Masafumi and {Matthews}, Brenda and {Moriarty-Schieven}, Gerald and {Nagata}, Tetsuya and {Nakamura}, Fumitaka and {Nakanishi}, Hiroyuki and {Ngoc}, Nguyen Bich and {Ohashi}, Nagayoshi and {Park}, Geumsook and {Parsons}, Harriet and {Peretto}, Nicolas and {Priestley}, Felix and {Pyo}, Tae-Soo and {Qian}, Lei and {Rao}, Ramprasad and {Rawlings}, Mark and {Rawlings}, Jonathan and {Retter}, Brendan and {Richer}, John and {Rigby}, Andrew and {Sadavoy}, Sarah and {Saito}, Hiro and {Savini}, Giorgio and {Seta}, Masumichi and {Shimajiri}, Yoshito and {Shinnaga}, Hiroko and {Tang}, Ya-Wen and {Tomisaka}, Kohji and {Tram}, Le Ngoc and {Tsukamoto}, Yusuke and {Viti}, Serena and {Wang}, Hongchi and {Wu}, Jintai and {Xie}, Jinjin and {Yang}, Meng-Zhe and {Yen}, Hsi-Wei and {Yoo}, Hyunju and {Yuan}, Jinghua and {Yun}, Hyeong-Sik and {Zenko}, Tetsuya and {Zhang}, Chuan-Peng and {Zhang}, Yapeng and {Zhang}, Guoyin and {Zhou}, Jianjun and {Zhu}, Lei and {de Looze}, Ilse and {Andr{\'e}}, Philippe and {Dowell}, C. Darren and {Eyres}, Stewart and {Falle}, Sam and {Robitaille}, Jean-Fran{\c{c}}ois and {van Loo}, Sven},
        title = "{The JCMT BISTRO-2 Survey: Magnetic Fields of the Massive DR21 Filament}",
      journal = {\apj},
         year = 2022,
        month = dec,
       volume = {941},
       number = {2},
          eid = {122},
        pages = {122},
          doi = {10.3847/1538-4357/ac9dfb},
archivePrefix = {arXiv},
       eprint = {2212.01981},
 primaryClass = {astro-ph.GA},
       adsurl = {https://ui.adsabs.harvard.edu/abs/2022ApJ...941..122C}
}

@ARTICLE{tu2024,
       author = {{Tu}, Yisheng and {Li}, Zhi-Yun and {Lam}, Ka Ho and {Tomida}, Kengo and {Hsu}, Chun-Yen},
        title = "{Protostellar discs fed by dense collapsing gravomagneto sheetlets}",
      journal = {\mnras},
         year = 2024,
        month = feb,
       volume = {527},
       number = {4},
        pages = {10131-10150},
          doi = {10.1093/mnras/stad3843},
archivePrefix = {arXiv},
       eprint = {2307.16774},
 primaryClass = {astro-ph.EP},
       adsurl = {https://ui.adsabs.harvard.edu/abs/2024MNRAS.52710131T}
}

@ARTICLE{schleuning1998,
       author = {{Schleuning}, D.~A.},
        title = "{Far-Infrared and Submillimeter Polarization of OMC-1: Evidence for Magnetically Regulated Star Formation}",
      journal = {\apj},
         year = 1998,
        month = jan,
       volume = {493},
       number = {2},
        pages = {811-825},
          doi = {10.1086/305139},
       adsurl = {https://ui.adsabs.harvard.edu/abs/1998ApJ...493..811S}
}

@ARTICLE{lin2016,
       author = {{Lin}, Yuxin and {Liu}, Hauyu Baobab and {Li}, Di and {Zhang}, Zhi-Yu and {Ginsburg}, Adam and {Pineda}, Jaime E. and {Qian}, Lei and {Galv{\'a}n-Madrid}, Roberto and {McLeod}, Anna Faye and {Rosolowsky}, Erik and {Dale}, James E. and {Immer}, Katharina and {Koch}, Eric and {Longmore}, Steve and {Walker}, Daniel and {Testi}, Leonardo},
        title = "{Cloud Structure of Galactic OB Cluster-forming Regions from Combining Ground- and Space-based Bolometric Observations}",
      journal = {\apj},
         year = 2016,
        month = sep,
       volume = {828},
       number = {1},
          eid = {32},
        pages = {32},
          doi = {10.3847/0004-637X/828/1/32},
archivePrefix = {arXiv},
       eprint = {1606.07645},
 primaryClass = {astro-ph.GA},
       adsurl = {https://ui.adsabs.harvard.edu/abs/2016ApJ...828...32L}
}

@ARTICLE{hennebelle2019,
       author = {{Hennebelle}, Patrick and {Inutsuka}, Shu-ichiro},
        title = "{The role of magnetic field in molecular cloud formation and evolution}",
      journal = {Frontiers in Astronomy and Space Sciences},
         year = 2019,
        month = mar,
       volume = {6},
          eid = {5},
        pages = {5},
          doi = {10.3389/fspas.2019.00005},
archivePrefix = {arXiv},
       eprint = {1902.00798},
 primaryClass = {astro-ph.GA},
       adsurl = {https://ui.adsabs.harvard.edu/abs/2019FrASS...6....5H}
}

@ARTICLE{zhang2025,
       author = {{Zhang}, Qizhou and {Liu}, Junhao and {Zeng}, Lingzhen and {Soler}, J.~D. and {Chen}, Huei-Ru Vivien and {Ching}, Tao-Chung and {Ho}, Paul T.~P. and {Girart}, Josep Miquel and {Koch}, Patrick M. and {Lai}, Shih-Ping and et al.},
        title = "{Impact of Gravity on Changing Magnetic Field Orientations in a Sample of Massive Protostellar Clusters Observed with ALMA}",
      journal = {\apj},
         year = 2025,
        month = oct,
       volume = {992},
       number = {1},
          eid = {103},
        pages = {103},
          doi = {10.3847/1538-4357/adfdcb},
archivePrefix = {arXiv},
       eprint = {2508.18538},
 primaryClass = {astro-ph.GA},
       adsurl = {https://ui.adsabs.harvard.edu/abs/2025ApJ...992..103Z}
}

@ARTICLE{shu1987,
       author = {{Shu}, Frank H. and {Adams}, Fred C. and {Lizano}, Susana},
        title = "{Star formation in molecular clouds: observation and theory.}",
      journal = {\araa},
         year = 1987,
        month = jan,
       volume = {25},
        pages = {23-81},
          doi = {10.1146/annurev.aa.25.090187.000323},
       adsurl = {https://ui.adsabs.harvard.edu/abs/1987ARA&A..25...23S}
}

@ARTICLE{crutcher2012,
       author = {{Crutcher}, Richard M.},
        title = "{Magnetic Fields in Molecular Clouds}",
      journal = {\araa},
         year = 2012,
        month = sep,
       volume = {50},
        pages = {29-63},
          doi = {10.1146/annurev-astro-081811-125514},
       adsurl = {https://ui.adsabs.harvard.edu/abs/2012ARA&A..50...29C}
}

@ARTICLE{crutcher1993,
       author = {{Crutcher}, R.~M. and {Troland}, T.~H. and {Goodman}, A.~A. and {Heiles}, C. and {Kazes}, I. and {Myers}, P.~C.},
        title = "{OH Zeeman Observations of Dark Clouds}",
      journal = {\apj},
         year = 1993,
        month = apr,
       volume = {407},
        pages = {175},
          doi = {10.1086/172503},
       adsurl = {https://ui.adsabs.harvard.edu/abs/1993ApJ...407..175C}
}

@ARTICLE{gk1981,
       author = {{Goldreich}, P. and {Kylafis}, N.~D.},
        title = "{On mapping the magnetic field direction in molecular clouds by polarization measurements}",
      journal = {\apjl},
         year = 1981,
        month = jan,
       volume = {243},
        pages = {L75-L78},
          doi = {10.1086/183446},
       adsurl = {https://ui.adsabs.harvard.edu/abs/1981ApJ...243L..75G}
}

@ARTICLE{seifried2011,
       author = {{Seifried}, D. and {Banerjee}, R. and {Klessen}, R.~S. and {Duffin}, D. and {Pudritz}, R.~E.},
        title = "{Magnetic fields during the early stages of massive star formation - I. Accretion and disc evolution}",
      journal = {\mnras},
         year = 2011,
        month = oct,
       volume = {417},
       number = {2},
        pages = {1054-1073},
          doi = {10.1111/j.1365-2966.2011.19320.x},
archivePrefix = {arXiv},
       eprint = {1106.4485},
 primaryClass = {astro-ph.SR},
       adsurl = {https://ui.adsabs.harvard.edu/abs/2011MNRAS.417.1054S}
}

@ARTICLE{koch2012a,
       author = {{Koch}, Patrick M. and {Tang}, Ya-Wen and {Ho}, Paul T.~P.},
        title = "{Magnetic Field Strength Maps for Molecular Clouds: A New Method Based on a Polarization-Intensity Gradient Relation}",
      journal = {\apj},
         year = 2012,
        month = mar,
       volume = {747},
       number = {1},
          eid = {79},
        pages = {79},
          doi = {10.1088/0004-637X/747/1/79},
archivePrefix = {arXiv},
       eprint = {1201.4263},
 primaryClass = {astro-ph.GA},
       adsurl = {https://ui.adsabs.harvard.edu/abs/2012ApJ...747...79K}
}

@ARTICLE{koch2013,
       author = {{Koch}, Patrick M. and {Tang}, Ya-Wen and {Ho}, Paul T.~P.},
        title = "{Interpreting the Role of the Magnetic Field from Dust Polarization Maps}",
      journal = {\apj},
         year = 2013,
        month = sep,
       volume = {775},
       number = {1},
          eid = {77},
        pages = {77},
          doi = {10.1088/0004-637X/775/1/77},
archivePrefix = {arXiv},
       eprint = {1308.6185},
 primaryClass = {astro-ph.GA},
       adsurl = {https://ui.adsabs.harvard.edu/abs/2013ApJ...775...77K}
}

@ARTICLE{soler2013,
       author = {{Soler}, J.~D. and {Hennebelle}, P. and {Martin}, P.~G. and {Miville-Desch{\^e}nes}, M. -A. and {Netterfield}, C.~B. and {Fissel}, L.~M.},
        title = "{An Imprint of Molecular Cloud Magnetization in the Morphology of the Dust Polarized Emission}",
      journal = {\apj},
         year = 2013,
        month = sep,
       volume = {774},
       number = {2},
          eid = {128},
        pages = {128},
          doi = {10.1088/0004-637X/774/2/128},
archivePrefix = {arXiv},
       eprint = {1303.1830},
 primaryClass = {astro-ph.GA},
       adsurl = {https://ui.adsabs.harvard.edu/abs/2013ApJ...774..128S}
}

@ARTICLE{lai2001,
       author = {{Lai}, Shih-Ping and {Crutcher}, Richard M. and {Girart}, Jos{\'e} M. and {Rao}, Ramprasad},
        title = "{Interferometric Mapping of Magnetic Fields in Star-forming Regions. I. W51 e1/e2 Molecular Cores}",
      journal = {\apj},
         year = 2001,
        month = nov,
       volume = {561},
       number = {2},
        pages = {864-870},
          doi = {10.1086/323372},
archivePrefix = {arXiv},
       eprint = {astro-ph/0107322},
 primaryClass = {astro-ph},
       adsurl = {https://ui.adsabs.harvard.edu/abs/2001ApJ...561..864L}
}

@ARTICLE{cortes2008,
       author = {{Cortes}, P.~C. and {Crutcher}, R.~M. and {Shepherd}, D.~S. and {Bronfman}, L.},
        title = "{Interferometric Mapping of Magnetic Fields: The Massive Star-forming Region G34.4+0.23 MM}",
      journal = {\apj},
         year = 2008,
        month = mar,
       volume = {676},
       number = {1},
        pages = {464-471},
          doi = {10.1086/524355},
archivePrefix = {arXiv},
       eprint = {0801.4316},
 primaryClass = {astro-ph},
       adsurl = {https://ui.adsabs.harvard.edu/abs/2008ApJ...676..464C}
}

@ARTICLE{girart2009,
       author = {{Girart}, Josep M. and {Beltr{\'a}n}, Maria T. and {Zhang}, Qizhou and {Rao}, Ramprasad and {Estalella}, Robert},
        title = "{Magnetic Fields in the Formation of Massive Stars}",
      journal = {Science},
         year = 2009,
        month = jun,
       volume = {324},
       number = {5933},
        pages = {1408},
          doi = {10.1126/science.1171807},
       adsurl = {https://ui.adsabs.harvard.edu/abs/2009Sci...324.1408G}
}

@ARTICLE{girart2013,
       author = {{Girart}, J.~M. and {Frau}, P. and {Zhang}, Q. and {Koch}, P.~M. and {Qiu}, K. and {Tang}, Y. -W. and {Lai}, S. -P. and {Ho}, P.~T.~P.},
        title = "{DR 21(OH): A Highly Fragmented, Magnetized, Turbulent Dense Core}",
      journal = {\apj},
         year = 2013,
        month = jul,
       volume = {772},
       number = {1},
          eid = {69},
        pages = {69},
          doi = {10.1088/0004-637X/772/1/69},
archivePrefix = {arXiv},
       eprint = {1305.6509},
 primaryClass = {astro-ph.GA},
       adsurl = {https://ui.adsabs.harvard.edu/abs/2013ApJ...772...69G}
}

@ARTICLE{hull2013,
       author = {{Hull}, Charles L.~H. and {Plambeck}, Richard L. and {Bolatto}, Alberto D. and {Bower}, Geoffrey C. and {Carpenter}, John M. and {Crutcher}, Richard M. and {Fiege}, Jason D. and {Franzmann}, Erica and {Hakobian}, Nicholas S. and {Heiles}, Carl and et al.},
        title = "{Misalignment of Magnetic Fields and Outflows in Protostellar Cores}",
      journal = {\apj},
         year = 2013,
        month = may,
       volume = {768},
       number = {2},
          eid = {159},
        pages = {159},
          doi = {10.1088/0004-637X/768/2/159},
archivePrefix = {arXiv},
       eprint = {1212.0540},
 primaryClass = {astro-ph.SR},
       adsurl = {https://ui.adsabs.harvard.edu/abs/2013ApJ...768..159H}
}

@ARTICLE{sridharan2014,
       author = {{Sridharan}, T.~K. and {Rao}, R. and {Qiu}, K. and {Cortes}, P. and {Li}, H. and {Pillai}, T. and {Patel}, N.~A. and {Zhang}, Q.},
        title = "{Hot Core, Outflows, and Magnetic Fields in W43-MM1 (G30.79 FIR 10)}",
      journal = {\apjl},
         year = 2014,
        month = mar,
       volume = {783},
       number = {2},
          eid = {L31},
        pages = {L31},
          doi = {10.1088/2041-8205/783/2/L31},
archivePrefix = {arXiv},
       eprint = {1312.2561},
 primaryClass = {astro-ph.GA},
       adsurl = {https://ui.adsabs.harvard.edu/abs/2014ApJ...783L..31S}
}

@ARTICLE{lihb2015,
       author = {{Li}, Hua-Bai and {Yuen}, Ka Ho and {Otto}, Frank and {Leung}, Po Kin and {Sridharan}, T.~K. and {Zhang}, Qizhou and {Liu}, Hauyu and {Tang}, Ya-Wen and {Qiu}, Keping},
        title = "{Self-similar fragmentation regulated by magnetic fields in a region forming massive stars}",
      journal = {\nat},
         year = 2015,
        month = apr,
       volume = {520},
       number = {7548},
        pages = {518-521},
          doi = {10.1038/nature14291},
archivePrefix = {arXiv},
       eprint = {1510.07094},
 primaryClass = {astro-ph.GA},
       adsurl = {https://ui.adsabs.harvard.edu/abs/2015Natur.520..518L}
}

@ARTICLE{beuther2020,
       author = {{Beuther}, Henrik and {Soler}, Juan D. and {Linz}, Hendrik and {Henning}, Thomas and {Gieser}, Caroline and {Kuiper}, Rolf and {Vlemmings}, Wouter and {Hennebelle}, Patrick and {Feng}, Siyi and {Smith}, Rowan and {Ahmadi}, Aida},
        title = "{Gravity and Rotation Drag the Magnetic Field in High-mass Star Formation}",
      journal = {\apj},
         year = 2020,
        month = dec,
       volume = {904},
       number = {2},
          eid = {168},
        pages = {168},
          doi = {10.3847/1538-4357/abc019},
archivePrefix = {arXiv},
       eprint = {2010.05825},
 primaryClass = {astro-ph.SR},
       adsurl = {https://ui.adsabs.harvard.edu/abs/2020ApJ...904..168B}
}

@ARTICLE{beuther2024,
       author = {{Beuther}, H. and {Gieser}, C. and {Soler}, J.~D. and {Zhang}, Q. and {Rao}, R. and {Semenov}, D. and {Henning}, Th. and {Pudritz}, R. and {Peters}, T. and {Klaassen}, P. and et al.},
        title = "{Density distributions, magnetic field structures, and fragmentation in high-mass star formation}",
      journal = {\aap},
         year = 2024,
        month = feb,
       volume = {682},
          eid = {A81},
        pages = {A81},
          doi = {10.1051/0004-6361/202348117},
archivePrefix = {arXiv},
       eprint = {2311.11874},
 primaryClass = {astro-ph.GA},
       adsurl = {https://ui.adsabs.harvard.edu/abs/2024A&A...682A..81B}
}

@ARTICLE{zhang2014,
       author = {{Zhang}, Qizhou and {Qiu}, Keping and {Girart}, Josep M. and {Liu}, Hauyu Baobab and {Tang}, Ya-Wen and {Koch}, Patrick M. and {Li}, Zhi-Yun and {Keto}, Eric and {Ho}, Paul T.~P. and {Rao}, Ramprasad and et al.},
        title = "{Magnetic Fields and Massive Star Formation}",
      journal = {\apj},
         year = 2014,
        month = sep,
       volume = {792},
       number = {2},
          eid = {116},
        pages = {116},
          doi = {10.1088/0004-637X/792/2/116},
archivePrefix = {arXiv},
       eprint = {1407.3984},
 primaryClass = {astro-ph.GA},
       adsurl = {https://ui.adsabs.harvard.edu/abs/2014ApJ...792..116Z}
}

@ARTICLE{rengarajan1996,
       author = {{Rengarajan}, T.~N. and {Ho}, P.~T.~P.},
        title = "{Search for Optically Thick H II Regions and Ionized Stellar Wind from Luminous Embedded Infrared Sources}",
      journal = {\apj},
         year = 1996,
        month = jul,
       volume = {465},
        pages = {363},
          doi = {10.1086/177425},
       adsurl = {https://ui.adsabs.harvard.edu/abs/1996ApJ...465..363R}
}

@ARTICLE{vdt2005,
       author = {{van der Tak}, F.~F.~S. and {Menten}, K.~M.},
        title = "{Very compact radio emission from high-mass protostars. II. Dust disks and ionized accretion flows}",
      journal = {\aap},
         year = 2005,
        month = jul,
       volume = {437},
       number = {3},
        pages = {947-956},
          doi = {10.1051/0004-6361:20052872},
archivePrefix = {arXiv},
       eprint = {astro-ph/0504026},
 primaryClass = {astro-ph},
       adsurl = {https://ui.adsabs.harvard.edu/abs/2005A&A...437..947V}
}

@ARTICLE{xu2023,
       author = {{Xu}, Feng-Wei and {Wang}, Ke and {Liu}, Tie and {Goldsmith}, Paul F. and {Zhang}, Qizhou and {Juvela}, Mika and {Liu}, Hong-Li and {Qin}, Sheng-Li and {Li}, Guang-Xing and {Tej}, Anandmayee and {Garay}, Guido and {Bronfman}, Leonardo and {Li}, Shanghuo and {Wu}, Yue-Fang and {G{\'o}mez}, Gilberto C. and {V{\'a}zquez-Semadeni}, Enrique and {Tatematsu}, Ken'ichi and {Ren}, Zhiyuan and {Zhang}, Yong and {Toth}, L. Viktor and {Liu}, Xunchuan and {Yue}, Nannan and {Zhang}, Siju and {Baug}, Tapas and {Issac}, Namitha and {Stutz}, Amelia M. and {Liu}, Meizhu and {Fuller}, Gary A. and {Tang}, Mengyao and {Zhang}, Chao and {Dewangan}, Lokesh and {Lee}, Chang Won and {Zhou}, Jianwen and {Xie}, Jinjin and {Jiao}, Wenyu and {Wang}, Chao and {Liu}, Rong and {Luo}, Qiuyi and {Soam}, Archana and {Eswaraiah}, Chakali},
        title = "{ATOMS: ALMA Three-millimeter Observations of Massive Star-forming regions - XV. Steady accretion from global collapse to core feeding in massive hub-filament system SDC335}",
      journal = {\mnras},
         year = 2023,
        month = apr,
       volume = {520},
       number = {3},
        pages = {3259-3285},
          doi = {10.1093/mnras/stad012},
archivePrefix = {arXiv},
       eprint = {2301.01895},
 primaryClass = {astro-ph.GA},
       adsurl = {https://ui.adsabs.harvard.edu/abs/2023MNRAS.520.3259X}
}

@ARTICLE{navarete2021,
       author = {{Navarete}, F. and {Damineli}, A. and {Steiner}, J.~E. and {Blum}, R.~D.},
        title = "{Principal component analysis tomography in near-infrared integral field spectroscopy of young stellar objects - I. Revisiting the high-mass protostar W33A}",
      journal = {\mnras},
         year = 2021,
        month = may,
       volume = {503},
       number = {1},
        pages = {270-291},
          doi = {10.1093/mnras/stab358},
archivePrefix = {arXiv},
       eprint = {2102.02803},
 primaryClass = {astro-ph.SR},
       adsurl = {https://ui.adsabs.harvard.edu/abs/2021MNRAS.503..270N}
}

@ARTICLE{chen2025,
       author = {{Chen}, Huei-Ru Vivien and {Zhang}, Qizhou and {Ching}, Tao-Chung and {Beuther}, H. and {Wang}, Kuo-Song},
        title = "{Pinched Magnetic Fields in the High-mass Protocluster W3 IRS5}",
      journal = {\apj},
         year = 2025,
        month = oct,
       volume = {992},
       number = {2},
          eid = {199},
        pages = {199},
          doi = {10.3847/1538-4357/adfbec},
archivePrefix = {arXiv},
       eprint = {2508.10128},
 primaryClass = {astro-ph.GA},
       adsurl = {https://ui.adsabs.harvard.edu/abs/2025ApJ...992..199C}
}

@ARTICLE{lada2003,
       author = {{Lada}, Charles J. and {Lada}, Elizabeth A.},
        title = "{Embedded Clusters in Molecular Clouds}",
      journal = {\araa},
         year = 2003,
        month = jan,
       volume = {41},
        pages = {57-115},
          doi = {10.1146/annurev.astro.41.011802.094844},
archivePrefix = {arXiv},
       eprint = {astro-ph/0301540},
 primaryClass = {astro-ph},
       adsurl = {https://ui.adsabs.harvard.edu/abs/2003ARA&A..41...57L}
}

@ARTICLE{zinnecker2007,
       author = {{Zinnecker}, Hans and {Yorke}, Harold W.},
        title = "{Toward Understanding Massive Star Formation}",
      journal = {\araa},
         year = 2007,
        month = sep,
       volume = {45},
       number = {1},
        pages = {481-563},
          doi = {10.1146/annurev.astro.44.051905.092549},
archivePrefix = {arXiv},
       eprint = {0707.1279},
 primaryClass = {astro-ph},
       adsurl = {https://ui.adsabs.harvard.edu/abs/2007ARA&A..45..481Z}
}

@ARTICLE{motte2018,
       author = {{Motte}, Fr{\'e}d{\'e}rique and {Bontemps}, Sylvain and {Louvet}, Fabien},
        title = "{High-Mass Star and Massive Cluster Formation in the Milky Way}",
      journal = {\araa},
         year = 2018,
        month = sep,
       volume = {56},
        pages = {41-82},
          doi = {10.1146/annurev-astro-091916-055235},
archivePrefix = {arXiv},
       eprint = {1706.00118},
 primaryClass = {astro-ph.GA},
       adsurl = {https://ui.adsabs.harvard.edu/abs/2018ARA&A..56...41M}
}

@ARTICLE{beuther2025,
       author = {{Beuther}, H. and {Kuiper}, R. and {Tafalla}, M.},
        title = "{Star Formation from Low to High Mass: A Comparative View}",
      journal = {\araa},
         year = 2025,
        month = aug,
       volume = {63},
       number = {1},
        pages = {1-44},
          doi = {10.1146/annurev-astro-013125-122023},
archivePrefix = {arXiv},
       eprint = {2501.16866},
 primaryClass = {astro-ph.GA},
       adsurl = {https://ui.adsabs.harvard.edu/abs/2025ARA&A..63....1B}
}

@ARTICLE{quarks2,
       author = {{Xu}, Fengwei and {Wang}, Ke and {Liu}, Tie and {Zhu}, Lei and {Garay}, Guido and {Liu}, Xunchuan and {Goldsmith}, Paul and {Zhang}, Qizhou and {Sanhueza}, Patricio and {Qin}, Shengli and {He}, Jinhua and {Juvela}, Mika and {Tej}, Anandmayee and {Liu}, Hongli and {Li}, Shanghuo and {Morii}, Kaho and {Zhang}, Siju and {Zhou}, Jianwen and {Stutz}, Amelia and {Evans}, Neal J. and {Kim}, Kee-Tae and {Liu}, Shengyuan and {Mardones}, Diego and {Li}, Guangxing and {Bronfman}, Leonardo and {Tatematsu}, Ken'ichi and {Lee}, Chang Won and {Lu}, Xing and {Mai}, Xiaofeng and {Jiao}, Sihan and {Chibueze}, James O. and {Su}, Keyun and {T{\'o}th}, Viktor L.},
        title = "{The ALMA-QUARKS Survey. II. The ACA 1.3 mm Continuum Source Catalog and the Assembly of Dense Gas in Massive Star-Forming Clumps}",
      journal = {Research in Astronomy and Astrophysics},
         year = 2024,
        month = jun,
       volume = {24},
       number = {6},
          eid = {065011},
        pages = {065011},
          doi = {10.1088/1674-4527/ad3dc3},
archivePrefix = {arXiv},
       eprint = {2404.02275},
 primaryClass = {astro-ph.GA},
       adsurl = {https://ui.adsabs.harvard.edu/abs/2024RAA....24f5011X}
}

@ARTICLE{hogge2018,
       author = {{Hogge}, Taylor and {Jackson}, James and {Stephens}, Ian and {Whitaker}, Scott and {Foster}, Jonathan and {Camarata}, Matthew and {Anish Roshi}, D. and {Di Francesco}, James and {Longmore}, Steven and {Loughnane}, Robert and {Moore}, Toby and {Rathborne}, Jill and {Sanhueza}, Patricio and {Walsh}, Andrew},
        title = "{The Radio Ammonia Mid-plane Survey (RAMPS) Pilot Survey}",
      journal = {\apjs},
         year = 2018,
        month = aug,
       volume = {237},
       number = {2},
          eid = {27},
        pages = {27},
          doi = {10.3847/1538-4365/aacf94},
archivePrefix = {arXiv},
       eprint = {1808.02533},
 primaryClass = {astro-ph.GA},
       adsurl = {https://ui.adsabs.harvard.edu/abs/2018ApJS..237...27H}
}

@ARTICLE{beltran2019,
       author = {{Beltr{\'a}n}, M.~T. and {Padovani}, M. and {Girart}, J.~M. and {Galli}, D. and {Cesaroni}, R. and {Paladino}, R. and {Anglada}, G. and {Estalella}, R. and {Osorio}, M. and {Rao}, R. and {S{\'a}nchez-Monge}, {\'A}. and {Zhang}, Q.},
        title = "{ALMA resolves the hourglass magnetic field in G31.41+0.31}",
      journal = {\aap},
         year = 2019,
        month = oct,
       volume = {630},
          eid = {A54},
        pages = {A54},
          doi = {10.1051/0004-6361/201935701},
archivePrefix = {arXiv},
       eprint = {1908.01597},
 primaryClass = {astro-ph.GA},
       adsurl = {https://ui.adsabs.harvard.edu/abs/2019A&A...630A..54B}
}

@ARTICLE{magmar7,
       author = {{Hwang}, Jihye and {Sanhueza}, Patricio and {Girart}, Josep Miquel and {Stephens}, Ian W. and {Beltr{\'a}n}, Maria T. and {Law}, Chi Yan and {Zhang}, Qizhou and {Liu}, Junhao and {Cort{\'e}s}, Paulo and {Olguin}, Fernando A. and {Koch}, Patrick M. and {Nakamura}, Fumitaka and {Saha}, Piyali and {Wang}, Jia-Wei and {Xu}, Fengwei and {Beuther}, Henrik and {Morii}, Kaho and {Fern{\'a}ndez L{\'o}pez}, Manuel and {Jiao}, Wenyu and {Kim}, Kee-Tae and {Li}, Shanghuo and {Zapata}, Luis A. and {Kim}, Jongsoo and {Choudhury}, Spandan and {Cheng}, Yu and {Pattle}, Kate and {Eswaraiah}, Chakali and {Sandhyarani}, Panigrahy and {Dewangan}, L.~K. and {Jadhav}, O.~R.},
        title = "{Magnetic Fields in Massive Star-forming Regions (MagMaR). VI. Magnetic Field Dragging in the Filamentary High-mass Star-forming Region G35.20--0.74N due to Gravity}",
      journal = {arXiv e-prints},
         year = 2025,
        month = oct,
          eid = {arXiv:2510.25078},
        pages = {arXiv:2510.25078},
          doi = {10.48550/arXiv.2510.25078},
archivePrefix = {arXiv},
       eprint = {2510.25078},
 primaryClass = {astro-ph.GA},
       adsurl = {https://ui.adsabs.harvard.edu/abs/2025arXiv251025078H}
}

@ARTICLE{urquhart2022,
       author = {{Urquhart}, J.~S. and {Wells}, M.~R.~A. and {Pillai}, T. and {Leurini}, S. and {Giannetti}, A. and {Moore}, T.~J.~T. and {Thompson}, M.~A. and {Figura}, C. and {Colombo}, D. and {Yang}, A.~Y. and {K{\"o}nig}, C. and {Wyrowski}, F. and {Menten}, K.~M. and {Rigby}, A.~J. and {Eden}, D.~J. and {Ragan}, S.~E.},
        title = "{ATLASGAL - evolutionary trends in high-mass star formation}",
      journal = {\mnras},
         year = 2022,
        month = mar,
       volume = {510},
       number = {3},
        pages = {3389-3407},
          doi = {10.1093/mnras/stab3511},
archivePrefix = {arXiv},
       eprint = {2111.12816},
 primaryClass = {astro-ph.GA},
       adsurl = {https://ui.adsabs.harvard.edu/abs/2022MNRAS.510.3389U}
}

@ARTICLE{mezger1967,
       author = {{Mezger}, P.~G. and {Henderson}, A.~P.},
        title = "{Galactic H II Regions. I. Observations of Their Continuum Radiation at the Frequency 5 GHz}",
      journal = {\apj},
         year = 1967,
        month = feb,
       volume = {147},
        pages = {471},
          doi = {10.1086/149030},
       adsurl = {https://ui.adsabs.harvard.edu/abs/1967ApJ...147..471M}
}

@ARTICLE{sternberg2003,
       author = {{Sternberg}, Amiel and {Hoffmann}, Tadziu L. and {Pauldrach}, A.~W.~A.},
        title = "{Ionizing Photon Emission Rates from O- and Early B-Type Stars and Clusters}",
      journal = {\apj},
         year = 2003,
        month = dec,
       volume = {599},
       number = {2},
        pages = {1333-1343},
          doi = {10.1086/379506},
archivePrefix = {arXiv},
       eprint = {astro-ph/0312232},
 primaryClass = {astro-ph},
       adsurl = {https://ui.adsabs.harvard.edu/abs/2003ApJ...599.1333S}
}

@ARTICLE{loren1984,
       author = {{Loren}, R.~B. and {Mundy}, L.~G.},
        title = "{The methyl cyanide hot and warm cores in Orion : statistical equilibrium excitation models of a symmetric-top molecule.}",
      journal = {\apj},
         year = 1984,
        month = nov,
       volume = {286},
        pages = {232-251},
          doi = {10.1086/162591},
       adsurl = {https://ui.adsabs.harvard.edu/abs/1984ApJ...286..232L}
}

@ARTICLE{araya2005,
       author = {{Araya}, Esteban and {Hofner}, Peter and {Kurtz}, Stan and {Bronfman}, Leonardo and {DeDeo}, Simon},
        title = "{CH$_{3}$CN Observations toward Southern Massive Star-forming Regions}",
      journal = {\apjs},
         year = 2005,
        month = apr,
       volume = {157},
       number = {2},
        pages = {279-301},
          doi = {10.1086/427187},
       adsurl = {https://ui.adsabs.harvard.edu/abs/2005ApJS..157..279A}
}

@ARTICLE{chen2006,
       author = {{Chen}, Huei-Ru and {Welch}, William J. and {Wilner}, David J. and {Sutton}, Edmund C.},
        title = "{A High-Mass Protobinary System in the Hot Core W3(H$_{2}$O)}",
      journal = {\apj},
         year = 2006,
        month = mar,
       volume = {639},
       number = {2},
        pages = {975-990},
          doi = {10.1086/499395},
archivePrefix = {arXiv},
       eprint = {astro-ph/0511294},
 primaryClass = {astro-ph},
       adsurl = {https://ui.adsabs.harvard.edu/abs/2006ApJ...639..975C}
}

@ARTICLE{beltran2016,
       author = {{Beltr{\'a}n}, M.~T. and {de Wit}, W.~J.},
        title = "{Accretion disks in luminous young stellar objects}",
      journal = {\aapr},
         year = 2016,
        month = jan,
       volume = {24},
          eid = {6},
        pages = {6},
          doi = {10.1007/s00159-015-0089-z},
archivePrefix = {arXiv},
       eprint = {1509.08335},
 primaryClass = {astro-ph.GA},
       adsurl = {https://ui.adsabs.harvard.edu/abs/2016A&ARv..24....6B}
}

@ARTICLE{girart1999,
       author = {{Girart}, Jos{\'e} M. and {Crutcher}, Richard M. and {Rao}, Ramprasad},
        title = "{Detection of Polarized CO Emission from the Molecular Outflow in NGC 1333 IRAS 4A}",
      journal = {\apjl},
         year = 1999,
        month = nov,
       volume = {525},
       number = {2},
        pages = {L109-L112},
          doi = {10.1086/312345},
       adsurl = {https://ui.adsabs.harvard.edu/abs/1999ApJ...525L.109G}
}

@ARTICLE{hwang2022,
       author = {{Hwang}, Jihye and {Kim}, Jongsoo and {Pattle}, Kate and {Lee}, Chang Won and {Koch}, Patrick M. and {Johnstone}, Doug and {Tomisaka}, Kohji and {Whitworth}, Anthony and {Furuya}, Ray S. and {Kang}, Ji-hyun and {Lyo}, A.-Ran and {Chung}, Eun Jung and {Arzoumanian}, Doris and {Park}, Geumsook and {Kwon}, Woojin and {Kim}, Shinyoung and {Tamura}, Motohide and {Kwon}, Jungmi and {Soam}, Archana and {Han}, Ilseung and {Hoang}, Thiem and {Kim}, Kyoung Hee and {Onaka}, Takashi and {Eswaraiah}, Chakali and {Ward-Thompson}, Derek and {Liu}, Hong-Li and {Tang}, Xindi and {Chen}, Wen Ping and {Matsumura}, Masafumi and {Hoang}, Thuong Duc and {Chen}, Zhiwei and {Le Gouellec}, Valentin J.~M. and {Kirchschlager}, Florian and {Poidevin}, Fr{\'e}d{\'e}rick and {Bastien}, Pierre and {Qiu}, Keping and {Hasegawa}, Tetsuo and {Lai}, Shih-Ping and {Byun}, Do-Young and {Cho}, Jungyeon and {Choi}, Minho and {Choi}, Youngwoo and {Choi}, Yunhee and {Jeong}, Il-Gyo and {Kang}, Miju and {Kim}, Hyosung and {Kim}, Kee-Tae and {Lee}, Jeong-Eun and {Lee}, Sang-Sung and {Lee}, Yong-Hee and {Lee}, Hyeseung and {Kim}, Mi-Ryang and {Yoo}, Hyunju and {Yun}, Hyeong-Sik and {Chen}, Mike and {Di Francesco}, James and {Fiege}, Jason and {Fissel}, Laura M. and {Franzmann}, Erica and {Houde}, Martin and {Lacaille}, Kevin and {Matthews}, Brenda and {Sadavoy}, Sarah and {Moriarty-Schieven}, Gerald and {Tahani}, Mehrnoosh and {Ching}, Tao-Chung and {Dai}, Y. Sophia and {Duan}, Yan and {Gu}, Qilao and {Law}, Chi-Yan and {Li}, Dalei and {Li}, Di and {Li}, Guangxing and {Li}, Hua-bai and {Liu}, Tie and {Lu}, Xing and {Qian}, Lei and {Wang}, Hongchi and {Wu}, Jintai and {Xie}, Jinjin and {Yuan}, Jinghua and {Zhang}, Chuan-Peng and {Zhang}, Guoyin and {Zhang}, Yapeng and {Zhou}, Jianjun and {Zhu}, Lei and {Berry}, David and {Friberg}, Per and {Graves}, Sarah and {Liu}, Junhao and {Mairs}, Steve and {Parsons}, Harriet and {Rawlings}, Mark and {Doi}, Yasuo and {Hayashi}, Saeko and {Hull}, Charles L.~H. and {Inoue}, Tsuyoshi and {Inutsuka}, Shu-ichiro and {Iwasaki}, Kazunari and {Kataoka}, Akimasa and {Kawabata}, Koji and {Kim}, Gwanjeong and {Kobayashi}, Masato I.~N. and {Nagata}, Tetsuya and {Nakamura}, Fumitaka and {Nakanishi}, Hiroyuki and {Pyo}, Tae-Soo and {Saito}, Hiro and {Seta}, Masumichi and {Shimajiri}, Yoshito and {Shinnaga}, Hiroko and {Tsukamoto}, Yusuke and {Zenko}, Tetsuya and {Chen}, Huei-Ru Vivien and {Duan}, Hao-Yuan and {Fanciullo}, Lapo and {Kemper}, Francisca and {Lee}, Chin-Fei and {Lin}, Sheng-Jun and {Liu}, Sheng-Yuan and {Ohashi}, Nagayoshi and {Rao}, Ramprasad and {Tang}, Ya-Wen and {Wang}, Jia-Wei and {Yang}, Meng-Zhe and {Yen}, Hsi-Wei and {Bourke}, Tyler L. and {Chrysostomou}, Antonio and {Debattista}, Victor and {Eden}, David and {Eyres}, Stewart and {Falle}, Sam and {Fuller}, Gary and {Gledhill}, Tim and {Greaves}, Jane and {Griffin}, Matt and {Hatchell}, Jennifer and {Karoly}, Janik and {Kirk}, Jason and {K{\"o}nyves}, Vera and {Longmore}, Steven and {van Loo}, Sven and {de Looze}, Ilse and {Peretto}, Nicolas and {Priestley}, Felix and {Rawlings}, Jonathan and {Retter}, Brendan and {Richer}, John and {Rigby}, Andrew and {Savini}, Giorgio and {Scaife}, Anna and {Viti}, Serena and {Diep}, Pham Ngoc and {Ngoc}, Nguyen Bich and {Tram}, Le Ngoc and {Andr{\'e}}, Philippe and {Coud{\'e}}, Simon and {Dowell}, C. Darren and {Friesen}, Rachel and {Robitaille}, Jean-Fran{\'c}ois},
        title = "{The JCMT BISTRO Survey: A Spiral Magnetic Field in a Hub-filament Structure, Monoceros R2}",
      journal = {\apj},
         year = 2022,
        month = dec,
       volume = {941},
       number = {1},
          eid = {51},
        pages = {51},
          doi = {10.3847/1538-4357/ac99e0},
archivePrefix = {arXiv},
       eprint = {2210.05937},
 primaryClass = {astro-ph.GA},
       adsurl = {https://ui.adsabs.harvard.edu/abs/2022ApJ...941...51H}
}

@ARTICLE{crutcher2019,
       author = {{Crutcher}, Richard M. and {Kemball}, Athol J.},
        title = "{Review of Zeeman Effect Observations of Regions of Star Formation K Zeeman Effect, Magnetic Fields, Star formation, Masers, Molecular clouds}",
      journal = {Frontiers in Astronomy and Space Sciences},
         year = 2019,
        month = oct,
       volume = {6},
          eid = {66},
        pages = {66},
          doi = {10.3389/fspas.2019.00066},
archivePrefix = {arXiv},
       eprint = {1911.06210},
 primaryClass = {astro-ph.GA},
       adsurl = {https://ui.adsabs.harvard.edu/abs/2019FrASS...6...66C}
}

@ARTICLE{koch2012b,
       author = {{Koch}, Patrick M. and {Tang}, Ya-Wen and {Ho}, Paul T.~P.},
        title = "{Quantifying the Significance of the Magnetic Field from Large-scale Cloud to Collapsing Core: Self-similarity, Mass-to-flux Ratio, and Star Formation Efficiency}",
      journal = {\apj},
         year = 2012,
        month = mar,
       volume = {747},
       number = {1},
          eid = {80},
        pages = {80},
          doi = {10.1088/0004-637X/747/1/80},
archivePrefix = {arXiv},
       eprint = {1201.4313},
 primaryClass = {astro-ph.GA},
       adsurl = {https://ui.adsabs.harvard.edu/abs/2012ApJ...747...80K}
}

@ARTICLE{koch2014,
       author = {{Koch}, Patrick M. and {Tang}, Ya-Wen and {Ho}, Paul T.~P. and {Zhang}, Qizhou and {Girart}, Josep M. and {Chen}, Huei-Ru Vivien and {Frau}, Pau and {Li}, Hua-Bai and {Li}, Zhi-Yun and {Liu}, Hau-Yu Baobab and {Padovani}, Marco and {Qiu}, Keping and {Yen}, Hsi-Wei and {Chen}, How-Huan and {Ching}, Tao-Chung and {Lai}, Shih-Ping and {Rao}, Ramprasad},
        title = "{The Importance of the Magnetic Field from an SMA-CSO-combined Sample of Star-forming Regions}",
      journal = {\apj},
         year = 2014,
        month = dec,
       volume = {797},
       number = {2},
          eid = {99},
        pages = {99},
          doi = {10.1088/0004-637X/797/2/99},
archivePrefix = {arXiv},
       eprint = {1411.3830},
 primaryClass = {astro-ph.GA},
       adsurl = {https://ui.adsabs.harvard.edu/abs/2014ApJ...797...99K}
}

@ARTICLE{tang2009,
       author = {{Tang}, Ya-Wen and {Ho}, Paul T.~P. and {Koch}, Patrick M. and {Girart}, Josep M. and {Lai}, Shih-Ping and {Rao}, Ramprasad},
        title = "{Evolution of Magnetic Fields in High-Mass Star Formation: Linking Field Geometry and Collapse for the W51 e2/e8 Cores}",
      journal = {\apj},
         year = 2009,
        month = jul,
       volume = {700},
       number = {1},
        pages = {251-261},
          doi = {10.1088/0004-637X/700/1/251},
archivePrefix = {arXiv},
       eprint = {0905.1996},
 primaryClass = {astro-ph.SR},
       adsurl = {https://ui.adsabs.harvard.edu/abs/2009ApJ...700..251T}
}

@ARTICLE{yen2020,
       author = {{Yen}, Hsi-Wei and {Zhao}, Bo and {Koch}, Patrick and {Krasnopolsky}, Ruben and {Li}, Zhi-Yun and {Ohashi}, Nagayoshi and {Shang}, Hsien and {Takakuwa}, Shigehisa and {Tang}, Ya-Wen},
        title = "{Transition from Ordered Pinched to Warped Magnetic Field on a 100 au Scale in the Class 0 Protostar B335}",
      journal = {\apj},
         year = 2020,
        month = apr,
       volume = {893},
       number = {1},
          eid = {54},
        pages = {54},
          doi = {10.3847/1538-4357/ab7eb3},
archivePrefix = {arXiv},
       eprint = {2003.04502},
 primaryClass = {astro-ph.SR},
       adsurl = {https://ui.adsabs.harvard.edu/abs/2020ApJ...893...54Y}
}

@ARTICLE{kirk2015,
       author = {{Kirk}, Helen and {Klassen}, Mikhail and {Pudritz}, Ralph and {Pillsworth}, Samantha},
        title = "{The Role of Turbulence and Magnetic Fields in Simulated Filamentary Structure}",
      journal = {\apj},
         year = 2015,
        month = apr,
       volume = {802},
       number = {2},
          eid = {75},
        pages = {75},
          doi = {10.1088/0004-637X/802/2/75},
archivePrefix = {arXiv},
       eprint = {1501.05999},
 primaryClass = {astro-ph.GA},
       adsurl = {https://ui.adsabs.harvard.edu/abs/2015ApJ...802...75K}
}

@ARTICLE{bertoldi1992,
       author = {{Bertoldi}, Frank and {McKee}, Christopher F.},
        title = "{Pressure-confined Clumps in Magnetized Molecular Clouds}",
      journal = {\apj},
         year = 1992,
        month = aug,
       volume = {395},
        pages = {140},
          doi = {10.1086/171638},
       adsurl = {https://ui.adsabs.harvard.edu/abs/1992ApJ...395..140B}
}

@ARTICLE{yang2026,
       author = {{Yang}, Kai and {Zhang}, Yichen and {Tanaka}, Kei E.~I. and {Liu}, Tie and {Sakai}, Nami and {Zhang}, Ziwei E. and {Lee}, Gyuho and {Kim}, Kee-Tae and {Ginsburg}, Adam and {Wang}, Lile and {Wang}, Yao and {Tang}, Yongzhi and {Cheng}, Yu and {Liu}, Hongli and {Jiao}, Wenyu and {Xu}, Fengwei and {Liu}, Xunchuan and {Mai}, Xiaofeng and {Yang}, Dongting},
        title = "{HOTDISK. Finding Massive Protostellar Disks with Water and Refractory Molecular Species}",
      journal = {arXiv e-prints},
         year = 2026,
        month = apr,
          eid = {arXiv:2604.19366},
        pages = {arXiv:2604.19366},
          doi = {10.48550/arXiv.2604.19366},
archivePrefix = {arXiv},
       eprint = {2604.19366},
 primaryClass = {astro-ph.GA},
       adsurl = {https://ui.adsabs.harvard.edu/abs/2026arXiv260419366Y}
}

@ARTICLE{liu2026,
       author = {{Liu}, Junhao and {Sanhueza}, Patricio and {Saha}, Piyali and {Morii}, Kaho and {Girart}, Josep Miquel and {Zhang}, Qizhou and {Nakamura}, Fumitaka and {Cortes}, Paulo C. and {Valdivia}, Valeska and {Commercon}, Benoit and {Koch}, Patrick M. and {Pattle}, Kate and {Lu}, Xing and {Karoly}, Janik and {Fernandez-Lopez}, Manuel and {Stephens}, Ian W. and {Chen}, Huei-Ru Vivien and {Law}, Chi-Yan and {Qiu}, Keping and {Li}, Shanghuo and {Beuther}, Henrik and {Chung}, Eun Jung and {Wang}, Jia-Wei and {Olguin}, Fernando A. and {Cheng}, Yu and {Hwang}, Jihye and {Panigrahy}, Sandhyarani and {Eswaraiah}, Chakali and {Beltran}, Maria T. and {Luo}, Qiuyi and {Choudhury}, Spandan and {Kang}, Ji-hyun and {Jiao}, Wenyu and {Zapata}, Luis A. and {Lyo}, A.-Ran},
        title = "{When turbulence beats magnetism: origin of massive star cluster seeds}",
      journal = {arXiv e-prints},
         year = 2026,
        month = mar,
          eid = {arXiv:2603.17254},
        pages = {arXiv:2603.17254},
          doi = {10.48550/arXiv.2603.17254},
archivePrefix = {arXiv},
       eprint = {2603.17254},
 primaryClass = {astro-ph.GA},
       adsurl = {https://ui.adsabs.harvard.edu/abs/2026arXiv260317254L}
}

@ARTICLE{giannetti2017,
       author = {{Giannetti}, A. and {Leurini}, S. and {K{\"o}nig}, C. and {Urquhart}, J.~S. and {Pillai}, T. and {Brand}, J. and {Kauffmann}, J. and {Wyrowski}, F. and {Menten}, K.~M.},
        title = "{Galactocentric variation of the gas-to-dust ratio and its relation with metallicity}",
      journal = {\aap},
         year = 2017,
        month = oct,
       volume = {606},
          eid = {L12},
        pages = {L12},
          doi = {10.1051/0004-6361/201731728},
archivePrefix = {arXiv},
       eprint = {1710.05721},
 primaryClass = {astro-ph.GA},
       adsurl = {https://ui.adsabs.harvard.edu/abs/2017A&A...606L..12G}
}

\begin{appendix}

\section{Velocity structures} \label{app:pv}

\begin{figure}[!h]
    \centering
    \includegraphics[width=\linewidth]{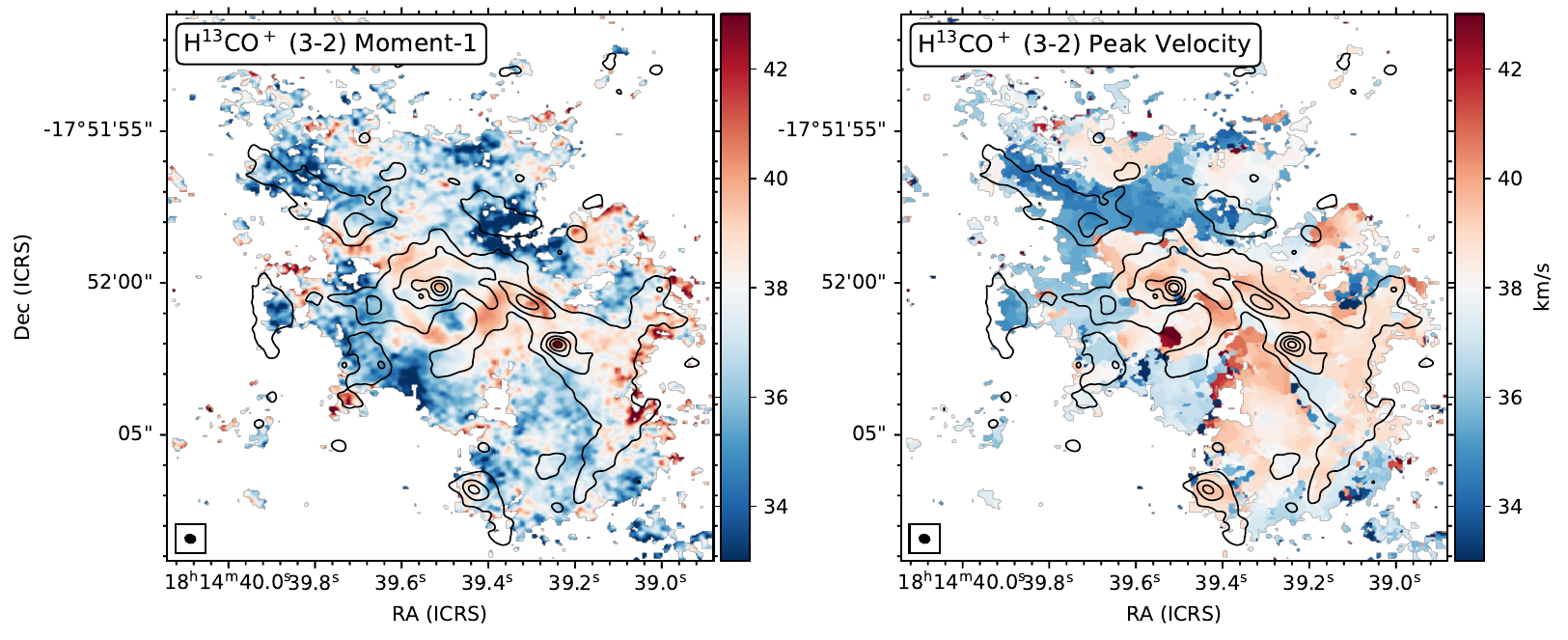}
    \caption{The moment-1 (left) and peak velocity (right) maps.}
    \label{fig:velocity}
\end{figure}

\begin{figure}[!h]
    \centering
    \includegraphics[width=\linewidth]{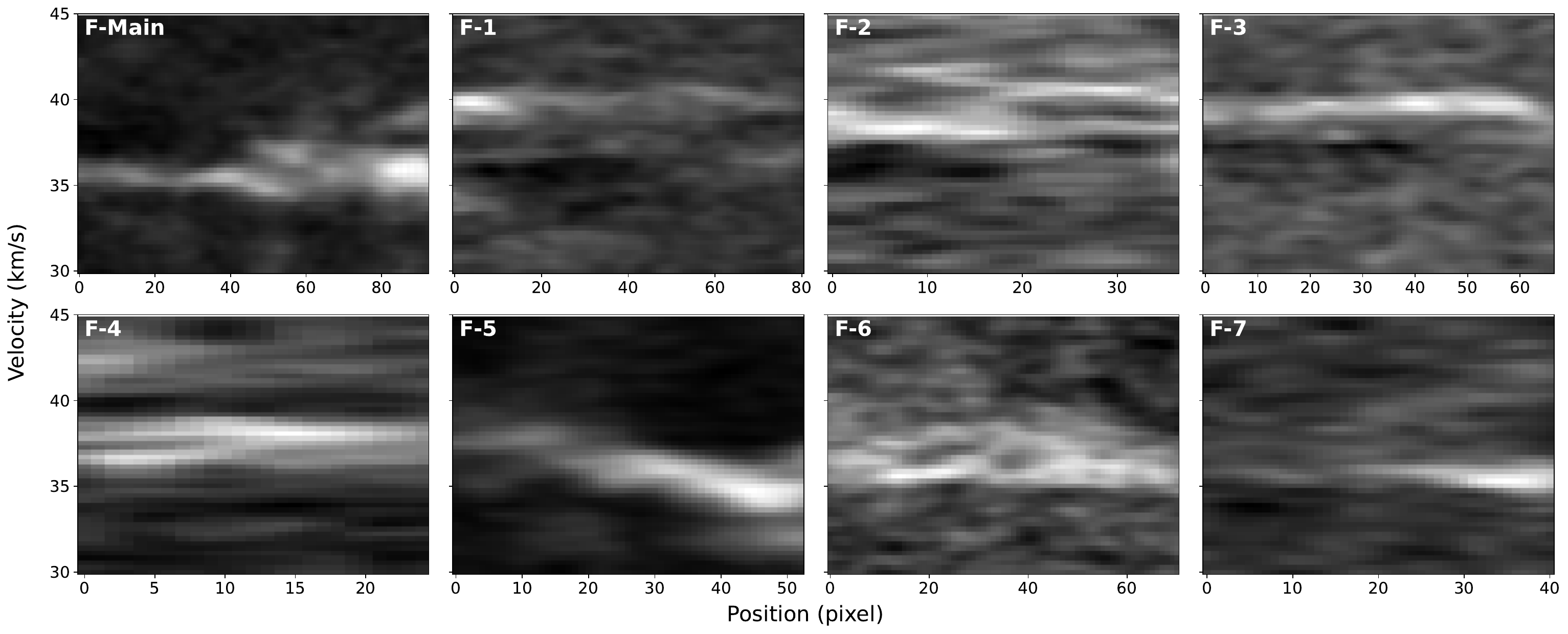}
    \caption{The position-velocity diagram of sliced along the filament spines.}
    \label{fig:pv}
\end{figure}

We present the moment-1 and peak velocity maps of the H$^{13}$CO$^+$~(3-2) spectral data cube to show the velocity structures in Fig.~\ref{fig:velocity}. To demonstrate the velocity coherence of the eight \textit{getsf} filaments, their position-velocity diagrams are presented in Fig~\ref{fig:pv}. 

\section{Polarization properties} \label{app:pol}

\begin{figure}
    \includegraphics[width=\linewidth]{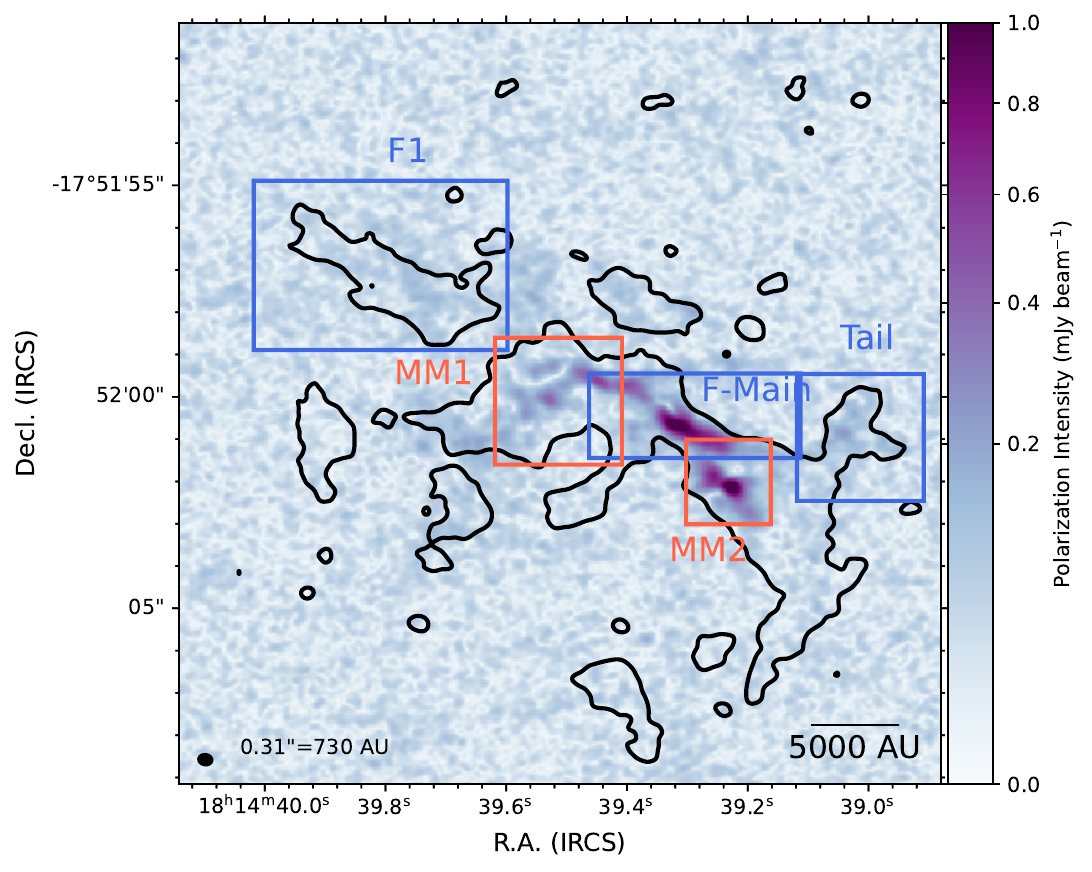}
    \includegraphics[width=\linewidth]{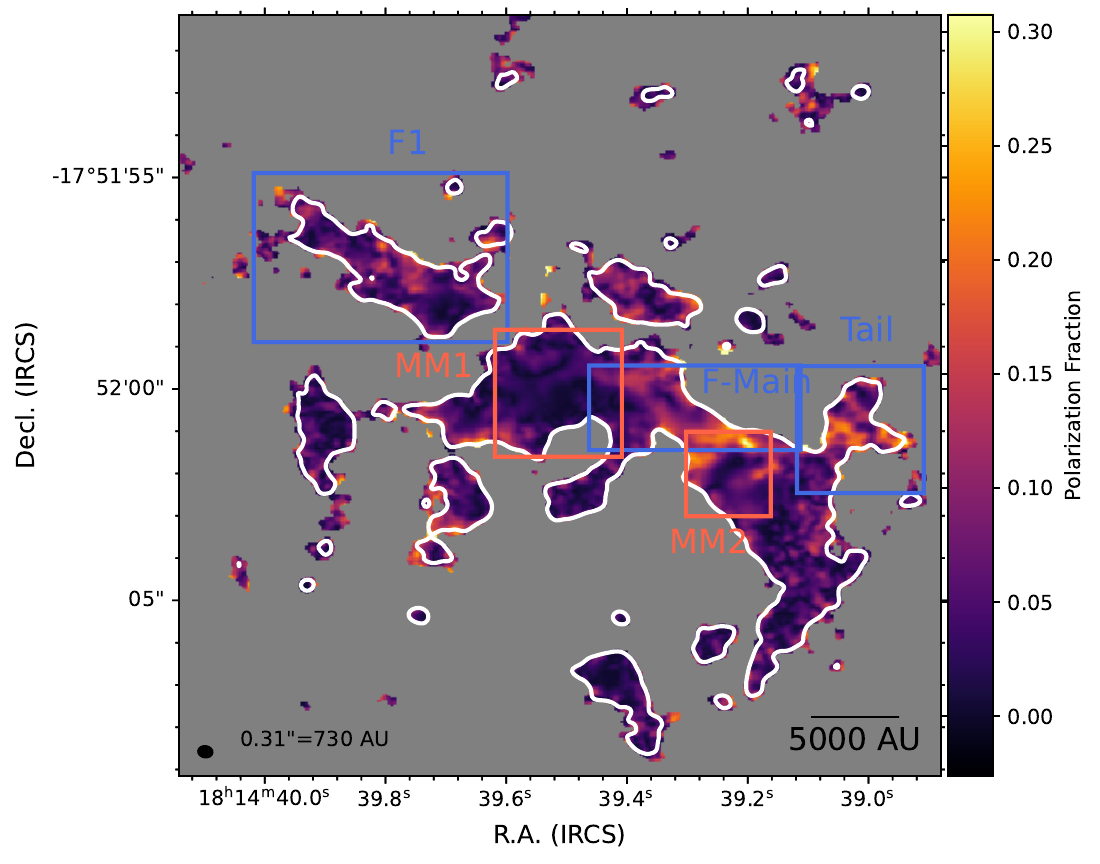}
    \caption{\textit{Upper}: The debiased polarization intensity $\hat{P_I}$. \textit{Lower}: The debiased polarization fraction $\hat{f}_{P}$. The black and white contours correspond to the 1.2~mm Stokes I continuum threshold of $4\sigma_I=0.52~\mjybeam$. The same rectangular regions are marked as Fig.~\ref{fig:bfield}. The synthesized beam and scale bar are shown in the bottom.}
    \label{fig:pol}
\end{figure}

Based on the debiased calculations in Eqs.~(\ref{eq:debiasp}--\ref{eq:debiasfp}), we present the polarization intensity $\hat{P_I}$ and polarization fraction $\hat{f}_{P}$ maps in Fig.~\ref{fig:pol}. 

\section{Methyl cyanide gas temperature} \label{app:ch3cn}

Methyl cyanide (CH$_3$CN) is a symmetric–top rotor, whose rotational spectrum is organized into $K$–ladders, where $K$ is the projection of the angular momentum onto the symmetry axis. For electric–dipole transitions the selection rule is $\Delta K=0$, so radiative coupling between different $K$–ladders is forbidden and population exchange among ladders occurs primarily via collisions. Consequently, the relative populations of the $K$ ladders within a given $J\!\rightarrow\!J\!-\!1$ transition respond sensitively to the kinetic temperature, making the relative line brightness among the $K$ components a widely used thermometer for dense interstellar gas \citep[e.g.]{loren1984, araya2005, chen2006}. In the frequency range of 256.69--257.53~GHz, CH$_3$CN has a set of closely spaced $K=0, 1, 3, \dots, 13$ components of the $J=14\rightarrow13$ transition with their upper level energies covering a large range of $93 - 1296 \; \mathrm{K}$. Their frequency, upper energy level and dipole-moment are queried from Cologne Database for Molecular Spectroscopy \citep[CDMS;][]{muller2001} by XCLASS toolbox \footnote{\url{https://xclass.astro.uni-koeln.de/}} \citep[eXtended CASA Line Analysis Software Suite][]{moller2017, moller2023} and listed in Table~\ref{tab:ch3cn_kladders}. 
Chemically, the abundance of CH$_3$CN is further enhanced in hot cores owing to thermal evaporation of grain mantles \citep{tielens1997, bisschop2007} and/or warm gas–phase synthesis initiated by nitrogen–bearing precursors \citep{charnley1992, charnley1995, millar1997}. This combination of temperature sensitivity and elevated abundance yields bright, closely spaced-in-frequency lines that are well suited to spectral modeling in hot molecular cores \citep[e.g.,][]{kalenskii2000}. 

\begin{figure*}
\centering
\includegraphics[height=0.45\linewidth]{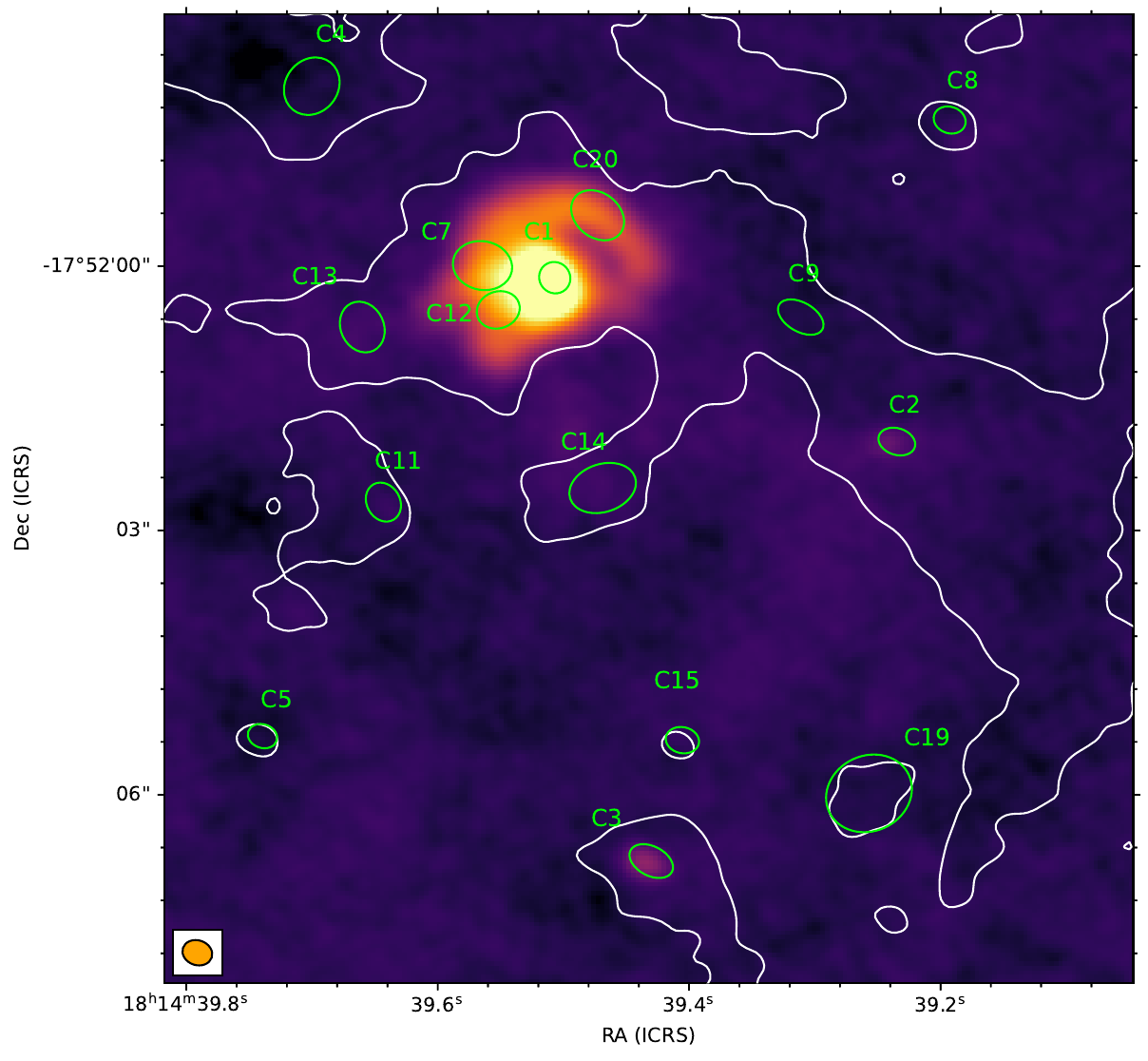}
\includegraphics[height=0.45\linewidth]{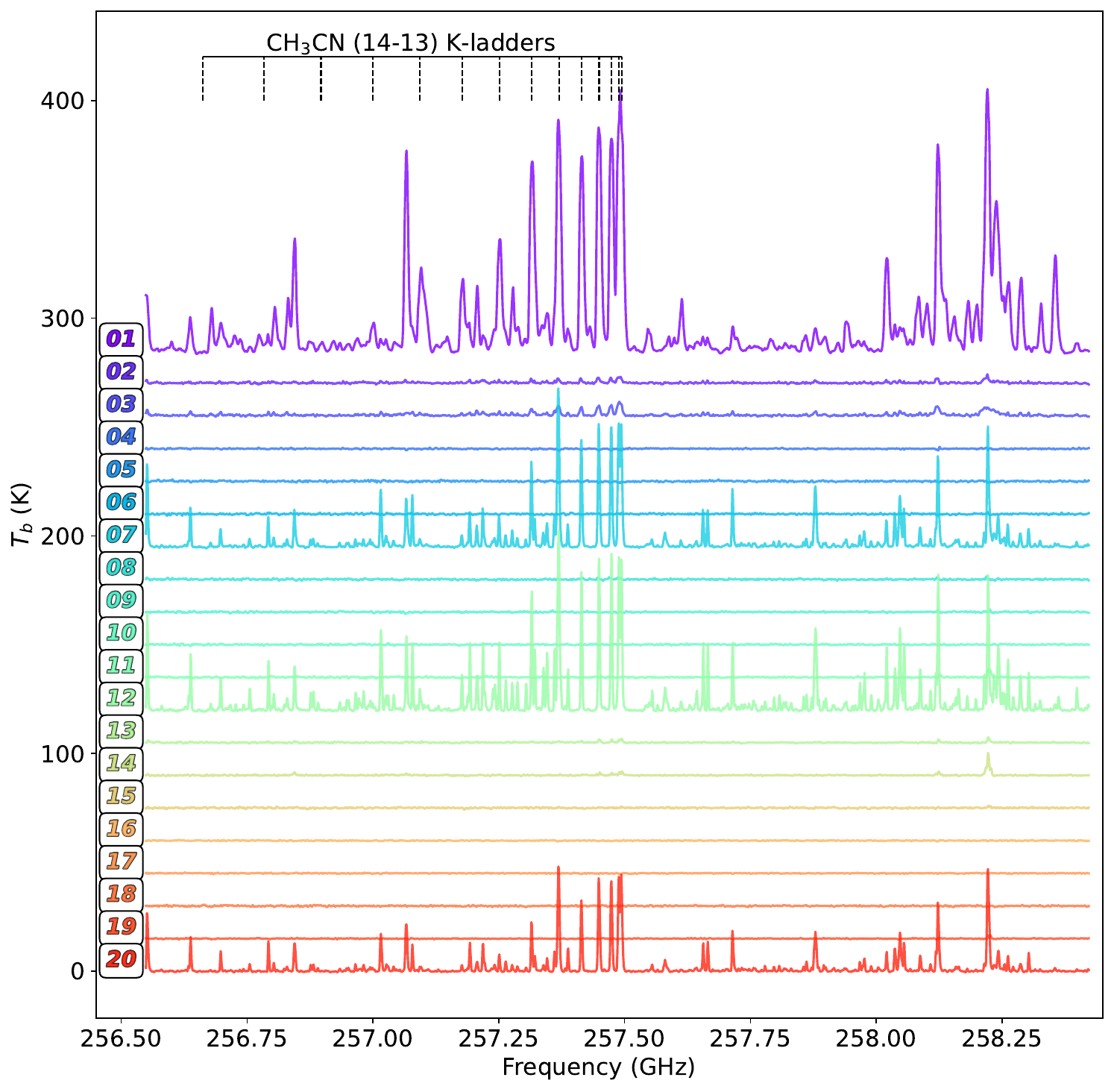}
\includegraphics[width=0.95\linewidth]{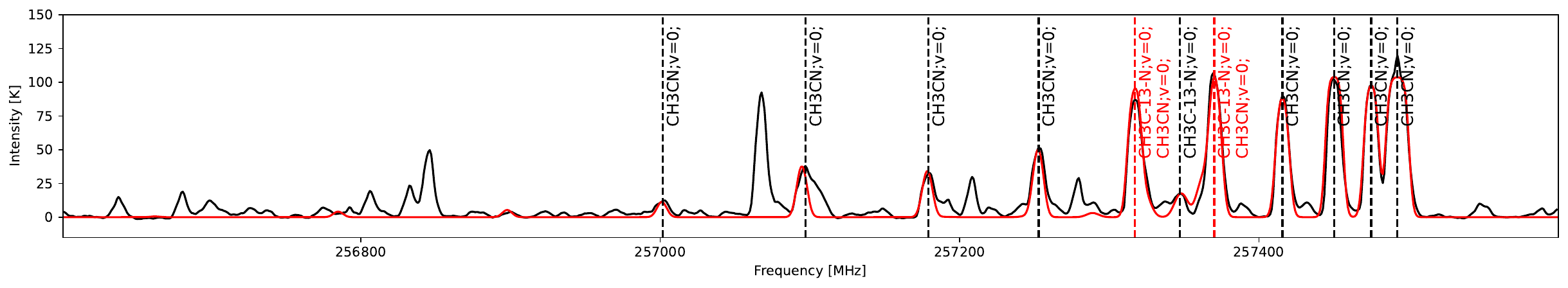}
\caption{\textit{Top left}: The integrated intensity map of CH$_3$CN 14(3)-13(3). The overlaid white contour shows the ALMA 1.2~mm continuum emission at $0.52~\mjybeam$. The green ellipses with labels respectively show the identified cores and their IDs. \textit{Top right}: The spectral window 256.55--258.42 GHz for the 20 dense cores, labeled by color scale. \textit{Bottom}: The observed (black) and modeled (red) spectra of core C1. CH$_3$CN and its isotope CH$_3\!^{13}$CN are shown in a 1-GHz spectral window. The gas excitation temperature is 274~K in this case. The red specie marker indicates substantial line blending. 
\label{fig:ch3cn}}
\end{figure*}

To trace the spatial distribution of CH$_3$CN and its correlation with dense cores, we investigated the $K=3$ spectra because the line is 1) not affected by line blending, and 2) intrinsically stronger  because of the larger statistical weight resulted from the  three-fold symmetry of the methyl group.
The integrated intensity map is shown in the top left panel of Fig.~\ref{fig:ch3cn} overlaid with dense cores. We also extract core-averaged spectra on the top right. Combining two, the cores 1, 3, 7, 12, and 20 show the detection with CH$_3$CN 14(3)-13(3) line emission and good spatial correlation with dust continuum. The dust temperature of these cores can be estimated through gas temperature diagnosis. 

Assuming local thermal equilibrium (LTE), we modeled the spectra of CH$_3$CN and its isotope CH$_3\!^{13}$CN within the frequency range of 256.6--257.6~GHz by \textit{spectuner} package\footnote{\url{https://spectuner.readthedocs.io/en/latest/index.html}}, an automated spectral line analysis \citep{qiu2025b}. We show an example of line fitting result in the bottom panel of Fig.~\ref{fig:ch3cn}. The obtained CH$_3$CN excitation temperature is assumed to be the dust temperature of the dense cores C1, C3, C7, C12, and C20, as listed in Table~\ref{tab:coreinfo}. 

\begin{table}
\caption{Transitions of the CH$_3$CN $K$-ladders.}
\label{tab:ch3cn_kladders}
\centering
\small
\begin{tabular}{lccc}
\hline\hline
QNs & Frequency (GHz) & $E_u$ (K) & $s_{ij}\mu^2$ (D$^2$) \\
\hline
14(13)--13(13)      & 256.69456 & 1296.2 & 59.3181 \\
14(12)--13($-$12)$^{*}$ & 256.81716 & 1118.7 & 114.2685 \\
14($-$12)--13(12)$^{*}$ & 256.81716 & 1118.7 & 114.2685 \\
14(11)--13(11)      & 256.93014 & 955.2  & 164.7974 \\
14(10)--13(10)      & 257.03344 & 805.8  & 210.9168 \\
14(9)--13($-$9)$^{*}$   & 257.12704 & 670.5  & 252.7305 \\
14($-$9)--13(9)$^{*}$   & 257.12704 & 670.5  & 252.7305 \\
14(8)--13(8)        & 257.21088 & 549.4  & 290.0232 \\
14(7)--13(7)        & 257.28494 & 442.4  & 322.9805 \\
14(6)--13($-$6)$^{*}$   & 257.34918 & 349.7  & 351.5789 \\
14($-$6)--13(6)$^{*}$   & 257.34918 & 349.7  & 351.5789 \\
14(5)--13(5)        & 257.40358 & 271.2  & 375.7534 \\
14(4)--13(4)        & 257.44813 & 207.0  & 395.4961 \\
14(3)--13($-$3)$^{*}$   & 257.48279 & 157.0  & 410.9243 \\
14($-$3)--13(3)$^{*}$   & 257.48279 & 157.0  & 410.9243 \\
14(2)--13(2)        & 257.50756 & 121.3  & 421.8698 \\
14(1)--13(1)        & 257.52243 & 99.8   & 428.5521 \\
14(0)--13(0)        & 257.52738 & 92.7   & 430.6699 \\
\hline
\end{tabular}
\tablefoot{
QNs: quantum numbers resolved to the (J,K) level, with hyperfine structure ignored.  
A superscript ${}^{*}$ denotes A nuclear-spin symmetry, consisting of two transitions ($A^+$/$A^-$) with identical frequencies and dipole moments; all other transitions correspond to E symmetry.
}
\end{table}

\section{Parabola curve fitting}
\label{app:parabola-fit}

The image plane in the Cartesian coordinates \((x,y)\), with position angle measured anticlockwise from \(+x\). Let the parabola symmetry axis has a position angle \(\alpha\). Define the mode unit segments
\begin{equation}
\mathbf{a} \equiv 
\begin{bmatrix}\cos\alpha\\ \sin\alpha\end{bmatrix},
\quad
\mathbf{p} \equiv 
\begin{bmatrix}-\sin\alpha\\ \cos\alpha\end{bmatrix}.
\end{equation}
So the rotation matrix writes,
\begin{equation}
\mathbf{B} = [\,\mathbf{p}\ \ \mathbf{a}\,] =
\begin{bmatrix}
-\sin\alpha & \ \cos\alpha\\
 \ \cos\alpha & \ \sin\alpha
\end{bmatrix}.
\end{equation}
Let the symmetric center be \(\mathbf{r}_0 \equiv (x_0,y_0)^\top\). For any image-plane position \(\mathbf{r}=(x,y)^\top\), 
\begin{equation}
\begin{bmatrix}X\\Y\end{bmatrix}
= \mathbf{B}^{\!\top}\,(\mathbf{r}-\mathbf{r}_0)
= \begin{bmatrix}\mathbf{p}^{\!\top}\\ \mathbf{a}^{\!\top}\end{bmatrix}(\mathbf{r}-\mathbf{r}_0),
\label{eq:img2model}
\end{equation}
while the inverse transformation writes,
\begin{equation}
\mathbf{r}
= \mathbf{r}_0 + \mathbf{B}
\begin{bmatrix}X\\Y\end{bmatrix}
= \mathbf{r}_0 + X\,\mathbf{p} + Y\,\mathbf{a}.
\label{eq:model2img}
\end{equation}

Following \citet{girart2006, qiu2014}, pinched B-field lines can be modeled by self-similar parabolic curves
\begin{equation}
Y \;=\; g\,C\,X^2 + g,
\label{eq:family}
\end{equation}
where \(C\ge 0\) controls the overall pinch and \(g\) labels individual streamlines.
Eliminating \(g\) yields a tangent slope in the model frame that is independent of \(g\):
\begin{equation}
m(X,Y) \equiv \frac{dY}{dX} \;=\; \frac{2\,C\,X\,Y}{C\,X^2 + 1}.
\label{eq:slope}
\end{equation}
The model-frame unit tangent in the image frame is
\begin{equation}
\widehat{\mathbf{t}}(\mathbf{r}\,|\,\alpha,C,\mathbf{r}_0)
\;=\;
\mathbf{B}\,
\frac{1}{\sqrt{1+m^2}}
\begin{bmatrix}1\\ m\end{bmatrix}.
\quad
\label{eq:tangent}
\end{equation}
Its model angle in our convention is \(\phi_{\rm mod}=\operatorname{tan}^{-1}(\widehat{t}_y,\widehat{t}_x)\), understood modulo \(180^\circ\).

Let the observed headless polarization orientation at \(\mathbf{r}_i\) be \(\phi_i\). Define unit segments
\(\widehat{\mathbf{u}}_i=(\cos\phi_i,\ \sin\phi_i)^\top\) and
\(\widehat{\mathbf{v}}_i=\widehat{\mathbf{t}}(\mathbf{r}_i)\).
The position angle residual in \((-90^\circ,90^\circ]\) writes
\begin{equation}
\Delta_i \;=\;
\operatorname{wrap}_{(-90,90]}\!\Big(
\operatorname{tan}^{-1}\big(\det[\widehat{\mathbf{u}}_i,\widehat{\mathbf{v}}_i],\ 
\widehat{\mathbf{u}}_i\!\cdot\!\widehat{\mathbf{v}}_i\big)
\Big),
\label{eq:residual}
\end{equation}
where \(\det[\mathbf{u},\mathbf{v}]=u_x v_y-u_y v_x\) and \(\operatorname{wrap}_{(-90,90]}\) folds by \(180^\circ\). We further estimate \(\Theta=\{\alpha,C,x_0,y_0\}\) by minimizing a robust loss over all pixels within mask \(i\):
\begin{equation}
\mathcal{L}(\Theta) \;=\; \sum^{\in\text{mask}}_i w_i\,\rho\!\left(\frac{\Delta_i}{\sigma}\right)
\ +\ \lambda\,\bigl[\log C\bigr]^2,
\label{eq:loss}
\end{equation}
where \(w_i\) are optional weights, \(\sigma\) sets the angular scale for robustness (we adopt \(\sigma\approx 10^\circ\), and \(\lambda\ll 1\) weakly regularizes \(C\). The second term is mainly for numerical stability when data are sparse or nearly straight. We optimize over \(\alpha\in[0, 180^\circ]\), \(\log C\in\mathbb{R}\), and vertex \((x_0,y_0)\). The loss in Eq.~(\ref{eq:loss}) is minimized. 

We estimate errors by bootstrap resampling of the fitted pixels: for each of \(N_{\rm boot}\) resamples (typically \(N_{\rm boot}=500\)), refit \(\Theta\) and quote 16 and 84 percentages as the $1\sigma$ credible intervals. Weight bootstrap by \(w_i\) if using S/N weights. The curvature of an individual member of parabolic curves at \(X=0\) is \(\kappa(0)=2|g|C\) (depends on \(g\)). The curvature radius is $R = 1/(2|g|C)$. 

\section{Smooth field removal}
\label{app:remove_smooth}

Let $(Q,U)$ denote the observed polarization maps, and $\mathcal{S}$ a Gaussian smoothing operator with a given kernel size $L$. The smoothed field is $(Q_0, U_0)$. So the residual Stokes parameters are then
\begin{equation}
Q' = Q - Q_0, \qquad
U' = U - U_0.
\end{equation}
The uniform field orientation and polarized intensity are obtained from
\begin{equation}
\psi_0 = \tfrac{1}{2}\,\mathrm{tan}^{-1}(U_0, Q_0), \qquad
P_0 = \sqrt{Q_0^2 + U_0^2}.
\end{equation}
The residual polarization angle which measures the 
local deviation from the smooth field, is computed through the Stokes inner product:
\begin{equation}
\Delta\psi = 
\tfrac{1}{2}\,
\mathrm{tan}^{-1}\!\big(UQ_0 - QU_0,\; QQ_0 + UU_0\big)
\;\in (-90^\circ, +90^\circ].
\end{equation}
The distribution of $\Delta\psi$ within a selected region (e.g.\ MM1) is then used to quantify the angular dispersion $\sigma_\phi$, which can directly enter into the DCF calculation.

\end{appendix}

\end{document}